\documentclass{ws-ijmpa}

\usepackage{bbm}

\newcommand{\UM}{\mathbbm 1}
\newcommand{\R}{\mathbbm R}

\newcommand{\Pl}{\mathbbm P}
\newcommand{\eqb}{\begin{equation}}
\newcommand{\eqe}{\end{equation}}
\newcommand{\dmb}{\begin{displaymath}}
\newcommand{\dme}{\end{displaymath}}
\newcommand{\pd}{\partial}

\newcommand{\eab}{\begin{eqnarray}}
\newcommand{\eae}{\end{eqnarray}}
\newcommand{\ra}{\right\rangle}
\newcommand{\la}{\left\langle}
\newcommand{\e}{\mbox{e}}
\newcommand{\be}{\begin{equation}}
\newcommand{\ee}{\end{equation}}

\newcommand{\La}{\Lambda}

\begin{document}

\markboth{Ralf Hofmann}{Nonperturbative approach to 
Yang-Mills thermodynamics}

%
\catchline{}{}{}{}{}
%

\title{Nonperturbative approach to 
Yang-Mills thermodynamics}

\author{Ralf Hofmann}

\address{Institut f\"ur Theoretische Physik\\ 
Johann Wolfgang Goethe -- Universit\"at\\ 
Max von Laue -- Str. 1\\ 
D-60438 Frankfurt am Main\\ 
Germany\\ 
r.hofmann@thphys.uni-heidelberg.de}

\maketitle

\begin{history}
\received{7 April 2005}
\revised{26 November 2006}
\end{history}

\begin{abstract}
An analytical and nonperturbative approach to SU(2) and SU(3) Yang-Mills thermodynamics is 
developed and applied. Each theory comes in 
three phases: A deconfining, a preconfining, and a 
confining one. We show how macroscopic and inert scalar fields emerge in each 
phase and how they determine the ground-state physics and 
the properties of the excitations. While the excitations in the deconfining and preconfining phase 
are massless or massive gauge modes of spin 1 the excitations in the confining phase 
are massless or massive spin-1/2 fermions. The nature of the two phase transitions 
is investigated for each theory. We compute the temperature evolution of 
thermodynamical quantities in the deconfining and 
preconfining phase, and we estimate the density of 
states in the confining phase. Some implications 
for particle physics and cosmology are discussed.

\end{abstract}

\ccode{PACS numbers: 11.15.Ex, 11.10.Wx, 11.15.Tk}

\section{Introduction\label{intro}}

The beauty, richness and usefulness of nonabelian gauge theories is 
generally appreciated. Yet, in a perturbative approach to gauge theories 
like the Standard Model of particle physics (SM) and its (non)supersymmetric extensions it is hard if not impossible 
to convincingly address a number of recent and not so recent experimental and 
observational results in particle physics and 
cosmology: Nondetection of the Higgs particle at LEP \cite{Higgs2000}, 
indications for a rapid thermalization and strongly 
collective behavior of the plasma that emerges in the early stage of an ultra-relativistic heavy-ion collision 
\cite{RHIC2003,ShuryakTeaney2001} despite the fact that the ideal hydrodynamical expansion essentially obeys 
a free-gas equation of state, dark energy and dark matter, a 
strongly favored epoch of cosmological inflation \cite{Cobe,WMAP2003I,Perlmutter1998,Schmidt1998} 
in the very early Universe \cite{Linde1982,Guth1982,Dymnikova} and today's 
accelerated cosmological expansion, a 
puzzling large-angle signal in the power spectrum of the 
cosmic microwave background for the cross correlation of 
electric-field polarization and temperature fluctuations \cite{WMAPPol}, 
the likely existence of intergalactic magnetic 
fields of so far unclarified origin \cite{Dai2002,IMF}, 
and the departure of $\sim 3\times 10^{43}$ protons p. a. (and hardly any negative charges) \cite{Manuel2004}
from the sun's surface (solar wind) which clearly is contradicting the 
charge conservation inherent in the SM. An analytical and 
nonperturbative approach to strongly interacting gauge 
theories may further our understanding of these phenomena. 

The objective of the present work 
is the thermodynamics of SU(2) and SU(3) Yang-Mills theories 
in four dimensions. It is difficult to gain 
insights in the dynamics of a strongly interacting 
four-dimensional gauge theory by analytical means if this dynamics is not severely 
constrained by certain global symmetries. 
We conjecture with Ref.\,\cite{Hagedorn1965} that a thermodynamical approach 
is an appropriate starting point for such an endeavor. 
On the one hand, this conjecture is reasonable since a strongly interacting 
system, being in equilibrium, communicates distortions almost instantaneously by 
rigid correlations, and thus a return to equilibrium takes place very rapidly. 
On the other hand, it turns out that 
the requirement of thermalization allows for an analytical and 
nonperturbative derivation of macroscopic ground states and the properties of their (quasiparticle) 
excitations in two of the three phases of each theory. A breakdown of equilibrium in a 
transition to the third phase is unproblematic since the dynamics 
then is uniquely determined by the remaining symmetry. 

Let us very briefly recall some aspects of the analytical 
approaches to thermal SU($N$) Yang-Mills theory as they are 
discussed in the literature. Because of asymptotic 
freedom \cite{GrossWilczek1973,Politzer1973} one would naively 
expect thermal perturbation theory to work well for temperatures 
$T$ much larger than the Yang-Mills scale $\La$ since the gauge coupling 
constant $\bar{g}(T)$ logarithmically approaches zero 
for $\frac{T}{\La}\to\infty$. 
It is known for a long time that this expectation is too optimistic since at 
any temperature perturbation theory 
is plagued by instabilities arising from the infrared sector (weakly screened, 
soft magnetic modes \cite{Linde1980}). As a consequence, the pressure $P$ can be computed 
perturbatively only up 
to (and including) order $\bar{g}^5$. The effects of resummations of 
one-loop diagrams (hard thermal loops), which rely on a 
scale separation furnished by the small value of 
the coupling constant $\bar{g}$, are summarized in terms of a 
nonlocal effective theory for soft and 
semi-hard modes \cite{BraatenPisarski1990}. In the computation of radiative corrections, 
based on this effective theory, infrared effects due to soft magnetic modes still appear in an 
uncontrolled manner. This has 
led to the construction of an effective theory where soft modes are collectively described in 
terms of classical fields whose dynamics is influenced by integrated 
semi-hard and hard modes \cite{Bodeker1998,Bodapps1998}. 
In Quantum Chromodynamics (QCD) a perturbative calculation of $P$ was pushed up to order 
$\bar{g}^6\log\,\bar{g}$, and an additive `nonperturbative' 
term at this order was fitted to lattice results \cite{Kajantie2002}. Within the 
perturbative orders a poor convergence of the expansion is observed for 
temperatures not much larger than the $\overline{\mbox{MS}}$ scale. While 
the work in \cite{Kajantie2002} is a computational masterpiece it could, by definition, 
not shed light on the missing, nonperturbative 
dynamics of the infrared sector. Screened perturbation theory, which relies 
on a split of the tree-level Yang-Mills action by the introduction of 
variational parameters, is a very interesting idea. Unfortunately, this approach generates 
temperature dependent ultraviolet divergences in its presently used 
form, see \cite{Blaizot2003} for a recent review. 

The purpose of the present work is to report, in a detailed way, on the development of 
a nonperturbative and analytical approach to the 
thermodynamics of SU(2) and SU(3) Yang-Mills theory 
(see \cite{Hofmann2000t2003} for intermediate stages). The reason why we consider only these 
two gauge groups is that for SU($N$) with $N\ge 4$ it is likely that the 
phase structure of the theory is not unique: In contrast to SU(2) and SU(3), 
which possess one confining (center), one
preconfining (magnetic), and one deconfining (electric) phase, more 
than three phases may exist for an SU($N$) Yang-Mills theory with $N\ge 4$. 

Our starting point is the derivation of a macroscopic ground state in 
the deconfining phase. This ground state originates 
from instantaneous, long-range correlations mediated by field configurations of 
topological charge $\pm 1$. Technically, this situation 
is described by a spatially homogeneous, quantum mechanically 
and statistically inert scalar field $\phi$, which transforms 
under the adjoint representation of the gauge group, and a 
pure-gauge configuration of trivial topology solving 
the Yang-Mills equations subject to a source term provided 
by $\phi$. Both the field $\phi$ and the 
pure-gauge configuration emerge after a spatial 
coarse-graining over quantum fluctuations 
down to a resolution corresponding to 
the length scale $|\phi|^{-1}$. 
While $\phi$ represents the spatial average over BPS saturated, 
topological defects, that is, `large' quantum fluctuations the 
pure-gauge configuration is a manifestation of averaged-over 
plane-wave quantum fluctuations.    

\begin{figure}
\begin{center}
\leavevmode
\leavevmode
\vspace{4.3cm}
\includegraphics{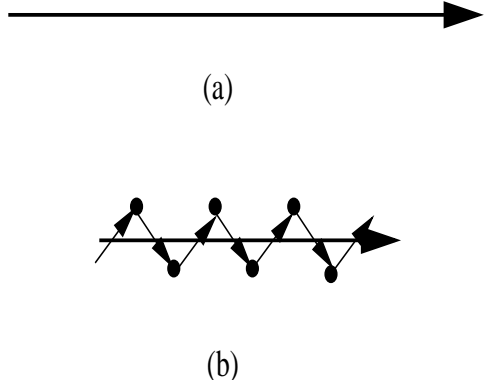}
\end{center}
\caption{\protect{Photon propagation without (a) and within (b) a 
Cooper-pair condensate.\label{Fig0}}}      
\end{figure}

Conceptually, our approach is similar to the 
macroscopic Landau-Ginzburg-Abrikosov theory for 
superconductivity in metals \cite{GinzburgLandau1950,Abrikosov1957}. 
Recall that this theory describes 
the existence of a condensate of Cooper 
pairs in terms of a nonvanishing expectation for a 
complex scalar field $\varphi$ (local order parameter), 
which is charged under the electromagnetic gauge group U(1), and in terms of a 
pure-gauge configuration. A nonzero value of $\varphi$ is enforced by a 
phenomenologically introduced potential $V$. 
As a consequence, coarse-grained U(1) gauge-field modes $\delta a_\mu$ (photons), 
which are deprived of the 
microscopic gauge-field fluctuations contributing to 
the formation of Cooper pairs and their subsequent condensation, 
acquire a mass. Microscopically, the generation of a photon mass can be 
visualized as a large sequence of elastic scattering processes off the 
electrons residing within individual Cooper pairs in the condensate. At a 
given photon momentum this slows down the effective velocity of propagation in comparison 
to a propagation without a Cooper-pairs condensate, see Fig.\,\ref{Fig0}. 
In the superconducting phase the U(1) gauge symmetry is 
spontaneously broken by the Cooper-pair condensate, 
and physical phenomena associated with this breakdown 
can be analysed in dependence of the parameters entering an 
effective action and in dependence of external conditions such as a 
magnetic field and/or temperature. 

When pursuing the idea of a dynamically generated, macroscopic 
ground state in each of the phases of an SU(2) or SU(3) Yang-Mills 
theory it turns out that a situation similar to superconductivity holds. Moreover, in a 
thermalized SU(2) or SU(3) Yang-Mills 
theory one is in the comfortable position of being able to {\sl derive} the dynamics 
of macroscopic scalar fields from first principles. Thus the 
stabilizing potentials for each scalar field are uniquely determined (up to an undetermined mass parameter -- 
the scale of the Yang-Mills theory). Each (gauge invariant) 
potential is specified by a unique microscopic 
definition of the scalar field's phase (in a suitably chosen gauge) 
and by the assumption that a dynamically generated, constant mass scale exists. 
This assumption is supported by 
perturbation theory \cite{GrossWilczek1973,Politzer1973} 
where the specification of the running of the 
gauge coupling requires such a boundary condition. The microscopic 
definition of the scalar field's phase is an average 
over an (nonlocal) operator saturated by noninteracting Bogomolnyi-Prasad-Sommerfield (BPS)\cite{PrasadSommerfield1975} 
saturated configurations of 
topological charge $\pm 1$. If a particular phase supports 
propagating gauge modes then, in a second step, 
interactions between the topological defects and microscopic 
radiative corrections are taken into account macroscopically by pure-gauge 
solutions to the equations of motion for gauge fields residing in the 
trivial-topology sector. The source term for these equations of motion is provided by 
the (inert) scalar fields.

More specifically, we have shown in \cite{HerbstHofmann2004} 
that at large temperatures upon spatial coarse-graining 
an adjoint scalar field $\phi$ emerges due to the spatial correlations mediated by 
trivial-holonomy calorons\cite{HarrigtonShepard1977}. (By large temperature 
we mean large as compared to the dynamically generated scale.) 
We discuss in Sec.\,\ref{eveffgc} why the critical 
temperature $T_P$ for the onset of $\phi$'s existence should be comparable to 
the cutoff-scale for a field theoretic description in four dimensions.
Trivial-holonomy calorons are BPS saturated (or selfdual) solutions \cite{PrasadSommerfield1975}
to the Euclidean Yang-Mills equations at finite temperature. (The time coordinate $\tau$ is 
compactified on a circle, $0\le\tau\le\frac{1}{T}$.) The topological charge of these 
configurations is integer. (Whenever we speak of a 
topological soliton in this section its antisoliton is also meant.) 
It will turn out that only calorons with topological charge 
one contribute to the moduli-space average defining $\phi$'s phase. The reason is that 
for higher-charge calorons the larger number of dimensionful moduli does not admit the 
definition of a dimensionless entity without the introduction of 
an explicit temperature dependence \cite{HerbstHofmann2004}. The latter, however, ought 
to be absent because of a temperature independent weight on the 
classical level. To understand macroscopic results generated by microscopic 
interactions between calorons and between their constituents an investigation 
of the properties of quantum corrected nontrivial-holonomy calorons 
is necessary \cite{Diakonov2004}.   

For a given SU($N$) gauge-field configuration nontrivial holonomy refers to the 
following property: Evaluate the Polyakov-loop on this configuration 
at spatial infinity and observe that the result is not an element 
\eqb
\label{noho}
\UM\,\exp\left[\frac{2\pi i}{N}k\right]\,\ \ \ (k=0,\cdots,N-1)
\eqe
of the center $Z_N$ of the gauge group. If this is the case then 
a mass scale exists in the configuration which determines the behavior 
$A_4(|\vec{x}|\to\infty)$. For a classical solution at finite temperature this 
mass scale must be temperature itself. The quantity $u\equiv T\int_0^\beta d\tau\,A_4(\tau,|\vec{x}|\to\infty)$ defines 
the holonomy of the configuration ($\beta\equiv\frac{1}{T}$). 
Due to a reflection symmetry $u\to -u$ one can restrict the values of $u$ as 
$0\le u\le 2\pi T$ for SU(2).         

Calorons with nontrivial holonomy are selfdual configurations that possess BPS magnetic monopole constituents 
\cite{Nahm1984,KraanVanBaalNPB1998,vanBaalKraalPLB1998,LeeLu1998,Brower1998}: 
For an SU($N$) caloron with no net magnetic charge 
there are $N$ constituent monopoles whose magnetic 
charges add up to zero. The masses of the monopoles are 
determined by the holonomy $u$ and thus are $\propto T$. 
By a recent heroic calculation the one-loop quantum weight for an SU(2) caloron 
with nontrivial holonomy was derived in \cite{Diakonov2004}.  

Since the one-loop effective action of a nontrivial-holonomy caloron scales 
with the spatial volume of the 
system one is tempted to conclude that these configurations do not contribute 
to the partition function of the theory in the 
thermodynamical limit and thus are irrelevant 
\cite{GrossPisarskiYaffe1981}. This conclusion, 
however, is not valid since one can show that 
nontrivial-holonomy calorons are {\sl instable} under 
quantum corrections\cite{Diakonov2004}. Moreover, if, on spatial average, interacting calorons are shown to be described by a 
quantum mechanically and statistically inert, 
{\sl macroscopic} adjoint scalar field $\phi$ and 
a pure-gauge configuration $a_\mu^{bg}$ \cite{HerbstHofmann2004} 
then the exponent of minus the associated effective action $S_{cl}$ can be factored out in the 
partition function and thus cancels in any physical average. 
This situation holds even if $S_{cl}$ scales with the spatial 
volume of the system and thus is infinite in the thermodynamical limit. On the microscopic level, the generation 
of (instable) nontrivial holonomy is due to interactions between 
trivial-holonomy calorons mediated by long-wavelength fields that 
reside in the topologically trivial sector of the theory.    

For the SU(2) case it was shown in \cite{Diakonov2004} that the one-loop fluctuations around 
calorons with a small holonomy generate an
{\sl attractive} potential between the two BPS monopole 
constituents. This implies that monopole and antimonopole 
annihilate shortly after they have been created. 
Thus the likelihood of such a process roughly is determined by the finite 
one-loop quantum weight of a trivial-holonomy caloron. Up to an additive correction, which depends on 
temperature and the caloron radius $\rho$ and which is finite, 
the effective action equals the classical action $S=\frac{8\pi^2}{g^2}$. 
For $g$ not too small the likelihood of generating a 
small holonomy is sizable. In the opposite case 
of a large holonomy a {\sl repulsive} potential between the 
monopole constituents arises due to one-loop fluctuation 
\cite{Diakonov2004}. Thus monopole and antimonopole 
separate back-to-back, and the caloron dissociates into a pair of an isolated monopole and an isolated 
antimonopole which are screened by intermediate caloron fluctuations 
\cite{KorthalsAltes,HoelbingRebbiRubakov2001}. Before 
screening, that is, on the level of the 
classical solution the mass of both monopole and 
antimonopole is much larger than 
temperature \cite{LeeLu1998}. We conclude that the generation and subsequent 
dissociation of a large-holonomy caloron is a very 
rare process due to an extreme Boltzmann suppression. Thus attraction between a monopole and its 
antimonopole is the dominating situation in the ground-state physics of an SU(2) or SU(3) Yang-Mills 
theory being in the electric phase. 
The macroscopic manifestation of monopole-antimonopole attraction, their subsequent
annhihilation, and recreation is a {\sl negative} 
ground-state pressure. Equating the exponent in 
the Boltzmann distribution of the monopole-antimonopole system before screening 
with the exponent in the one-loop quantum weight of the caloron allows for an estimate of the typical distance 
between a monopole and an antimonopole at 
a given temperature. We will see that, on the scale of 
the inverse temperature, isolated and screened 
monopoles are very dilute. The case of SU(2) has a 
straight-forward generalization to SU(3): No qualitative changes 
take place in the above discussion when going from SU(2) to SU(3).
  
From the selfduality or BPS saturation of the caloron it follows 
that its energy-momentum tensor vanishes identically. Since the 
macroscopic field $\phi$ originates from the spatial correlations of noninteracting 
trivial-holonomy calorons of topological charge 
one \cite{HerbstHofmann2004} $\phi$'s 
macroscopic energy-momentum tensor vanishes in the absence of a 
coupling to the topologically trivial sector 
of the theory. This is a derived condition which needs to be 
imposed to fix some of the ambiguities which emerge when calculating $\phi$'s phase. Namely, one insists 
on a BPS saturation of the $\tau$ dependence 
of this phase: A linear 
second-order equation of motion for $\phi$'s phase can be derived 
from a microscopic definition involving a moduli-space average over a 
two-point correlator, and the requirement of BPS saturation determines the 
solution up to an irrelevant global gauge choice and an 
irrelevant constant phase shift \cite{HerbstHofmann2004}. 

Subsequently, a potential 
$V_E$ for the canonically normalized field $\phi$ is derived by appealing to the derived 
information on $\phi$'s phase and to the assumptions 
that a dynamically generated scale $\La_E$ exists and that 
the right-hand side of $\phi$'s BPS equation 
is analytic in $\phi$. (Here the 
subscript $E$ refers to the electric phase.) 
As a consequence, $\phi$'s modulus is $\propto T^{-1/2}$ and $V_E\propto T$. 
Thus the caloron sector contributes to thermal quantities 
in a power-suppressed way at large temperatures. Moreover, 
we will show that the ground state in the electric phase is degenerate with 
respect to a global electric $Z_2$ (SU(2)) and a global electric $Z_3$ (SU(3)) 
symmetry. Therefore, the electric phase is {\sl deconfining}. 

In the coarse-grained theory a useful decomposition of 
field configurations $a_\mu$ with trivial topology (only those 
ones appear as explicit gauge fields) is       
\eqb
\label{fludec}
a_\mu=a_\mu^{bg}+\delta a_\mu\,
\eqe
where $a_\mu^{bg}$ denotes a pure-gauge configuration 
belonging to the ground state, and $\delta a_\mu$ 
is a finite-curvature fluctuation. In unitary gauge, where $a_\mu^{bg}=0$ and 
thus no coupling between $\delta a_\mu$ and $a_\mu^{bg}$ 
exists, a fluctuation $\delta a_\mu$ acquires a mass by the adjoint Higgs mechanism 
if $[\phi,\delta a_\mu]\not=0$. Since the field $\phi$ dynamically breaks 
the gauge symmetries SU(2)$\to$U(1)and SU(3)$\to$U(1)$^2$, two and six directions in the three and 
eight dimensional adjoint color space acquire mass, respectively. 

We shall discuss 
why the scale $\La_E$, which measures the strength 
of apparent gauge-symmetry breaking by calorons at a given temperature, is physically 
set at a temperature scale $T_P$ where any four-dimensional 
gauge theory fails to describe reality. Common belief is that $T_P$ is comparable to 
the Planck mass $M_P\sim 10^{19}\,$GeV. 

How does the existence of the temperature dependent 
scale $|\phi|$ influence the 
propagation of gauge modes in the 
infrared and ultraviolet? Calorons induce
quasiparticle masses on tree level 
in the effective theory which are 
of the order $e|\phi|$. Here $e$ denotes 
the {\sl effective} gauge coupling. This coupling measures the 
interaction strength between the topologically trivial 
off-Cartan fluctuations (in unitary gauge) and the coarse-grained 
manifestation of nontrivial topology. Furthermore, it is a measure for 
the screening of the magnetic charge of a monopole. (As far as thermodynamical quantities are concerned, essentially 
all excitations are free (quasi)particles \cite{HerbstHofmannRohrer2004}.). 
The evolution of $e$ with temperature follows from the 
requirement of thermodynamical selfconsistency 
of this interaction. Except for a small range in temperature 
to the right of the electric-magnetic transition
at $T_{c,E}$, where 
\eqb
\label{coupldiv}
e(T)\sim -\log(T-T_{c,E})\,,
\eqe
the coupling $e$ is constant: A manifestation of the existence 
of screened\footnote{The screening of the monopoles is due to surrounding 
caloron fluctuations of small holonomy and not due to Cartan {\sl excitations} 
since the latter are not capable of screening 
static magnetic fields \cite{Linde1980}.}, isolated and conserved magnetic charges generated 
by dissociating large-holonomy calorons. 

Infrared cutoffs $\sim e|\phi|$ arise 
from a reduction of propagation speed for off-Cartan fluctuations by their 
interactions with calorons, compare with the analogous situation for a superconducting material in 
Fig.\,\ref{Fig0}. Due to the existence of these cutoffs in the loop expansions of 
thermodynamical quantities the problem of a 
magnetic instability, as encountered in perturbation theory \cite{Linde1980}, is resolved 
\cite{Linde1980,HerbstHofmannRohrer2004}. 
In the ultraviolet, $|\phi|$ acts as a compositeness constraint by setting a maximal 
scale for the off-shellness of quantum fluctuations. This is a 
consistent requirement since {\sl all} gauge modes, on-shell or off-shell, 
originate from the nontrivial ground state and thus ought 
not be capable of destroying it. Moreover, plane-wave quantum fluctuations of an off-shellness 
larger than $|\phi|^2$ are contained, in a coarse-grained form, 
in the pure-gauge configuration $a_\mu^{bg}$. Compositeness constraints are 
implemented in a physical gauge with respect to the unbroken 
subgroups U(1) or U(1)$^2$. The usual renormalization 
program, needed to make sense of ultraviolet 
divergences in thermal perturbation theory, is 
abandoned in the effective theory: The ground state itself provides 
for a {\sl physical} ultraviolet cutoff. 

As a consequence of the simultaneous existence of both an 
ultraviolet and an infrared cutoff the contributions of higher loops in the 
expansion of thermodynamical quantities are very small. 
(Technically, an evaluation of two-loop corrections to the pressure is 
quite involved \cite{HerbstHofmannRohrer2004}.) Obviously, the situation outlined so far differs 
substantially from the idea of a Wilsonian flow for 
nonabelian gauge theories in its usual 
implementation: One derives effective dynamics in dependence 
of an externally set scale\footnote{This scale either is continuous, 
see \cite{LitimPawlowski1999} for a review on gauge theories, or it reflects a scale 
hierarchy originating from the assumed smallness of the 
coupling constant $\bar{g}$ at a large temperature $T$, for example $k=\bar{g}T$ 
\cite{BraatenPisarski1990}.} $k$ 
by integrating plane-wave modes harder than $k$ into couplings 
that appear in an ansatz for an effective action. This effective 
action describes the dynamics of gauge 
modes of maximal hardness $k$. The dynamics of theses modes is, 
however, not only influenced by integrated-out 
high-momentum fluctuations but also by (spatially) small-scale 
fluctuations of nontrivial topology. Recall that the later can 
not be expanded in terms of the former because of an 
essential singularity in their weight at a vanishing value of 
the fundamental gauge coupling. Thus we propose an approach which is the 
converse of the usual picture: Integrate the 
topological sector first and determine subsequently, that is, after spatial coarse-graining, 
what its average effect on the trivial-topology fluctuations is. 

We have already mentioned that this approach to SU(2) or SU(3) Yang-Mills thermodynamics 
implies the existence of three rather than two phases. 
In the magnetic phase, where the isolated and 
screened magnetic monopoles of the electric phase are massless and thus 
condensed and where off-Cartan modes are thermodynamically 
decoupled, the dual gauge symmetries U(1)$_D$ or U(1)$^2_D$ are broken 
dynamically, and the global electric center 
symmetries $Z_2$ or $Z_3$ are restored in the ground state. We will show that 
the ground state in the magnetic phase is a Bose condensate 
of monopole-antimonopole systems. Each condensate is described in terms of a macroscopic and inert 
complex scalar field and a pure-gauge configuration. 

A monopole condensate confines fundamentally 
charged, fermionic and static test charges. At the same 
time, dual and massive gauge modes {\sl propagate} in the magnetic phase. 
Thus it is appropriate to refer to the magnetic phase as a preconfining phase. 
The magnetic coupling $g$, 
which measures the interaction strength between dual gauge modes 
and condensed magnetic monopoles on the one hand and the screening 
of center-vortex loops on the other hand, is zero at $T_{c,E}$ and 
rises rapidly into a logarithmic divergence of the same 
form as in Eq.\,(\ref{coupldiv}) at a 
temperature $T_{c,M}$. For SU(2) and SU(3) we 
have $T_{c,M}=0.835\times\,T_{c,E}$ and 
$T_{c,M}=0.877\times\,T_{c,E}$, respectively. Thus the magnetic 
phase occupies only a small region in the phase diagram of either theory. This and the fact 
that the monopole condensates possess infinite correlation 
lengths $\sim \mbox{(monopole mass)}^{-1}$ are the reasons 
why the magnetic phase has escaped its direct 
detection by simulations on finite-size lattices. We will 
show though that it is possible to observe the existence of the 
magnetic phase in a lattice simulations of the infrared 
insensitive entropy density when using the 
so-called differential method. The latter provides the 
best-controlled circumvention of the infrared problem in simulations on finite-size 
lattices \cite{Brown1988}.   

\begin{figure}
\begin{center}
\leavevmode
\leavevmode
\vspace{4.3cm}
\includegraphics{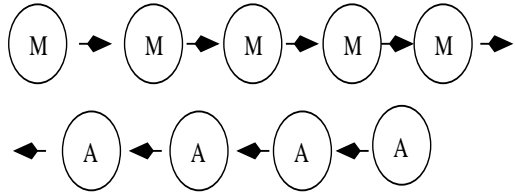}
\end{center}
\caption{Microscopics of the core of a center vortex. Monopoles (M) and antimonopoles (A) move in 
opposite directions.\label{Fig0b}}      
\end{figure}

At $T_{c,M}$, where $g$ diverges logarithmically, 
another phase transition takes place. All dual gauge modes 
decouple and thus the monopole condensates, macroscopically described by 
nonfluctuating, BPS saturated complex scalar fields $\varphi$ (SU(2)) and 
$\varphi_1,\varphi_2$ (SU(3)) and pure gauges, dominate the 
thermodynamics. As a consequence, the 
entropy density vanishes at $T_{c,M}$ and the 
equation of state is 
\eqb
\label{eosintro}
\rho=-P\,.
\eqe
Just like magnetic monopoles are isolated defects 
in the electric phase there are isolated and closed 
magnetic flux lines in the magnetic phase\footnote{The fact that 
only closed loops occur is explained by the absence of isolated 
magnetic charges in a monopole condensate.} which, 
however, collapse as soon as they are created if the magnetic coupling $g$ is finite: 
Abrikosov-Nielsen-Olesen (ANO) vortex loops \cite{NielsenOlesen1973}. 
Only one unit of flux with respect to U(1)$_D$ (SU(2)) or either factor 
in U(1)$^2_D$ (SU(3)) is carried by a 
given vortex loop since in the electric phase 
only charge-one calorons dissociate into magnetic monopoles 
with one unit of magnetic charge\footnote{The core of a vortex line, where U(1)$_D$ (SU(2)) or one factor in 
U(1)$^2_D$ (SU(3)) is restored, can be pictured as a directed motion of 
magnetic monopoles (to the right) and antimonopoles (to the left) 
in the rest frame of the heatbath\cite{Olejnik1997}, see Fig.\,\ref{Fig0b}. The magnetic flux, 
which penetrates a spatial 
hyperplane perpendicular to 
the direction of monopole or 
antimonopole motion, is by Stoke's theorem 
measured by a line integral $g\oint_{{\cal C}} dz_\mu A^D_{\mu}$ along a circular 
curve ${\cal C}$ with infinite radius lying in this plane. Here $A^D_\mu$ 
denotes the gauge field with respect to the dual gauge group U(1) (SU(2)) or either factor 
in U(1)$^2$ (SU(3)) generated by the moving chains of 
monopoles and antimonopoles. If we choose to evaluate the 
line integral in a covariant gauge then the contribution to $dz_\mu A^D_\mu$ 
of each moving monopole or antimonopole is that of a 
static monopole or antimonopole since the perpendicular part 
of the gauge field is invariant under boosts 
along the vortex axis. Thus the state of 
motion of each monopole and antimonopole 
is irrelevant for its effect on the total magnetic flux 
carried by the vortex as long as the 
net motion of all monopoles and antimonopoles in a given segment 
defines the direction of the vortex axis: The only thing that determines 
the magnetic flux of the vortex line is the {\sl charge} of a monopole.}.  
This allows for an interpretation of ANO vortex loops as center-vortex loops. For SU($N$) 
the magnetic flux of the latter is determined by the differences 
in phase modulo $N$ of two center elements, see Eq.\,(\ref{noho}). There is one unit of center 
flux for SU(2), and there are two separate units of center flux for SU(3). 

The core-size of a center vortex is determined 
by the length $l_g\sim (g|\varphi|)^{-1}$ (SU(2)) or 
$l_{g,1}\sim (g|\varphi_1|)^{-1},l_{g,2}\sim (g|\varphi_2|)^{-1}$ (SU(3)) of penetration of the vortex' gauge 
field into a direction perpendicular to the vortex. While center vortices are 
thick close to $T_{c,E}$ their core size vanishes at $T_{c,M}$. Since the energy of a typical 
center-vortex loop\cite{NielsenOlesen1973} is $\propto g^{-1}$ spin-0 systems composed of a vortex loop and its 
antivortex loop condense at $T_{c,M}$. We will show that there is a parameter with discrete values 
characterizing the possible values of a macroscopic, 
complex scalar field $\Phi$ which describes the vortex condensate. 

The transition to the center phase is 
of the Hagedorn type and thus nonthermal. An order parameter for this transition is the 
expectation of the 't Hooft loop whose modulus measures the 
strength of center-vortex condensation. If the 't Hooft 
loop does not vanish then the {\sl magnetic} $Z_2$ (SU(2)) or $Z_3$ (SU(3)) 
symmetry is dynamically broken. These center symmetries are 
{\sl local} in four-dimensional spacetime. Under large U(1)$_D$ (SU(2)) 
or large  U(1)$^2_D$ (SU(3)) gauge transformations 
the macroscopic field $\Phi$ transforms by multiplications 
with center elements. 

In the course of the Hagedorn transition the ground state of the magnetic 
phase decays through creation of single and 
self-intersecting center-vortex loops. In contrast to the magnetic phase, 
center-vortex loops are stable in the center phase, thus are 
particle-like, and possess a density of states which is 
over-exponentially rising with energy. 
There are precisely two polarization states for each self-intersecting or single 
loop (spin-1/2 fermions). After the decay of the monopole condensate is completed 
the new ground state is a Cooper-pair condensate of systems composed of a 
massless single vortex and antivortex 
loop. The energy-momentum tensor on this ground state {\sl vanishes identically}. This result 
is protected against radiative corrections. 

The center phase is truly confining: A pair of electric, static, oppositely and fundamentally charged, and 
fermionic test charges forms a confining electric flux tube because of 
the presence of condensed electric dipoles 
(single center-vortex loops) in the 
ground state, and all gauge modes are infinitely 
heavy. Notice that the absence of propagating gauge modes implies that 
only contact interactions are possible between (thin) center fluxes and the 
self-intersection points of their vortex loops. 

Thermal lattice simulations fail to produce physical results for infrared sensitive quantities 
at temperatures shortly below $T_{c,E}$: The center 
phase as well as the magnetic phase possess infinite correlation lengths. In the center phase 
this correlation length is given by the inverse (vanishing) mass of a single center-vortex loop. Thus the fermionic 
nature of excitations and their over-exponentially rising density of states in the 
center phase escapes simulations performed on finite-size lattices.         

It is self-evident that what was said above has implications 
for particle physics and cosmology. We only would like to 
mention a few questions that are likely to 
be answered by SU(2) Yang-Mills (thermo)dynamics alone: 
Electroweak symmetry breaking, namely, 
the origin of the masses of $Z_0$ and $W^\pm$; the mass hierarchy 
between the two members of a lepton family; the question of whether 
the neutrino is Dirac or Majorana; the smallness of the 
cosmological constant on particle physics scales; the nature of 
cosmological and clustering dark matter; cosmic coincidence; baryon and lepton asymmetries; 
intergalactic magnetic fields; and large-angle 
anomalies in some of the power spectra of fluctuations in the cosmic 
microwave background. 

The article is organized as follows. In the first part of Sec.\,\ref{EP} 
we provide prerequisites on caloron physics. The classical solutions of trivial and nontrivial holonomy 
and their behavior under quantum corrections are discussed for SU(2). As an aside we 
estimate the separation of isolated and screened monopoles in terms of 
the inverse temperature in the electric phase. The second part of Sec.\,\ref{EP} 
is devoted to the derivation of the macroscopic, adjoint 
field $\phi$ in terms of a moduli-space and $S_3$ integral over the spatial two-point 
correlations in a caloron-anticaloron system. Subsequently, we discuss the full ground-state physics 
in the electric phase. The existence of a dynamically generated scale $\La_E$ needs to be assumed 
to determine $\phi$'s modulus at a given temperature. We show that a degeneracy 
with respect to a global electric 
$Z_2$ (SU(2)) and $Z_3$ (SU(3)) symmetry occurs which proves that 
this phase is deconfining. The last part of Sec.\,\ref{EP} addresses the full 
thermodynamics of the electric phase, including excitations. An evolution equation 
for the effective gauge coupling $e$ is derived and solved numerically. It is observed 
that an ultraviolet-infrared decoupling is manifest in this evolution. 
We present analytical results for the electric screening mass, associated 
with the massless mode in the SU(2) case, and for the two-loop 
correction to the pressure.            

In Sec.\,\ref{MP} we investigate the magnetic phase. Prerequisites on 
the BPS monopole are given. Considering the average magnetic flux, generated by a 
screened monopole-antimonopole system in a thermal environment, 
through an $S_2$ with infinite radius, where the monopole-antimonopole system is 
located outside of this $S_2$, a continuous parameter (proportional to the Euclidean time) 
is derived. This parameter governs 
the temporal winding of a spatially homogeneous, complex, and 
inert scalar field $\varphi$ in the limit where monopole mass and spatial 
momentum of the system vanish (Bose condensation). For SU(3) two such 
fields exist, each for every independent monopole species. Assuming the existence 
of a dynamically generated scale $\La_M$, the modulus of $\varphi$ is derived. We then discuss 
the ground-state physics in the magnetic phase. It is shown that the electric $Z_2$ (SU(2)) and electric 
$Z_3$ (SU(3)) degeneracy of the ground state, as observed in the electric phase, gives 
way to a unique ground state in the magnetic phase. Thus we derive test-charge 
confinement in the magnetic phase. Evolution equations for the effective 
magnetic coupling $g$ are derived and solved numerically. A discussion of the full 
Polyakov-loop expectation (including excitations) and an analysis of the critical behavior at the 
electric-magnetic phase boundary are presented. We find that this transition is 
second order with mean-field exponents both for SU(2) and SU(3) with the difference that the 
peak in the specific heat is about three times smaller in the former as compared to the latter case.

In Sec.\,\ref{CVCM} we investigate the center phase. We start by providing prerequisites 
on the ANO vortex. In particular, we emphasize the fact that an ANO vortex generates 
negative pressure at finite magnetic coupling $g$. While a vortex-loop is a particle-like 
excitation at $g=\infty$ it collapses as soon as it is created for $g<\infty$. Collapsing ANO or 
center-vortex loops dominate the (negative) pressure inside the 
magnetic phase where $g<\infty$. In analogy to the monopole condensate we determine a parameter (mean center 
flux through an $S_1$ of infinite radius) which governs the expectation of a macroscopic, 
complex scalar field  $\Phi$ describing the Bose condensate 
of massless vortex-antivortex-loop pairs. The values of this parameter are discrete: Two possible values 
for SU(2) and three possible values for SU(3). At $T_{c,M}$, where $g$ diverges in a logarithmic
way and where dual gauge modes acquire an infinite mass, the center-vortex 
condensate starts to form under (spin-1/2) 
particle creation. We construct an effective potential $V_C$ for $\Phi$ involving a scale $\La_C$. We 
check $V_C$'s uniqueness, and 
discuss how vortex-loop creation takes place by center jumps of $\Phi$'s phase. An estimate for the 
density of fermion states, created by $\Phi$'s relaxation to zero energy density and 
pressure, is provided. As a result, we analytically establish 
that the center-magnetic transition is of the Hagedorn type. 

In Sec.\,\ref{MCC} we discuss how the scales $\La_E$ and $\La_M$ are related 
by the continuity of the pressure across the electric-magnetic phase boundary. We also provide 
an approximate relation between $\La_M$ and $\La_C$. 

In Sec.\,\ref{PEEN} we present numerical results for the temperature dependence 
of thermodynamical quantities thoughout
the electric and the magnetic phase. The following quantities are 
discussed: Pressure, energy density, interaction measure,
specific heat per volume, and entropy density. While the former quantities are very 
sensitive to the ground-state physics at low temperatures, which is determined by 
very large spatial correlations and thus is inaccessible to finite-size lattices, 
the entropy density is only 
sensitive to the excitations. This fact makes a comparison of our results for 
the entropy density with those obtained on lattices 
(employing the differential method) useful, all other quantities exhibit quantitative 
disagreements with their lattice-obtained values at low temperatures. 

In Sec.\,\ref{Apps} we discuss implications of our work 
for particle physics and cosmology. We start by addressing the 
cosmic-coincidence and the old cosmological-constant 
problem in view of a Planck-scale axion, originating from dynamically generated and subsequently 
integrated spin-1/2 fermions at the Planck scale, and in view of an SU(2) gauge theory of 
Yang-Mills scale close to the temperature of the cosmic microwave background (CMB) 
(SU(2)$_{\tiny\mbox{CMB}}$). This theory is at the electric-magnetic 
phase boundary, and its only massless and unscreened excitation is the photon. Throughout cosmological 
evolution the axion mass is provided by the axial anomaly involving nonconfining 
SU(N) gauge theories of Yang-Mills scales lower than the Planck scale. The presence of 
a Planck-scale axion may explain the particle-number asymmetries and 
CP violation in the weak interactions. Some of the phenomenology of the 
electroweak sector of the SM is addressed in view of leptons being 
stable solitons in the center phase of various 
SU(2) Yang-Mills theories. These solitons are embedded into instable higher-charge excitations with 
an over-exponentially rising density of states which protect their apparent 
structurelessness seen in scattering experiments 
(the photon couples to the lepton because of mixing) 
up to center-of-mass energies comparable to the mass 
of the $Z$ boson (with exceptions at momenta comparable to the lepton masses). We postdict the mass ratio 
$\frac{m_{\nu_e}}{m_e}$ in terms of the mass ratio $\frac{m_{e}}{m_Z}$. 
Finally, we present some ideas on how fractionally charged light 
quarks and their confinement may arise in Quantum Chromodynamics (QCD) in terms 
of electric-magnetically dual SU(3) gauge dynamics and the 
fractional Quantum Hall effect. 

The last section of the present work briefly summarizes our results.

\section{The electric phase\label{EP}}

\subsection{Prerequisites}

\subsubsection{The Harrington-Shepard solution (trivial holonomy)\label{HSS}}

Calorons of trivial holonomy are the field configurations which 
enter the definition of the phase of the macroscopic adjoint scalar 
field $\phi$. We only need to consider the SU(2) 
case since the SU(3) ground-state thermodynamics can be 
derived from a 'democratic' embedding of SU(2) calorons. We use the nonperturbative 
definition of the gauge field where the coupling 
constant is absorbed into the field. 

(Anti)Calorons are (anti)selfdual, that is, their 
field strength is up to a sign equal to 
their dual field strength
\eqb
\label{BPSCal}
F_{\mu\nu}[A^{(C,A)}]=\pm\tilde{F}_{\mu\nu}[A^{(C,A)}]\,
\eqe
where the superscript $C(A)$ refers to caloron (anticaloron). 
Only calorons of topological charge one (minus one) 
enter the definition of $\phi$'s phase and 
thus we will focus on this case only. 

The Harrington-Shepard solutions \cite{HarrigtonShepard1977} are given as  
\eab
\label{HS}
A^C_\mu(\tau,\vec{x})&=&\bar{\eta}_{a\mu\nu}\frac{\lambda^a}{2}\pd_{\nu}\ln \Pi(\tau,\vec{x})\,\ \ \ \mbox{or}\nonumber\\ 
A^A_\mu(\tau,\vec{x})&=&\eta_{a\mu\nu}\frac{\lambda^a}{2}
\pd_{\nu}\ln \Pi(\tau,\vec{x})\,
\eae
where the 't Hooft symbols $\eta_{a\mu\nu}$ and $\bar{\eta}_{a\mu\nu}$ 
are defined by
\eab
\label{tHooftsym}
\eta_{a\mu\nu} &=& \epsilon_{a\mu\nu} + \delta_{a\mu}\delta_{\nu4} - \delta_{a\nu}\delta_{\mu4} \nonumber\\ 
\bar\eta_{a\mu\nu} &=& \epsilon_{a\mu\nu} - \delta_{a\mu}\delta_{\nu4} + \delta_{a\nu}\delta_{\mu4}\,.
\eae
In Eq.\,(\ref{HS}) $\lambda^a$, ($a=1,2,3$), 
denote the Pauli matrices. The periodic solutions in Eq.\,(\ref{HS}), $A^{C,A}_\mu(0,\vec{x})=
A^{C,A}_\mu(1/T,\vec{x})\,,$ 
are generated by a temporal mirror sum of 
the 'pre'potential 
\eqb
\label{PsI}
\Pi_0=1+\frac{\rho^2}{x^2}
\eqe
of a single BPST (anti)instanton of scale $\rho$ \cite{BPST} 
in singular gauge \cite{Atiyah1978}. 
Here $x^2\equiv\tau^2+\vec{x}^2$. The scalar function $\Pi(\tau,\vec{x})$ 
in Eq.\,(\ref{HS}) is given as 
\eab
\label{Pi}
\Pi(\tau,\vec{x})&=&\sum_{n=-\infty}^{\infty}\frac{\rho^2}{(\tau-n\beta,\vec{x})^2}\nonumber\\ 
&=&\bar{\Pi}(\tau,r)\equiv1+\frac{\pi\rho^2}{\beta r}
\frac{\sinh\left(\frac{2\pi r}{\beta}\right)}{\cosh\left(\frac{2\pi r}{\beta}\right)-
\cos\left(\frac{2\pi\tau}{\beta}\right)}\,
\eae
where $r\equiv|\vec{x}|$ and $\beta\equiv 1/T $. Evaluating the integral 
of the Chern-Simons current over a 
small three-sphere $S_3$, centered at the singular point 
$(\tau=0,\vec{x}=0)$, one obtains plus (or minus) one unit of 
topological charge. For a given value of 
$\rho$ the solutions in Eq.\,(\ref{HS}) can be generalized by 
shifting the center from $z=0$ to $z=(\tau_z,\vec{z})$ by the (quasi) 
translational invariance of the classical action. (The temporal shift $\tau_z$ is 
restricted to $0\le\tau_z\le\beta$ because of periodicity.) In addition, the color orientation of each 
solution can be rotated by global gauge transformations. 

Computing the Polyakov loop at spatial infinity on either of 
the configurations $A^C_\mu(\tau,\vec{x})$ and $A^A_\mu(\tau,\vec{x})$ yields 
the following result
\eqb
\label{Polyainf}
{\bf P}(|\vec x|\to\infty)=
{\cal P}\exp\left[i\int_0^\beta d\tau A^{C,A}_4(\tau,|\vec x|\to\infty)
\right]=\UM\,.
\eqe
Thus the Harrington-Shepard solutions possess trivial holonomy.

\subsubsection{The Lee-Lu-Kraan-van Baal 
solution (nontrivial holonomy)}

For a discussion of (anti)selfdual SU(2) configurations 
with nontrivial holonomy and topological charge one (minus one) 
we use the conventions and closely follow the presentation 
of \cite{LeeLu1998} which to our taste makes the 
magnetic monopole content most explicit. (In \cite{Diakonov2004} the constituents 
of nontrivial-holonomy calorons 
are referred to as dyons because 
the nonabelian magnetic and electric field 
of each constituent is equal and Coulomb-like for large 
distances away from a given monopole core. This property, however, follows 
from the selfduality of the caloron 
configuration. With respect to the unbroken 
U(1) the charge of a constituent monopole 
is purely magnetic (and not dyonic as in \cite{JuliaZee1975}) 
since the $A_4$ field serves as a Higgs field and not as 
the gauge potential for the electric field.) The existence of these solutions was shown 
by Nahm \cite{Nahm1984}. Explicit analytical constructions 
were independently performed by Lee and Lu 
\cite{LeeLu1998} and Kraan and 
van Baal \cite{KraanVanBaalNPB1998,vanBaalKraalPLB1998}. 

Lee and Lu use antihermitian generators and parametrize 
the holonomy $u$ as
\eqb
\label{holonomy}
A^C_4(\tau,|\vec x|\to\infty)=-i\frac{u}{2}\lambda_3\,
\eqe
where $0\le u\le \frac{2\pi}{\beta}$. Using the 
Nahm data for a monopole coexisting with an antimonopole as an 
input to the Atiyah-Drinfeld-Hitchin-Manin-Nahm (ADHMN) 
equations (subject to a normalization condition), a 
selfdual field configuration with monopole-antimonopole constituents 
was constructed in \cite{LeeLu1998}. It reads
\eab
\label{caloronnth}     
A_\mu(\vec{x},\tau)&=&C_1^\dagger V_\mu(\vec{y}_1;u)C_1+
C_2^\dagger V_\mu(\vec{y}_2;\frac{2\pi}{\beta}-u)C_2+\nonumber\\ 
&&C^\dagger_1\pd_\mu C_1+C^\dagger_1\pd_\mu C_1+
C^\dagger_2\pd_\mu C_2+S^\dagger\pd_\mu S\,
\eae
\begin{figure}
\begin{center}
\leavevmode
\leavevmode
\vspace{5.0cm}
\includegraphics{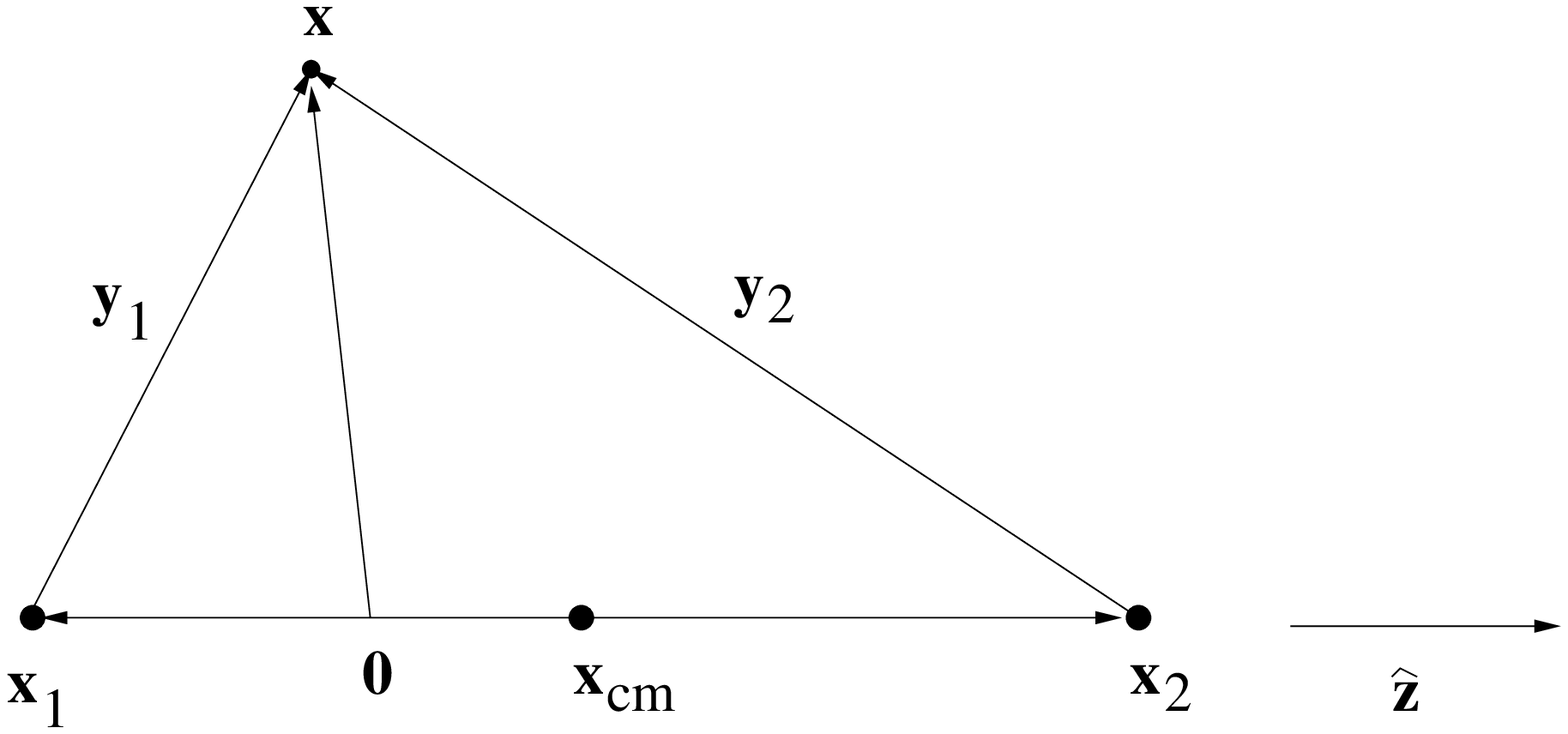}
\end{center}
\caption{Meaning of the spatial arguments $\vec{y}_1,\vec{y}_2$ entering the 
solution in Eq.\,(\protect\ref{caloronnth}). The points $\vec{x}_1,\vec{x}_2$ are the 
core positions of the monopole and the antimonopole. At the 
point $\tau=0,\vec{x}_{\tiny\mbox{cm}}$ the solution is singular.\label{Fig-1b}}      
\end{figure}
where
\eab
\label{Vmu}
V_4(\vec{x};u)&=&\frac{\lambda_a}{2i}\hat{x}_a\left(\frac{1}{|\vec{x}|}-
\frac{u}{\coth(u|\vec{x}|)}\right)\,,\nonumber\\ 
V_i(\vec{x};u)&=&\frac{\lambda_a}{2i}\epsilon_{aij}\hat{x}_j\left(\frac{1}{|\vec{x}|}-
\frac{u}{\sinh(u|\vec{x}|)}\right)\,.
\eae
Interpreting $V_4(\vec{x};u)$ as an adjoint Higgs field, Eqs.\,(\ref{Vmu}) represent 
the BPS magnetic monopole \cite{PrasadSommerfield1974}. The matrices $C_1,C_2$ 
in Eq.\,(\ref{caloronnth}) are given as
\eab
\label{C1C2}
C_1&=&\sqrt{\frac{2D N_1}{{\cal N}}}\frac{B_1^\dagger}{{\cal M}}\left[\exp\left(-\frac{\vec{\lambda}}{2}
\cdot \vec{s}_2\right)Q_++\exp\left(\frac{\vec{\lambda}}{2}
\cdot \vec{s}_2\right)Q_-\right]\exp\left(-i\frac{\pi}{\beta}\tau\lambda_3\right)\,,\nonumber\\ 
C_2&=&\sqrt{\frac{2D N_2}{{\cal N}}}\frac{B_2^\dagger}{{\cal M}}\left[\exp\left(\frac{\vec{\lambda}}{2}
\cdot \vec{s}_1\right)Q_++\exp\left(-\frac{\vec{\lambda}}{2}
\cdot \vec{s}_2\right)Q_-\right]
\eae
where $Q_\pm=\frac{1}{2}(1\pm\lambda_3)$ are projection operators. The 
matrices $B_1,B_2$ are 
\eab
\label{B1B2}
B_1&=&\exp\left[i\frac{\pi}{\beta}\tau\right]\exp\left[-\frac{\vec{\lambda}}{2}
\cdot\vec{s_1}\right]\exp\left[-\frac{\vec{\lambda}}{2}
\cdot\vec{s_2}\right]-\nonumber\\ 
&&\exp\left[-i\frac{\pi}{\beta}\tau\right]\exp\left[\frac{\vec{\lambda}}{2}
\cdot\vec{s_1}\right]\exp\left[\frac{\vec{\lambda}}{2}
\cdot\vec{s_2}\right]\,,\nonumber\\ 
B_2&=&\exp\left[i\frac{\pi}{\beta}\tau\right]\exp\left[-\frac{\vec{\lambda}}{2}
\cdot\vec{s_2}\right]\exp\left[-\frac{\vec{\lambda}}{2}
\cdot\vec{s_1}\right]-\nonumber\\ 
&&\exp\left[-i\frac{\pi}{\beta}\tau\right]\exp\left[\frac{\vec{\lambda}}{2}
\cdot\vec{s_2}\right]\exp\left[\frac{\vec{\lambda}}{2}
\cdot\vec{s_1}\right]\,
\eae
and the scalar ${\cal M}$ is defined as
\eqb
\label{M}
{\cal M}=2\left(\cosh s_1\cosh s_2+\hat{y}_1\cdot\hat{y}_2\sinh s_1\sinh
s_2-\cos\left[\frac{2\pi}{\beta}\tau\right]\right)\,.
\eqe
In addition, one defines
\eqb
\label{Ni}
N_i=\frac{1}{y_i}\sinh s_i\,,\ \ \ \ \ (i=1,2)\,,
\eqe
and 
\eqb
\label{N}
{\cal N}=1+\frac{2D}{{\cal M}}\left(N_1(\cosh s_2-(\hat{y}_2)_3\sinh s_2)+
N_2(\cosh s_1+(\hat{y}_1)_3\sinh s_1)\right)\,,
\eqe
and 
\eqb
\label{S}
S=\frac{1}{\sqrt{\cal N}}\,\exp\left[-i\frac{u}{2}\tau\lambda_3\right]\,.
\eqe
The spatial arguments of the configuration in Eq.\,(\ref{caloronnth}), 
compare with Fig.\,\ref{Fig-1b}, are 
defined as
\eab
\label{geom}
\vec{y}_i&=&\vec{x}-\vec{x}_i\,,\ \ \ \ \ y_i=|\vec{y}_i|\,,\ \ \ \ \ \ \ \ s_i=|\vec{s}_i|\,\ \ \ \ \ \ \ (i=1,2)\,, \nonumber\\ 
\vec{s}_1&=&u\vec{y}_1\,,\ \ \ \ \ \ \ \ \vec{s}_2=\left(\frac{2\pi}{\beta}-u\right)\vec{y}_2\,,\ \ \ \ 
D=|\vec{x}_2-\vec{x}_1|\,.
\eae
(A $\hat{\mbox{}}$ -sign indicates a unit vector.) Kraan and van Baal show that the distance $D$ 
between the two BPS monopoles can be expressed by the scale $\rho$ 
of a trivial-holonomy caloron which is deformed 
to nontrivial holonomy \cite{vanBaalKraalPLB1998}. One has
\eqb
\label{relrhoD}
D=\frac{\pi}{\beta}\rho^2\,.
\eqe
\begin{figure}
\begin{center}
\leavevmode
\leavevmode
\vspace{4.3cm}
\includegraphics{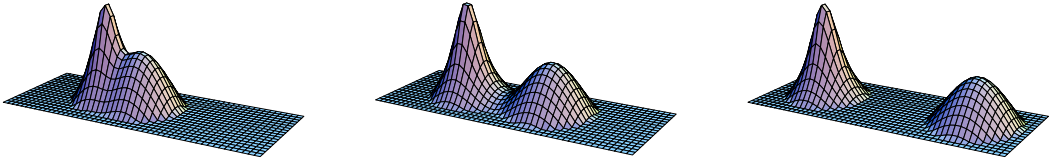}
\end{center}
\caption{Action density of an SU(2) caloron with nontrivial holonomy plotted on a two-dimensional spatial slice. 
The caloron radius $\rho$ and therefore the separation $D$ (Eq.\,(\protect\ref{relrhoD})) 
increases from left to right while temperature and holonomy are fixed. Figures are taken 
from a paper by Kraan and van Baal. The peaks of the action density 
coincide with the core positions of the constituent BPS monopoles.\label{instrad}}      
\end{figure}
It was shown in \cite{LeeLu1998} 
that for $y_1\ll D$ 
\eqb
\label{asy}
C_2,S\sim \frac{1}{\sqrt{D}}
\eqe
and 
\eqb
\label{C1asy}
C_1=\frac{\lambda_3\cosh\frac{s_1}{2}-\vec{\lambda}\cdot\hat{y}_1\sinh\frac{s_1}{2}}{\sqrt{\cosh s_1-
(\hat{y}_1)_3\sinh s_1}}+{\cal O}(1/D)\,.
\eqe
Thus $C_1$ is a single-valued unitary matrix, and for 
$y_1\ll D$ the configuration in Eq.\,(\ref{caloronnth}) is an approximate 
gauge transform of a BPS monopole. Similarily, for $y_2\ll D$ $C_2$ 
is a unitary matrix. The difference as compared with $C_1$ for $y_1\ll D$ 
is that for $y_2\ll D$ the matrix $C_2$ induces a large gauge rotation due to an extra 
factor $\exp\left[-i\frac{\pi}{\beta}\tau\lambda_3\right]$. This gauge transformation 
inverts the charge of the BPS monopole at $\vec{x}_2$ as compared to the charge of 
the BPS monopole at $\vec{x}_1$. There is a singularity of the 
solution at the point $(\tau=0,\vec{x}_{\tiny\mbox{cm}})$ where 
\eqb
\label{cmpoint}
\vec{x}_{\tiny\mbox{cm}}=\frac{\beta u}{2\pi}\vec{x}_1+
\left(1-\frac{\beta u}{2\pi}\right)\vec{x}_2\,.
\eqe
This point carries one unit of topological 
charge. One can show this by expanding the solution about 
$(\tau=0,\vec{x}_{\tiny\mbox{cm}})$ and by 
performing the integral of 
the Chern-Simons current over a 
small $S_3$ centered at this point. A plot of the action 
density of a nontrivial-holonomy caloron with varying radius $\rho$ at a fixed temperature and a fixed 
holonomy is presented in Fig.\,\ref{instrad}.   

On the classical level the masses of the monopoles 
at $\vec{x}_1$ and $\vec{x}_2$ are given as
\eqb
\label{massesMonLL}
m_1=4\pi u\,\ \ \ \ \ \ m_2=4\pi \left(\frac{2\pi}{\beta}-u\right)\,,
\eqe
respectively. For a large holonomy, that is $u\sim\frac{\pi}{\beta}$, 
we have $m_1\sim m_2\sim 4\pi^2\,T\sim 40\,T$. Thus large holonomy is extremely Boltzmann 
suppressed.

\subsubsection{One-loop quantum weights}

In this section we first present the results for the one-loop 
effective action of a trivial-holonomy caloron, 
which was obtained by Gross, Pisarski, and Yaffe 
\cite{GrossPisarskiYaffe1981} by appealing to the 
results obtained in \cite{BrownCarlitzLee1977,BrownCarlitzCreamerLee1978,BrownCreamer1978} and the 
pioneering work of 't Hooft \cite{Hooft1976}. 
Subsequently, we sketch the results obtained recently by 
Diakonov, Gromov, Petrov, and Slizovskiy \cite{Diakonov2004} for the 
one-loop quantum weight of a caloron with nontrivial holonomy. 
Both results are important for a grasp of 
the microcopics of the ground-state 
physics in the electric phase.\vspace{0.1cm}\\ 

\noindent\underline{Harrington-Shepard solution:}\vspace{0.1cm}\\ 

The functional determinant around a 
Harrington-Shepard caloron was 
calculated in \cite{GrossPisarskiYaffe1981}. The result for the 
quantum weight $\exp[-S_{\tiny\mbox{eff}}]$ 
of this configuration is given in terms of the effective action as
\eqb
\label{effactthcal}
S_{\tiny\mbox{eff}}=\frac{8\pi^2}{\bar{g}^2}+\frac{4}{3}\sigma^2+16\,A(\sigma)\,
\eqe
where the dimensionless quantity $\sigma$ is defined as 
$\sigma\equiv \pi\frac{\rho}{\beta}$ and 
\eqb
\label{A(alpha)}
A(\sigma)\equiv\frac{1}{12}\left[\int_0^\beta d\tau \frac{d^3x}{16\pi^2}
\left(\frac{(\pd_\mu \Pi)^2}{\Pi^2}\right)^2-\int \frac{d^4x}{16\pi^2}
\left(\frac{(\pd_\mu \Pi_0)^2}{\Pi_0^2}\right)^2\right]\,.
\eqe
The weight $\exp[-S_{\tiny\mbox{eff}}]$ is relevant for 
the integration over the classical moduli space in the presence of one-loop quantum fluctuations. 
(The nonflat metric is the same as in the zero-temperature situation.)

In Eq.\,(\ref{A(alpha)}) the scalar quantities $\Pi$ and $\Pi_0$ are 
defined in Eqs.\,(\ref{Pi}) and (\ref{PsI}), respectively. The first 
integral in Eq.\,(\ref{A(alpha)}) is over $S_1\times \R^3$ while the second integral 
is over $\R^4$. It is worth mentioning how $A(\sigma)$ behaves in the 
high- and low-temperature limits $\sigma\to\infty$ and $\sigma\to 0$:
\eqb
\label{hlTL}
A(\sigma)\to -\frac{1}{6}\log\sigma\,,\ \ \ (\sigma\to\infty)\,,\ \ \ \ 
A(\sigma)\to -\frac{\sigma^2}{36}\,,\ \ \ (\sigma\to 0)\,.
\eqe
At a given caloron radius $\rho$ the correction to the classical action 
$\frac{8\pi^2}{\bar{g}^2}$ thus is large in the 
high-temperature regime, indicating that the contribution of 
trivial-holonomy calorons to the partition function is 
suppressed, while it is small at low temperatures, 
implying the increasing importance of calorons as the 
temperature of the system drops. The distinction between high and 
low temperatures is made by a dynamically generated scale $\Lambda_E$ which also 
determines the $\rho$ dependence of the 
coupling constant $\bar{g}$ in Eq.\,(\ref{effactthcal}). The latter, by zero-temperature 
one-loop renormalization-group running \cite{GrossWilczek1973,Politzer1973}, 
estimatedly becomes larger than unity for $\rho^{-1}\sim\Lambda_E$ and is 
logarithmically small for $\rho^{-1}\gg\Lambda_E$. \vspace{0.1cm}\\     

\noindent\underline{Lee-Lu-Kraan-van Baal 
solution}\vspace{0.1cm}\\ 

The calculation of the one-loop quantum weight for a caloron of 
nontrivial holonomy is much harder than for the trivial case. This explains why this result only 
appeared in the literature \cite{Diakonov2004} more than six years 
after the analytical form of the nontrivial-holonomy solution was published in 
\cite{LeeLu1998,vanBaalKraalPLB1998}. The expressions 
are so involved that the contribution ${\cal Z}_{\tiny\mbox{n.h.}}$ 
of an isolated, quantum-blurred caloron to the total 
partition function ${\cal Z}$ of the theory has so far only been stated in closed analytical form 
in the limit 
\eqb
\label{limitDiakonov}
\frac{D}{\beta}=\pi\left(\frac{\rho}{\beta}\right)^2\gg 1\,.
\eqe
This, however, is 
the relevant physical situation, see 
Sec.\,\ref{EC} where is is shown 
that (trivial-holonomy) calorons with $\rho\gg \beta$ 
dominate the phase of the macroscopic adjoint 
scalar field $\phi$. As we shall see, the 
ground-state physics in the electric phase is dominated by small-holonomy deformations of 
the trivial case.          
 
Apart from the restriction in Eq.\,(\ref{limitDiakonov}) 
the result obtained in \cite{Diakonov2004} for ${\cal Z}_{\tiny\mbox{n.h.}}$ is valid 
for any value of the holonomy, $0\le u\le \frac{2\pi}{\beta}$. After the (trivial) 
integrations of the overall color orientation 
and time translations are performed one 
obtains \cite{Diakonov2004}
\eab
\label{Zcalonh}
{\cal Z}_{\tiny\mbox{n.h.}}&=C\beta^{-6}&\int d^3 x_1\int d^3 x_2\,
\left(\frac{8\pi^2}{\bar{g}^2}\right)^4\left(\frac{\La\e^{\gamma_E}\beta}{4\pi}\right)^{22/3}
\left(\frac{\beta}{D}\right)^{5/3}\times\nonumber\\ 
&&(2\pi+\beta u\bar{u} D)(uD+1)^{\frac{4}{3\pi}u\beta-1}(\bar{u}D+1)^{\frac{4}{3\pi}\bar{u}\beta-1}\times\nonumber\\ 
&&\exp[-V\,P(u)-2\pi DP^{\prime\prime}(u)]\,,\ \ \ \ \ (D\gg\beta).
\eae
(The number of independent integration 
variables in Eq.\,(\ref{Zcalonh}) is four because $\int d^3 x_1\int d^3 x_2=4\pi\int d^3x\int dD D^2$ 
where $\vec{x}=\frac{1}{2}\left(\vec{x}_1+\vec{x}_2\right)$). 
In Eq.\,(\ref{Zcalonh}) $C\sim 1.0314$, $\vec{x}_1$ and $\vec{x}_2$ are the core positions 
of the monopoles in the classical solution 
(compare with Fig.\,\ref{Fig-1b}), $\bar{u}\equiv\frac{2\pi}{\beta}-u$, $\gamma_E$ is the Euler constant, 
$V$ denotes the typical spatial volume belonging to the one-caloron system, 
and $\La$ is a scale which is a one-loop 
renormalization group invariant (dimensional transmutation). The functions $P(u)$ 
and $P^{\prime\prime}(u)$ are given as
\eab
\label{PPdP}
P(u)&=&\frac{\beta}{12\pi^2}u^2\bar{u}^2\,,\nonumber\\ 
P^{\prime\prime}(u)&=&\frac{\beta}{\pi^2}\left[\frac{\pi}{\beta}\left(1-\frac{1}{\sqrt{3}}\right)-u\right]
\left[\bar{u}-\frac{\pi}{\beta}\left(1-\frac{1}{\sqrt{3}}\right)\right]\,.
\eae
The function $P(u)$ is always positive for $u\not=0,\frac{2\pi}{\beta}$. 
The occurrence of the spatial volume $V$ in the exponent 
in Eq.\,(\ref{Zcalonh}) would mean total suppression of nontrivial holonomy 
in the naive thermodynamical limit $V\to\infty$. This, however, is not a valid conclusion since 
nontrivial-holonomy calorons are unstable: They either dissociate into a pair of BPS 
monopoles (large holonomy) or they collapse back onto trivial holonomy by 
an annihilation of their BPS monopole constituents. This can be checked by investigating 
the second contribution to the exponent in Eq.\,(\ref{Zcalonh}). 
  
\begin{figure}
\begin{center}
\leavevmode
\leavevmode
\vspace{4.3cm}
\includegraphics{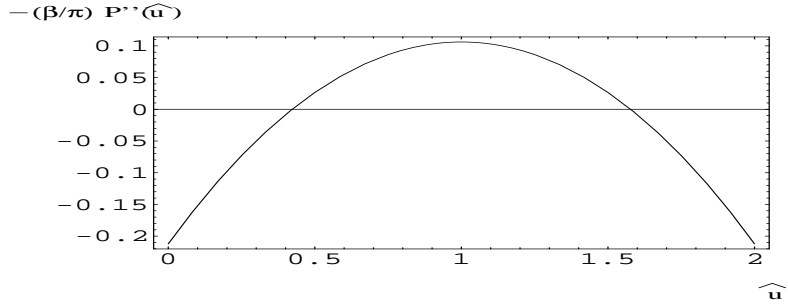}
\end{center}
\caption{The quantity $-\frac{\beta}{\pi}P^{\prime\prime}(\hat{u})$, compare with Eq.\,(\protect\ref{Zcalonh}), 
as a function of the dimensionless holonomy $\hat{u}\equiv\frac{u}{\pi T}$.\label{Fig1c}}      
\end{figure}

\begin{figure}
\begin{center}
\leavevmode
\leavevmode
\vspace{4.3cm}
\includegraphics{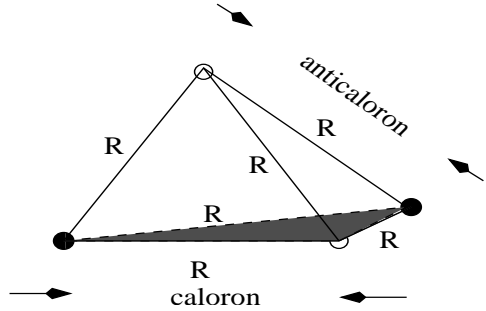}
\end{center}
\caption{The typical volume spanned by two pairs of BPS monopoles created 
by the dissociation of two calorons whose large holonomy was generated 
by the interaction of trivial-holonomy calorons.\label{Fig1d}}      
\end{figure}

In Fig.\,\ref{Fig1c} a plot of $-\frac{\beta}{\pi}P^{\prime\prime}(\hat{u})$ 
is shown. For $0\le u\le \frac{\pi}{\beta}(1-\frac{1}{\sqrt{3}})$ and 
for $\frac{\pi}{\beta}(1+\frac{1}{\sqrt{3}})\le u\le 2\,\frac{\pi}{\beta}$ the quantity 
$P^{\prime\prime}(u)$ is positive (small holonomy) while it is negative in the complementary range (large holonomy). 
According to Eq.\,(\ref{Zcalonh}) this means that in the former (latter)
case the BPS monopoles experience a linear attractive (repulsive) potential. 
Let us now make an estimate of the typical size of an equilateral 
tetrahedron whose corners are the positions of 
(screened) magnetic monopoles, see Fig.\,\ref{Fig1d}, which are generated by 
the dissociation of a caloron and an anticaloron whose large holonomy 
was created by their interaction.  
The edge length $R$ of the tetrahedron is the 
typical maximal distance between two BPS 
monopoles generated by a caloron with a 
holonomy close to maximally nontrivial, 
$u_{\tiny\mbox{max}}=\frac{\pi}{\beta}$. Once a large holonomy 
has been created the dissociation of the caloron generates the 
stabilized distance $R$ with probability one. (Once a monopole is at rest with 
respect to the heat bath there is no screening of its magnetic charge 
by Cartan fluctuations \cite{Linde1980} but only by small-holonomy calorons 
in its surroundings. Thus the screening of magnetic 
charge is not described by Eq.\,(\ref{Zcalonh}).) Thus it is 
appropriate to equate the probability for reaching the 
distance $R$, where monopoles are sufficiently screened to be at rest, governed 
by Eq.\,(\ref{Zcalonh}), with the thermal 
probability for exciting the monopoles in a caloron of large holonomy 
to start with. Since monopoles also are at rest shortly after being created the latter probability is 
roughly given as
\eqb
\label{Boltzmann}
\exp[-\beta(m_1+m_2)]\,,\ \ \ \ \ m_1\sim m_2\sim \frac{4\pi^2}{\beta}\,,
\eqe
see Eq.\,(\ref{massesMonLL}). Taking only the exponentially sensitive part of the 
caloron weight into account and substituting for $V$ the volume of the 
tetrahedron, $V=\frac{1}{6\sqrt{2}}\,R^3$, this 
translates into the following condition:
\eqb
\label{conddoms}
-\frac{\pi^2}{72\sqrt{2}}\left(\frac{R}{\beta}\right)^3+\frac{2}{3}\pi\frac{R}{\beta}+8\pi^2=0\,.
\eqe
There exists only a single real and positive solution to this equation. Numerically, we 
obtain $R\sim 10.1\,\beta$. So on the scale of 
the inverse temperature the gas of screened magnetic monopoles 
is dilute. This fits nicely with the lattice results 
obtained in \cite{KorthalsAltes,HoelbingRebbiRubakov2001}.   

While the (extremely small) 
likelihood for the generation 
of large-holonomy calorons depends on the 
value of the holonomy only (and not on the distance $D$) this is 
not true for a caloron with holonomy close to trivial. Since the latter configuration 
always collapses back onto trivial holonomy the likelihood for its generation 
is determined by the caloron weight $\exp[-S_{\tiny\mbox{eff}}]$ with $S_{\tiny\mbox{eff}}$ 
given in Eq.\,(\ref{effactthcal}). A strong dependence of $S_{\tiny\mbox{eff}}$ 
on $D$ (or $\rho$) at a given temperature exists. In contrast to $\exp[-\beta(m_1+m_2)]\sim \exp[-8\pi^2]$ 
the weight $\exp[-S_{\tiny\mbox{eff}}]$ is sizable at 
$S_{\tiny\mbox{eff}}$'s minimum $\sigma_{\tiny\mbox{min}}$. 

We conclude that {\sl attraction} between a BPS monopole and its antimonopole 
(small holonomy), which are in equilibrium with respect of their 
creation and annihilation, by far dominates the ground-state physics as compared 
to the case where monopole and antimonopole {\sl repulse} one another 
(large holonomy). Macroscopically, this situation expresses itself by a {\sl negative} pressure 
of the ground state. We shall compute the temperature dependence of this 
pressure in Sec.\,\ref{pomod}.

\subsubsection{Microscopic definition for the phase 
of an adjoint and macroscopic scalar field $\phi$\label{MDms}}

The results that were discussed in the last two subsections are important 
for an understanding of the infrared physics 
in the electric phase. The detailed microscopic 
dynamics is very complicated and, as it seems, it 
is impossible to derive macroscopic quantities such as the pressure or the 
energy density or the mass of thermal quasiparticles 
by performing literal ensemble averages on the microscopic level. 
What turns out 
to be feasible and thermodynamically exhaustive 
is to compute the spatial average (spatial coarse-graining) 
over the physics generated by the 
topologically nontrivial sector. This procedure introduces the concept of a 
macroscopic, thermal ground state. As far as thermodynamics is 
concerned one still obtains 
exact results this way. The advantage of such an approach is that the 
complications of a microscopic calculation are avoided. Once the ground-state physics 
is understood and quantitatively described its effect on the propagation of 
trivial-topology modes can be investigated.    

If the ground state is to be characterized by a macroscopic field other than a pure-gauge 
configuration then, by spatial isotropy, this macroscopic field must be a Lorentz 
scalar $\phi$. Moreover, in a pure Yang-Mills theory, where all local fields 
transform under the adjoint representation of the gauge group, the composite 
field $\phi$ needs to transform in an adjoint way under the 
remnants of a microscopic, spacetime dependent gauge transformations. Since space 
dependent gauge transformations are constant on the macroscopic level 
(due to the spatial average) no space dependence of 
$\phi$ occurs in any chosen gauge. Apart from its modulus, which is governed 
by a dynamically emerging scale $\Lambda_E$ and temperature, 
the only nontrivial information on $\phi$ is the $\tau$ dependence 
of its color orientation in a given gauge. In the following we will refer 
to $\phi$'s color orientation as $\phi$'s phase.   

Let us imagine a (hypothetical) Yang-Mills world where the 
only field configurations allowed to contribute to the partition function 
are classical and noninteracting caloron configurations of trivial holonomy. 
We will show in Sec\,\ref{EC} that it is consistent 
to adopt this point of view in the derivation of the {\sl macroscopic} 
ground-state physics. Since the Yang-Mills 
scale $\Lambda_E$ can not be computed we focus on the computation of 
$\phi$'s phase first. Because $\phi$'s phase is a ratio of the field and 
its modulus and hence dimensionless the associated measure for the $\rho$-average is flat.     

We closely follow the presentation 
in \cite{HerbstHofmann2004} for the remainder of this section 
and for Sec.\,\ref{EC}. Due to the selfduality of calorons 
any local definition of $\phi's$ phase 
yields the trivial result zero. Thus we start by defining:
\eab
\label{defphi}
\frac{\phi^a}{|\phi|}(\tau)&\sim &\mbox{tr}\Bigg[\nonumber\\ 
&&\beta^0 1!\int d^3x\,\int d\rho\, \nonumber\\ 
&&\frac{\lambda^a}{2} F_{\mu\nu}[A_\alpha(\rho,\beta)]\left((\tau,0)\right)\,
\left\{(\tau,0),(\tau,\vec{x})\right\}[A_\alpha(\rho,\beta)]\times
\nonumber\\ 
&&F_{\mu\nu}[A_\alpha(\rho,\beta)]\left((\tau,\vec{x})\right)\,
\left\{(\tau,\vec{x}),(\tau,0)\right\}[A_\alpha(\rho,\beta)]+
\nonumber\\ 
&&\beta^{-1} 2!\int d^3x\int d^3y\,\int d\rho\, 
\nonumber\\ 
&&\frac{\lambda^a}{2} F_{\mu\lambda}[A_\alpha(\rho,\beta)]\left((\tau,0)\right)\,
\left\{(\tau,0),(\tau,\vec{x})\right\}[A_\alpha(\rho,\beta)]\times
\nonumber\\ 
&&\,F_{\lambda\nu}[A_\alpha(\rho,\beta)]\left((\tau,\vec{x})\right)\,
\left\{(\tau,\vec{x}),(\tau,\vec{y})\right\}[A_\alpha(\rho,\beta)]\times
\nonumber\\ 
&&F_{\nu\mu}[A_\alpha(\rho,\beta)]\left((\tau,\vec{y})\right) \left\{(\tau,\vec{y}),(\tau,0)\right\}
[A_\alpha(\rho,\beta)]+
\nonumber\\ 
&&\beta^{-2} 3!\int d^3x\,\int d^3y\,\int d^3u\,\int d\rho\, 
\nonumber\\ 
&&\frac{\lambda^a}{2} F_{\mu\lambda}[A_\alpha(\rho,\beta)]\left((\tau,0)\right)\,
\left\{(\tau,0),(\tau,\vec{x})\right\}[A_\alpha(\rho,\beta)]\times
\nonumber\\ 
&&\,F_{\lambda\nu}[A_\alpha(\rho,\beta)]\left((\tau,\vec{x})\right)\,
\left\{(\tau,\vec{x}),(\tau,\vec{y})\right\}[A_\alpha(\rho,\beta)]\times
\nonumber\\ 
&&F_{\nu\kappa}[A_\alpha(\rho,\beta)]\left((\tau,\vec{y})\right) 
\left\{(\tau,\vec{y}),(\tau,\vec{u})\right\}[A_\alpha(\rho,\beta)] 
F_{\kappa\mu}[A_\alpha(\rho,\beta)]\left((\tau,\vec{u})\right)\times 
\nonumber\\ 
&&\left\{(\tau,\vec{u}),(\tau,0)\right\}[A_\alpha(\rho,\beta)]+\cdots\Bigg]\,.
\eae
A number of comments are in order: 
The dots in (\ref{defphi}) stand for the contributions of higher $n$-point functions and for reducible, 
that is, factorizable contributions with respect to the spatial integrations. The factors $(n-1)!$ in front of the $n$-point contribution measures 
the multiplicity of the corresponding integral. Factors $\beta^{n-2}$ are needed 
to make the contribution dimensionless. The argument $A_\alpha(\rho,\beta)$ 
(spacetime dependence suppressed) 
refers to either a caloron or an anticaloron configuration, the Harrington-Shepard 
solutions of Sec.\,\ref{HSS}. Moreover, the following definitions apply: 
\eab
\label{defdefphi}
|\phi|&\equiv&\frac{1}{2}\,\mbox{tr}\,\phi^2\,
,\nonumber\\ 
\left\{(\tau,0),(\tau,\vec{x})\right\}[A_\alpha]&\equiv& {\cal P}\,
\exp\left[i\int_{(\tau,0)}^{(\tau,\vec{x})}dy_\beta\,A_\beta(y,\rho)\right]
\,,\nonumber\\ 
\left\{(\tau,\vec{x}),(\tau,0)\right\}[A_\alpha]&\equiv& {\cal P}\,
\exp\left[-i\int_{(\tau,0)}^{(\tau,\vec{x})}dy_\beta\,A_\beta(y,\rho)\right]\,
\eae
where ${\cal P}$ is the path-ordering symbol. 
\noindent Under a microscopic gauge transformation $\Omega(y)$ we have:
\eab
\label{micrOm}
\left\{(\tau,0),(\tau,\vec{x})\right\}[A_\alpha]&\rightarrow & \Omega^\dagger ((\tau,0))\,
\left\{(\tau,0),(\tau,\vec{x})\right\}[A_\alpha]\,
\Omega((\tau,\vec{x}))\,,\nonumber\\ 
\left\{(\tau,\vec{x}),(\tau,0)\right\}[A_\alpha]&\rightarrow & \Omega^\dagger 
((\tau,\vec{x}))\,\left\{(\tau,\vec{x}),(\tau,0)\right\}[A_\alpha]\,
\Omega((\tau,0))\,,\nonumber\\ 
F_{\mu\nu}[A_\alpha]\left((\tau,\vec{x})\right)&\rightarrow &
\Omega^\dagger((\tau,\vec{x}))\,F_{\mu\nu}[A_\alpha]((\tau,\vec{x}))\,\Omega((\tau,\vec{x}))\,,\nonumber\\ 
F_{\mu\nu}[A_\alpha]\left((\tau,0)\right)&\rightarrow &
\Omega^\dagger((\tau,0))\,F_{\mu\nu}[A_\alpha]((\tau,0))\,\Omega((\tau,0))\,.
\eae
As a consequence of Eq.\,(\ref{micrOm}) the right-hand side of (\ref{defphi}) transforms as 
\eqb
\label{phitrans}
\frac{\phi^a}{|\phi|}(\tau)\rightarrow R_{ab}(\tau)\,\frac{\phi^b}{|\phi|}(\tau)
\eqe
where the SO(3) matrix $R_{ab}(\tau)$ is defined as
\eqb
\label{Rdef}
R^{ab}(\tau)\lambda^b=\Omega((\tau,0))\,\lambda^a\,\Omega^\dagger((\tau,0))\,.
\eqe
Thus we have defined an adjointly transforming 
scalar in (\ref{defphi}). In addition, we have just 
shown that only the time-dependent part of a microscopic gauge transformation 
survives on the macroscopic level. (Shifting the spatial part of the argument 
$(\tau,0)\to(\tau,\vec{z})$ in (\ref{defphi}) 
introduces a finite {\sl parameter} $\vec{z}$ to the gauge rotation 
$R^{ab}$: $R^{ab}(\tau)\to R^{ab}(\tau,\vec{z})$. Such a shift, however,  
introduces an arbitrary but finite mass scale $|\vec{z}|^{-1}$ 
into the definition of $\phi$'s phase 
which, on the classical level, is absent. Also, a finite value of $|\vec{z}|$ would introduce an explicit breaking 
or rotational symmetry into the definition (\ref{defphi}). Thus we have $\vec{z}=0$. Moreover, the integration
path connecting the points $(\tau,0)$ with $(\tau,\vec{x})$ 
in Eq.\,(\ref{defdefphi}) ought to be a straight line since a spatial 
curvature would imply the existence of a mass scale other than 
temperature on the classical level.) Integrations over shifts $\tau\to\tau+\tau_s$ ($0\le\tau_s\le\beta$) project a nontrivial (periodic) 
$\tau$ dependence of $\phi$'s phase onto zero and thus are forbidden. Integrations 
over global gauge rotations are forbidden for the same reason. 
Spatial shifts $\vec{x}\to\vec{x}+\vec{x}_s,\vec{y}\to\vec{y}+\vec{x}_s,\cdots$ 
leave the integrals in (\ref{defphi}) invariant. These averages are 
already performed. Thus the only admissible integration over moduli-space parameters is over 
$\rho$ with a flat measure. 

In (\ref{defphi}) the $\sim$ sign indicates that both left- and right-hand sides satisfy the same 
homogeneous evolution equation in $\tau$  
\eqb
\label{deffequationhomo}
{\cal D}\left[\frac{\phi}{|\phi|}\right]=0\,.
\eqe
Here ${\cal D}$ is 
a differential operator such that Eq.\,(\ref{deffequationhomo}) represents a 
homogeneous differential equation. 
As it will turn out, Eq.\,(\ref{deffequationhomo}) is a {\sl linear} 
second-order equation which, up to global gauge rotations and a choice of winding sense, 
determines the first-order or BPS equation whose 
solution $\phi$'s phase is. (The ambiguities in the evaluation 
of the right-hand side span the solution space of Eq.\,(\ref{deffequationhomo}), and thus ${\cal D}$ 
is uniquely determined by (\ref{defphi}).) 

We now discuss why $n$-point functions with $n>2$ do not contribute to the 
right-hand side of (\ref{defphi}). Since the classical (anti)caloron action 
$S=\frac{8\pi^2}{g^2}$ and the classical moduli-space metric are 
independent of temperature we conclude that 
no {\sl explicit} dependence on $\beta$ may occur in the definition of $\phi$'s phase. 
For $n>2$, however, explicit $\beta$ dependences do occur, see (\ref{defphi}). We conclude that 
these contributions to $\phi$'s phase do not exist. 

What about calorons of higher topological charge? Some of these solutions have been 
constructed, see for example \cite{Actor1983,Chakrabarti1987}. 
The essential difference to the charge-one case is that 
more dimensionful moduli occur than just the parameter 
$\rho$. For example, for charge-two configurations 
there is a spatial core separation between the two seed 
instantons and an additional instanton 
radius $\rho^\prime$. The reader may now 
convince himself that along the lines of (\ref{defphi}) 
a nonlocal definition of $\phi$'s {\sl dimensionless} phase, which would also 
have to include integrations over the additional moduli of dimension 
length, is impossible for higher-charge calorons. (This is certainly 
true for the integral over two-point functions. For every increment in 
$n$ there is an increase in power of length scale by one unit arising from an additional 
$d^3x\times F_{\mu\nu}$. This makes the situation even worse in comparison 
to the two-point case.)

\subsubsection{Essentials of the calculation\label{EC}}

Before we dive into the essential parts of the 
calculation, which will lead to the unique determination 
of the operator ${\cal D}$ in 
Eq.\,(\ref{deffequationhomo}), we would like 
to discuss a condition which severely constrains 
the possible solutions to this equation. 

By (anti)selfduality the energy-momentum tensor vanishes 
identically on a caloron or an 
anticaloron,
\eqb
\label{thetacal}
\theta_{\mu\nu}[A^{(C,A)}_\alpha]\equiv 0\,.
\eqe
Since $\phi$'s phase is 
obtained by an average over (the admissible part of) 
the moduli space of a caloron-anticaloron system (no interactions) the macroscopic energy-momentum tensor 
$\bar{\theta}_{\mu\nu}[\phi]$ should vanish identically as well, 
\eqb
\label{thetabar}
\bar{\theta}_{\mu\nu}[\phi]\equiv 0\,.
\eqe
In a thermal equilibrium situation, described by Euclidean dynamics, 
this is true if and only if the $\tau$ dependence of $\phi$ (or $\phi$'s phase) 
is BPS saturated. Thus $\phi$ solves the first-order equation
\eqb
\label{BPSphi}
\pd_\tau\phi=V_E^{(1/2)}\,
\eqe
where $V_E^{(1/2)}$ denotes the 'square-root' of a suitable 
potential. (The fact that an ordinary and not a 
covariant derivative appears in Eq.\,(\ref{BPSphi}) is, of course, tied to our specific gauge 
choice. If we were to leave the (singular) 
gauge for the seed (anti)instanton, in which the solutions of Eq.\,(\ref{HS}) are constructed, 
by a time-dependent gauge rotation $\bar{\Omega}
(\tau)$ then a pure-gauge configuration 
$A^{p.g.}_\mu(\tau)=i\delta_{\mu4}\bar{\Omega}^\dagger\pd_{\tau}\bar{\Omega}$ 
would appear in a {\sl covariant} derivative on the left-hand side 
of Eq.\,(\ref{BPSphi}). Also recall the fact 
that the heat bath breaks boost invariance. 
This is encoded in the noninvariance of Eq.\,(\ref{BPSphi}) 
under O(4) rotations.) $V_E\equiv\mbox{tr}\,\left(V_E^{(1/2)}\right)^\dagger\,V_E^{(1/2)}$. 
As we will see below, the right-hand side of Eq.\,(\ref{BPSphi}) is 
determined only up to a global gauge rotation and a choice of winding sense.

Let us now discuss essential details of the 
calculation of the right-hand side of (\ref{defphi}) which, 
after what was said in Sec.\,\ref{MDms}, reduces to 
\eab
\label{reddefphi}
\frac{\phi^a}{|\phi|}(\tau)&\sim &\mbox{tr}\int d^3x\,\int d\rho\, \frac{\lambda^a}{2}\,F_{\mu\nu}[A_\alpha(\rho,\beta)]\left((\tau,0)\right)\,
\left\{(\tau,0),(\tau,\vec{x})\right\}[A_\alpha(\rho,\beta)]\times\nonumber\\ 
&&F_{\mu\nu}[A_\alpha(\rho,\beta)]\left((\tau,\vec{x})\right)\,
\left\{(\tau,\vec{x}),(\tau,0)\right\}[A_\alpha(\rho,\beta)]\,
\eae
where a sum over the contributions of a trivial-holonomy caloron and an anticalorons 
is to be performed. 

Since the integrand in the exponent of the Wilson line $\left\{(\tau,0),(\tau,\vec{x})\right\}
[A_\alpha^{C,A}]$ is a hedgehog the path-ordering 
prescription can be omitted. For the caloron contribution one obtains
\eab
\label{3DUP}
\left.\frac{\phi^a}{|\phi|}\right|_{\tiny{C}}&\sim&
i\int d\rho \int d^3x\,\frac{x^a}{r}\times\nonumber\\ 
&&\left[\frac{\left(\partial_4\Pi(\tau+\tau_C,0)\right)^2}{\Pi^2(\tau+\tau_C,0)}-\frac23
\frac{\partial^2_4\Pi(\tau+\tau_C,0)}{\Pi(\tau+\tau_C,0)}\right]\Big\{4\cos(2g(\tau+\tau_C,r))\times 
\nonumber\\  
&&\left.\left[\frac{\pd_r\pd_4\Pi(\tau+\tau_C,r)}{\Pi(\tau+\tau_C,r)}-
2\frac{\left(\pd_r\Pi(\tau+\tau_C,r)\right)\left(\pd_4\Pi(\tau+\tau_C,r)\right)}
{\Pi^2(\tau+\tau_C,r)}\right]+\right.\nonumber\\ 
&&\left.\sin(2g(\tau+\tau_C,r))\left[4\,\frac{\left(\pd_4\Pi(\tau+\tau_C,r)\right)^2-
\left(\pd_r\Pi(\tau+\tau_C,r)\right)^2}{\Pi^2(\tau+\tau_C,r)}+\right.\right.\nonumber\\ 
&&\left.\left. 2\,\frac{\pd^2_r\Pi(\tau+\tau_C,r)-\pd^2_4\Pi(\tau+\tau_C,r)}{\Pi(\tau+\tau_C,r)}
\right]\right\}\, 
\eae
where 
\eqb
\label{grtau}
g(\tau+\tau_C,r)\equiv \int_0^1 ds\, \frac{r}{2} \, \pd_4 \ln \Pi(\tau+\tau_C,sr)\,,
\eqe
the function $\Pi(\tau,r)$ is defined in Eq.\,(\ref{Pi}), and $\tau_C$ refers 
to a constant but arbitrary temporal shift of the caloron center ($0\le\tau_C\le\beta$). It is worth mentioning 
that the integrand in Eq.\,(\ref{grtau}) is proportional to $\delta(s)$ 
for $r\gg\beta$. The dependences on $\rho$ and $\beta$ are suppressed in the 
integrands of (\ref{3DUP}) and Eq.\,(\ref{grtau}). 

As compared to the 
contribution of the caloron there are ambiguities in the contribution 
of the anticaloron: First, the $\tau$ dependence of the contribution of the anticaloron may be shifted by $\tau_A$ 
($0\le\tau_A\le\beta$). Second, the color orientation of caloron and anticaloron 
contributions may be different. Third, the normalization of the two 
contributions may be different. 
To see that this is true, we need to investigate the convergence properties of the radial 
integration in (\ref{3DUP}). It is easily checked that 
all terms give rise to a converging $r$ integration 
except for the following one: 
\eqb
\label{nonconvr}
2\,\frac{x^a}{r}\,\sin(2g(\tau+\tau_C,r))\,\frac{\pd^2_r\Pi(\tau+\tau_C,r)}{\Pi(\tau+\tau_C,r)}\,.
\eqe
Namely, for $r>R\gg\beta$ (\ref{nonconvr}) goes over in
\eqb
\label{nonconvrlarger}
4\,\frac{x^a}{r}\frac{\pi\rho^2\sin(2g(\tau+\tau_C,r))}{\beta r^3}\,.
\eqe
Thus the $r$-integral of the term in (\ref{nonconvr}) is 
logarithmically divergent in the infrared: (The integral converges for $r\to 0$.)
\eqb
\label{nocI}
4\,\frac{\pi\rho^2}{\beta}\int_R^\infty\frac{dr}{r}\frac{x^a}{r}\,\sin(2g(\tau+\tau_C,r))\,\,.
\eqe
Recall, that $g(\tau+\tau_C,r)$ behaves like a constant 
in $r$ for $r>R$. The angular 
integration, on the other hand, would 
yield zero if the radial integration was regular. 
Thus a logarithmic divergence can be cancelled 
by the angular integral to yield some 
finite and real answer. To investigate this in more detail, both angular and radial 
integration need to regularized.

\begin{figure}
\begin{center}
\leavevmode
\leavevmode
\vspace{4.8cm}
\includegraphics{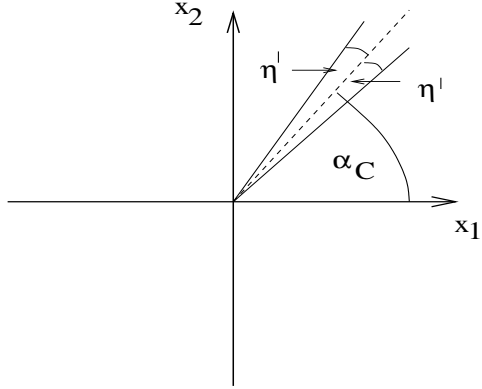}
\end{center}
\caption{The axis for the regularized azimuthal integration.\label{Fig1e}}      
\end{figure}

\begin{figure}
\begin{center}
\vspace{5.3cm}
\includegraphics{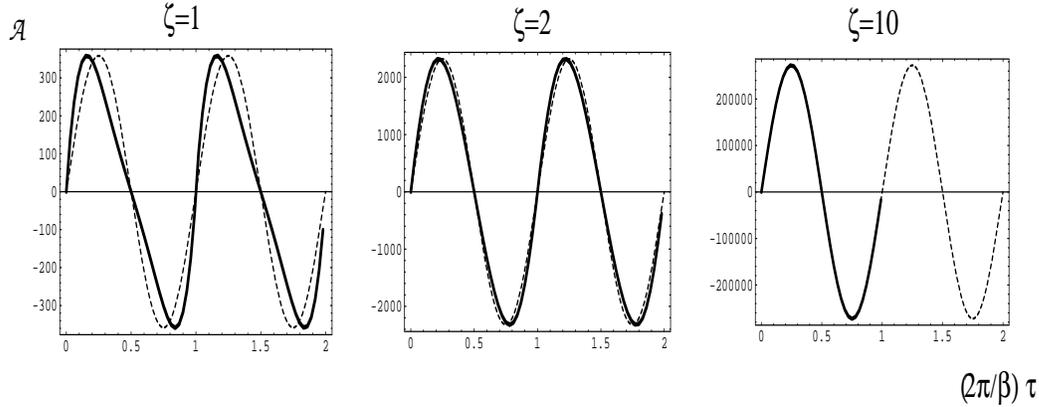}
\end{center}
\caption{${\cal A}$ as a function of $\frac{2\pi}{\beta}\tau$ for $\zeta=1,2,10$. For each case 
the dashed line is a plot of $\mbox{max}\,{\cal A}\times\sin\left(\frac{2\pi}{\beta}\tau\right)$. 
We have fitted the asymptotic dependence on $\zeta$ of the amplitude of ${\cal A}$ as 
${\cal A}\left(\frac{2\pi}{\beta}\tau=\frac{\pi}{2},\zeta\right)=272\,\zeta^3,\,(\zeta>10)$. 
The fit is stable under variations of the
fitting interval. For the case $\zeta=10$ the difference 
between the two curves can not be resolved anymore.\label{Fig1f}}      
\end{figure}

One introduces a regularization, conveniently we have chosen 
dimensional regularization in \cite{HerbstHofmann2004} 
with a dimensionless regularization parameter $\eta_C>0$, 
for the $r$-integral 
in Eq.\,(\ref{nocI}) while the angular integration can be 
regularized by introducing defect (or surplus) angles 
$2\eta^\prime_C$ in the $\theta$ integration 
(azimuthal angle in the $x_1x_2$ plane). Any other plane for the azimuthal 
angular integration could have been chosen. Moreover, the value of $\alpha_C$ is 
determined by a (physically irrelevant)
initial condition, as we will show below, 
see Fig.\,\ref{Fig1e}. Together, the choice of the
regularization plane and of the angle $\alpha_C$ amount 
to a global choice of gauge: an apparent breaking of rotational symmetry by the 
angular regularization is nothing but a gauge choice.)   
Without restriction of generality (global gauge choice)
we may also for the contribution of the anticaloron 
use an axis for the angular regularization which lies in the
$x_1x_2$ plane, but with a different angle $\alpha_A$. Then we have
\eab
\label{totalc}
\frac{\phi^a}{|\phi|}
&=&\left.\frac{\phi^a}{|\phi|}\right|_{\tiny{C}}+
\left.\frac{\phi^a}{|\phi|}\right|_{\tiny{A}}\nonumber\\ 
&=&
\pm\Xi_C\,\left(\delta_{a1}\cos\alpha_C+\delta_{a2}\sin\alpha_C\right) {\cal A}
\left(\frac{2\pi(\tau+\tau_C)}{\beta}\right)
\nonumber\\ 
&&\pm\Xi_A\,\left(\delta_{a1}\cos\alpha_A+\delta_{a2}\sin\alpha_A\right) {\cal A}
\left(\frac{2\pi(\tau+\tau_A)}{\beta}\right)
\nonumber\\ 
&\neq&0\,, 
\eae
where $\Xi_C$ , $\Xi_A$, $\tau_C$, $\tau_A$, $\alpha_C$, and $\alpha_A$ 
are undetermined. ($\Xi_C$, $\Xi_A$ are the ratios of $\eta_{C,A}$ and $\eta^\prime_{C,A}$, respectively.) 
The function ${\cal A}\left(\frac{2\pi(\tau)}{\beta}\right)$ 
is defined as
\eab
\label{calAdef}
{\cal A}
\left(\frac{2\pi\tau}{\beta}\right)&\equiv&
\frac{32}{3}\,\frac{\pi^7}{\beta^3}\,
\int d\rho\,\left[\lim_{r\to\infty}\sin(2g(\tau,r))\right]\times\nonumber\\ 
&&\rho^4\frac{\pi^2 \rho^2 + \beta^2 \left(2 + \cos \left(\frac{2\pi\tau}{\beta}\right)\right)}
{\left[2\pi^2 \rho^2 + \beta^2 \left(1-\cos \left(\frac{2\pi\tau}{\beta}\right)\right)\right]^2}\,.
\eae
Eq.\,(\ref{totalc}) and Eq.\,(\ref{calAdef}) provide the basis for 
fixing the operator $\cal D$ in Eq.\,(\ref{deffequationhomo}). 
To evaluate the function ${\cal A}\left(\frac{2\pi\tau}{\beta}\right)$ 
in Eq.\,(\ref{calAdef}) numerically, we introduce the same cutoff for the $\rho$ integration 
in the caloron and anticaloron case 
as follows:
\eqb
\label{cutoffrho}
\int d\rho\to \int_0^{\zeta \beta} d\rho\,,\ \ \ \ \ \ (\zeta>0)\,.
\eqe
This introduces an additional dependence of ${\cal A}$ on $\zeta$. 
In Fig.\,\ref{Fig1f} the $\tau$ dependence of ${\cal A}$ for 
various values of $\zeta$ is depicted. 
Therefore we have 
\eab
\label{finalres}
\frac{\phi^a}{|\phi|} &\sim& 272\,\zeta^3\,
\Bigg(
\Xi_C 
\left(\delta_{a1}\cos\alpha_C+\delta_{a2}\sin\alpha_C\right)\,
\sin\left(\frac{2\pi}{\beta}(\tau+\tau_C)\right)
\nonumber \\ &&
+
\Xi_A 
\left(\delta_{a1}\cos\alpha_A+\delta_{a2}\sin\alpha_A\right)\,
\sin\left(\frac{2\pi}{\beta}(\tau+\tau_A)\right)
\Bigg)
\nonumber\\
&\equiv& \hat \phi^a
\,.
\eae
Just like the numbers $\Xi_C$ and $\Xi_A$ are 
undetermined on the classical level due to the invariance of the classical action under 
spatial scale transformations so is the number $\zeta$. It is clear, however, 
from Eq.\,(\ref{finalres}) that the integral in ${\cal A}$ 
is strongly dominated by $\rho$-values close to the upper integration limit.   
Let us now discuss the physical content of (\ref{finalres}). For fixed values of
the parameters $\zeta^3\,\Xi_C$, $\zeta^3\,\Xi_A$, $\frac{\tau_C}{\beta}$ and $\frac{\tau_A}{\beta}$ 
the right-hand side of Eq.\,(\ref{finalres}) resembles an elliptic polarization 
in the $x_1x_2$ plane of adjoint color space. 
For a given polarization plane the two independent numbers (normalization and phase-shift) 
for each of the two oscillations parametrize the solution space of the second-order linear differential equation 
\eqb\label{2order}
{\cal D} \hat\phi=0\,.
\eqe
From (\ref{finalres}) we observe that the operator $\cal D$ is
\eqb
{\cal D} = \partial_\tau^2 + \left( \frac{2\pi}{\beta} \right)^2\,.
\eqe
The ambiguities in Eq.\,(\ref{finalres}) parameterize the 
solution space of Eq.\,(\ref{2order}) for a 
given polarization plane which depends on a global choice of gauge. Thus the 
differential operator ${\cal D}$ is {\sl uniquely} 
determined by Eq.\,(\ref{finalres}). What is needed to assure the 
validity of Eq.\,(\ref{thetabar}) is a BPS 
saturation of the solution to the linear Eq.\,(\ref{2order}) since 
the modulus of $\phi$ may not depend on $\tau$ in thermal equilibrium. 

Thus we need to find first-order equations whose solutions are traceless, hermitian and 
solve the second-order equation (\ref{2order}).
The relevant two first-order equations are 
\eqb\label{1orderbps}
\pd_\tau \hat{\phi}=\pm\frac{2\pi i}{\beta}\lambda_3\,\hat{\phi}\,.
\eqe
Obviously, the right-hand sides of Eqs.\,(\ref{1orderbps}) are subject to a 
global gauge ambiguity (associated with the choice of polarization 
plane in which the regularization of the azimuthal angular
integration is carried out) and a choice of sign: Any normalized generator other 
than $\pm\lambda_3$ could have appeared. Moreover, a solution to either of the two equations (\ref{1orderbps})
also solves Eq.\,(\ref{2order}) for a given polarization plane,
\eqb
\partial_\tau^2 \hat\phi=\pm \frac{2\pi i}{\beta}\lambda_3\, \partial_\tau \hat \phi=
\frac{2\pi i}{\beta}\lambda_3\,\frac{2\pi i}{\beta}\lambda_3\,\hat\phi=-
\left( \frac{2\pi}{\beta} \right)^2\hat\phi\,.
\eqe
Solutions to Eqs.\,(\ref{1orderbps}) are given as
\eqb\label{solutiona}
\hat{\phi} = C \, \lambda_1 \, \exp\left(\mp \frac{2\pi i}{\beta} \lambda_3 (\tau-\tau_0)  \right)
\eqe
where $C$ and $\tau_0$ denote real integration constants which
both are undetermined. We set $\tau_0=0$ in what follows. 
The solutions in Eq.\,(\ref{solutiona}) represent a circular polarization
in the $x_1x_2$ plane of adjoint color space
and thus indicate that the field $\phi$ winds along an $S_1$
on the group manifold $S_3$ of SU(2). Both winding senses appear but 
can not be distinguished physically: A change in winding 
sense does not affect the potential nor does it affect 
the admissibility of the transformation to 
unitary gauge, see Sec.\,\ref{gtug}. 
 
\subsubsection{$\phi$'s modulus and potential\label{pomod}}

\noindent\underline{SU(2) case:}
\vspace{0.1cm}\\ 
The information in Eq.\,(\ref{solutiona}) on $\phi$'s 
phase can be used to infer its modulus once the 
existence of an externally given mass scale $\Lambda_E$ is 
assumed. (The scale $\Lambda_E$ determines the typical distance between caloron centers at a
given temperature.) As long as no interactions between trivial-holonomy calorons 
are allowed for this is consistent since the BPS saturation of $\phi$ 
forbids the occurrence of (gravitationally) 
measurable effects: The macroscopic energy-momentum 
tensor $\bar{\theta}_{\mu\nu}$ vanishes identically, and thus assuming the existence of the 
scale $\Lambda_E$ does not yet influence the ground-state physics. We have 
\eqb
\label{tafel}
\phi = |\phi|(\beta,\Lambda_E)\,\hat{\phi}\left(\frac{\tau}{\beta}\right)\,.
\eqe
In order to reproduce the phase in Eq.\,(\ref{solutiona}) a {\sl linear} dependence on $\phi$
must appear on the right-hand side of the BPS equation (\ref{BPSphi}). Moreover, 
this right-hand side ought not depend on $\beta$ explicitly and must be 
analytic in $\phi$. The former requirement derives from the fact that $\phi$ and its potential $V$ are 
obtained by functionally integrating over the (admissible part of the) moduli space of a 
caloron-anticaloron system with no interactions. The associated part of the partition function 
does not exhibit an explicit $\beta$ dependence since 
the action and thus the weight are $\beta$ independent on the moduli space. 
Thus a $\beta$ dependence of $V$ or $V^{(1/2)}$ can only be generated 
via the periodicity of $\phi$ itself. The latter requirement 
derives from the demand that the thermodynamics at temperature $T + \delta T$ to any given accuracy 
must be smoothly derivable from the thermodynamics at temperature $T$ for $\delta T$ sufficiently 
small provided no phase transition occurs at $T$. This is done 
by expanding the right-hand side of the
BPS equation (finite radius of convergence) 
which, in turn, is the starting point for a 
perturbative treatment with expansion parameter $\frac{\delta T}{T}$. 

Linearity, analyticity, and no explicit 
dependence of $\beta$ only allow the BPS equation for $\phi$ to be one 
the two following possibilities:
\eqb\label{bps13}
\pd_\tau\phi=\pm i\,\Lambda_E\,\lambda_3\,\phi 
\eqe
or 
\eqb\label{bps14}
\pd_\tau\phi=\pm i\,\Lambda_E^3\,\lambda_3\,\phi^{-1} 
\eqe
where $\phi^{-1}\equiv \frac{\phi}{|\phi|^2}$. Recall that
\eqb
\phi^{-1} = \phi_0^{-1} \sum_{n=0}^{\infty} (-1)^n \phi_0^{-n} \left(\phi-\phi_0\right)^n
\eqe
has a finite radius of convergence. According to Eqs.\,(\ref{tafel}) and 
(\ref{solutiona}) we may write
\eqb
\label{ansatzBPS}
\phi = |\phi|(\beta,\Lambda_E)\, \times\,\lambda_1\, \exp\left( \mp \frac{2\pi i}{\beta} \lambda_3 \tau\right) \,.
\eqe
Substituting Eq.\,(\ref{ansatzBPS}) into 
Eq.\,(\ref{bps13}) yields
\eqb
\label{contraBPS}
\Lambda_E=\frac{2\pi}{\beta}
\eqe
which is unacceptable since $\Lambda_E$ is a constant scale. For the other possibility 
Eq.\,(\ref{bps14}), we obtain
\eqb
\label{nocontraBPS}
|\phi|(\beta,\Lambda_E)=\sqrt{\frac{\beta\Lambda_E^3}{2\pi}}=\sqrt{\frac{\Lambda_E^3}{2\pi\,T}}\,
\eqe
when substituting Eq.\,(\ref{ansatzBPS}) into Eq.\,(\ref{bps14}). 
This is acceptable and indicates 
that at $T\gg \Lambda_E$ $\phi$'s modulus is small. 
The right-hand side of Eq.\,(\ref{bps14}) defines the 'square-root' $V^{(1/2)}$ of a potential
$V(|\phi|)\equiv\mbox{tr}\,\left(V^{(1/2)}\right)^\dagger\,V^{(1/2)}=\Lambda_E^6 \, \mbox{tr} \, \phi^{-2}$.
The equation of motion (\ref{bps14})
can be derived from the following action:
\eqb    \label{actionphi}
S_{\phi} = \mbox{tr} \, \int_0^\beta d\tau \int d^3x 
\left( \partial_\tau \phi \partial_\tau \phi + \Lambda_E^6 \phi^{-2}  \right)  \,.
\eqe
Notice that due to BPS saturation it is not possible to 
add a constant to the potential in Eq.\,(\ref{actionphi}) without 
changing the ground-state physics. (In fact, adding a constant, the modified 
BPS equation would not admit periodic solutions anymore.) The scale $|\phi|$ must be 
interpreted as the maximal resolution that remains after the spatial coarse-graining over 
calorons and anticalorons is performed. As we shall show later, a 
critical temperature $2\pi T_{c,E}=13.867\,\Lambda_E$ exists. 
Thus, expressing the critical cutoff $|\phi|^{-1}=\sqrt{\frac{2\pi}{\Lambda_E^3\beta_{c,E}}}$ in units of $\beta_{c,E}$, yields 
$8.22$; for $T>T_{c,E}$ this number grows as $(T/T_{c,E})^{3/2}$. But cutting off the $\rho$- and $r$-integration at $>8.22\,\beta$ perfectly 
represents the infinite-volume limit in Eq.\,(\ref{calAdef})!   

The ratios of the mass-squared of $\phi$-field fluctuations, $\pd^2_{|\phi|}\,V(|\phi|)$, and 
the compositeness scale $|\phi|$ squared or $T^2$ are given as 
\eqb
\label{ratiosCC}
\frac{\pd^2_{|\phi|}V_E}{|\phi|^2}=12\,\lambda_E^3\,,\ \ \ \ \ \ \ \ 
\frac{\pd^2_{|\tilde{\phi}_l|}V_E}{T^2}=48\,\pi^2\,\,,
\,
\eqe
where $\lambda_E\equiv\frac{2\pi T}{\La_E}$. We will show in 
Sec.\,\ref{eveffgc} that $\lambda_E\ge 13.867$ in the 
electric phase. Thus both ratios in 
Eq.\,(\ref{ratiosCC}) are much larger than unity: The field 
$\phi$ is quantum mechanically and statistically inert. 
It represents a {\sl background} for the dynamics of 
the topologically trivial sector after spatial coarse-graining. As a consequence, our 
assumption that only noninteracting calorons of trivial 
holonomy contribute to the average in Eq.\,(\ref{reddefphi}) 
is consistent.\vspace{0.1cm}\\ 
\noindent\underline{SU(3) case:}\vspace{0.1cm}\\ 
For SU(3) we write three sets of SU(2) generators as
\eqb
\label{lanot}
{\lambda}_1=\left(\begin{array}{ccc}0&1&0\\ 
1&0&0\\ 
0&0&0\end{array}\right)\,,\ \ \ 
{\lambda}_2=\left(\begin{array}{ccc}0&-i&0\\ 
i&0&0\\ 
0&0&0\end{array}\right)\,,\ \ \ 
{\lambda}_3=\left(\begin{array}{ccc}1&0&0\\ 
0&-1&0\\ 
0&0&0\end{array}\right)\,,
\eqe
and
\eqb
\label{barla}
\bar{\lambda}_1=\left(\begin{array}{ccc}0&0&1\\ 
0&0&0\\ 
1&0&0\end{array}\right)\,,\ \ \ 
\bar{\lambda}_2=\left(\begin{array}{ccc}0&0&-i\\ 
0&0&0\\ 
i&0&0\end{array}\right)\,,\ \ \ 
\bar{\lambda}_3=\left(\begin{array}{ccc}1&0&0\\ 
0&0&0\\ 
0&0&-1\end{array}\right)\,,
\eqe
and
\eqb
\label{tildela}
\tilde{\lambda}_1=\left(\begin{array}{ccc}0&0&0\\ 
0&0&1\\ 
0&1&0\end{array}\right)\,,\ \ \ 
\tilde{\lambda}_2=\left(\begin{array}{ccc}0&0&0\\ 
0&0&-i\\ 
0&i&0\end{array}\right)\,,\ \ \ 
\tilde{\lambda}_3=\left(\begin{array}{ccc}0&0&0\\ 
0&1&0\\ 
0&0&-1\end{array}\right)\,.
\eqe
One generator is dependent. This just reflects the fact that the group manifold 
of SU(3) locally is not $S_3\times S_3\times S_3$ but $S_3\times S_5$ 
\cite{Aguilar1999,Steenrod1951}. A set of independent generators is obtained by 
replacing the two matrices 
$\bar{\lambda}_3$ and $\tilde{\lambda}_3$ by the single matrix
\eqb
\label{GellMann}
\lambda_8=\frac{1}{\sqrt{3}}\left(\bar{\lambda}_3+\tilde{\lambda}_3\right)=
\frac{1}{\sqrt{3}}\left(\begin{array}{ccc}1&0&0\\ 
0&1&0\\ 
0&0&-2\end{array}\right)\, 
\eqe
and by keeping the other matrices. The result is the familiar set 
of Gell-Mann matrices generating the group SU(3). 

For the case of SU(3) the field $\phi$ may 
wind in each of the above SU(2) algebras. Except for the 
points $\tau=0,\frac{\beta}{3},\frac{2\beta}{3}$, where it jumps into a 
new algebra, a solution to the 
BPS equation 
\eqb
\label{BPSSU(3)}
\pd_\tau\phi=\pm i\,\Lambda_E^3\,\left\{\begin{array}{c}\lambda_3\,\frac{\phi}{|\phi|^2}\,,
\ \ \ \ (0\le\tau<\frac{\beta}{3})\nonumber\\ 
\bar{\lambda}_3\,\frac{\phi}{|\phi|^2}\,,\ \ \ \ (\frac{\beta}{3}\le\tau<\frac{2\beta}{3})\nonumber\\ 
\tilde{\lambda}_3\,\frac{\phi}{|\phi|^2}\,,\ \ \ \ (\frac{2\beta}{3}\le\tau<\beta)\,\end{array}\right. 
\eqe
is given as
\eqb
\label{solBPSSU(3)}
\phi(\tau)=\sqrt{\frac{\La_E^3}{2\pi T}}\,\left\{
\begin{array}{c}
\lambda_1\, \exp\left( \mp \frac{2\pi i}{\beta} \lambda_3 \tau\right)\,, \,\ \ \ \ \ \ \ \ \ \ \ \ (0\le\tau<\frac{\beta}{3})\nonumber\\ 
\bar{\lambda}_1\, \exp\left( \mp \frac{2\pi i}{\beta} \bar{\lambda}_3 (\tau-\frac{\beta}{3})\right)\,,\ \ \ \ (\frac{\beta}{3}\le\tau<\frac{2\beta}{3})\nonumber\\ 
\tilde{\lambda}_1\, \exp\left( \mp \frac{2\pi i}{\beta} \tilde{\lambda}_3 (\tau-\frac{2\beta}{3})
\right)\,,\ \ \ \ (\frac{2\beta}{3}\le\tau<\beta)\,.
\end{array}\right.\,
\eqe
Notice that the potential $V_E=2\frac{\Lambda_E^6}{|\phi|^2}$ is the same 
on the configuration $\phi(\tau)$ in Eq.\,(\ref{solBPSSU(3)}) as for the SU(2) case and that by the 
same calculation one shows its quantum mechanical and statistical 
inertness.

\subsection{A macroscopic ground state\label{macgs}}

The action Eq.\,(\ref{actionphi}) governs the dynamics of 
$\phi$. We have not yet included caloron interactions, mediated by the topologically trivial sector, 
which change the holonomy of calorons and induce interactions between 
their (BPS monopole) constituents. This is the objective of the present section.  

\subsubsection{Pure-gauge configuration\label{PGC}}

The action Eq.\,(\ref{actionphi}) can be extended to include 
topologically trivial configurations $a_\mu$. This is accomplished by a minimal
coupling $\partial_\tau \phi \to \partial_\mu \phi + ie[\phi,a_\mu] \equiv D_\mu \phi$
and by adding a kinetic term for these configurations. Here $e$ denotes the {\sl effective} gauge coupling.
The total Yang-Mills action $S$, governing the electric phase, can thus be rewritten as
\eqb \label{actiontotal}
S = \mbox{tr}\, \int_0^\beta d\tau \int d^3x \left( \frac12 G_{\mu\nu}G_{\mu\nu} +
D_\mu \phi D_\mu \phi + \Lambda_E^6 \phi^{-2} \right)
\,,
\eqe
where $G_{\mu\nu}=G^a_{\mu\nu} \frac{\lambda^a}{2}$ and
$G^a_{\mu\nu}=\pd_\mu a^a_\nu-\pd_\nu a^a_\mu+e\,f^{abc}a^b_\mu a^c_\nu$. 
The classical equation of motion for $a_\mu$, 
derived from the action (\ref{actiontotal}), reads
\eqb
\label{gfeom}
D_\mu G_{\mu\nu}=ie[\phi,D_\nu \phi] \,
\eqe
There is a pure-gauge solution $a_\mu^{bg}$ to Eq.\,(\ref{gfeom}) with $D_\nu \phi=0$. 
Thus the total action density of the ground state $(\phi,a_\mu^{bg})$ 
is the potential $V_E=4\pi\,\La_E^3\,T$, corresponding to an energy-momentum tensor 
$\bar{\theta}_{\mu\nu}=V_E\,\delta_{\mu\nu}$ or $P^{gs}=-\rho^{gs}=-4\pi\,\La_E^3\,T$ ($P^{gs},\rho^{gs}$ 
the ground-state pressure and energy density, respectively): The so-far hidden scale
 $\Lambda_E$ becomes (gravitationally) measurable by coarse-grained interactions between calorons.\vspace{0.1cm}\\ 
\noindent\underline{SU(2) case:}\vspace{0.1cm}\\ 
In the background 
\eqb
\label{backsu2}
\phi(\tau)=\sqrt{\frac{\Lambda_E^3}{2\pi\,T}}\,
\lambda_1\, \exp\left( \mp \frac{2\pi i}{\beta}\lambda_3 \tau\right)
\eqe
we have 
\eqb
\label{puregaugesu2}
a_\mu^{bg}=\pm\delta_{\mu 4}\frac{\pi}{e\beta}\,\lambda_3\,.
\eqe
$\mbox{}$\vspace{0.1cm}\\ 
\noindent\underline{SU(3) case:}\vspace{0.1cm}\\ 
In the background $\phi$ of Eq.\,(\ref{solBPSSU(3)}) the pure-gauge 
solution to Eq.\,(\ref{gfeom}) with $D_\nu \phi=0$ reads
\eqb
\label{puregaugesu3}
a_\mu^{bg}=\pm\delta_{\mu 4}\frac{\pi}{e\beta}\,\left\{
\begin{array}{c}\lambda_3\,,\ \ \ \ (0\le\tau<\frac{\beta}{3})\nonumber\\ 
\ \bar{\lambda}_3\,,\ \ \ \ \ (\frac{\beta}{3}\le\tau<\frac{2\beta}{3})\nonumber\\ 
\ \tilde{\lambda}_3\,,\ \ \ \ \ (\frac{2\beta}{3}\le\tau<\beta)\,.
\end{array}\right.
\eqe

\subsubsection{Polyakov loop and rotation to unitary gauge\label{gtug}}

Here we would like to investigate whether the 
ground state, described by $(\phi,a_\mu^{bg})$, is degenerate 
with respect to the global electric $Z_2$ (SU(2)) or 
$Z_3$ (SU(3)) symmetry under which the 
Polyakov loop ${\bf P}$ transforms as ${\bf P}\to Z {\bf P}$ 
where $Z\in Z_2$ (SU(2)) or $Z\in Z_3$ (SU(3)). We will refer to 
the gauge, where $\phi$'s phase is $\tau$ dependent 
as in Eq.\,(\ref{ansatzBPS}) or in Eq.\,(\ref{solBPSSU(3)}), 
as winding gauge. (Microscopically, this is the 
singular gauge for an instanton in which the Harrington-Shepard solution is 
constructed.) \vspace{0.1cm}\\    

\noindent\underline{SU(2) case:}\vspace{0.1cm}\\   
Evaluating the Polyakov loop on the configuration $a^{bg}_\mu$ of Eq.\,(\ref{puregaugesu2}), 
we have
\eqb
\label{Polsu2}
{\bf P}[a^{bg}_\mu]=\exp\left[\pm i\pi\lambda_3\right]=-\UM_2\,.
\eqe
We are searching a gauge transformation $\tilde{\Omega}\in$SU(2) to the unitary 
gauge $\phi=|\phi|\lambda_3$ and $a_\mu^{bg}=0$. A periodic but not 
differentiable gauge transformation $\tilde{\Omega}$ doing this 
can be obtained from a nonperiodic but smooth gauge transformation 
$\Omega$ by multiplication with a local center transformation 
$Z$ and by multiplication with a global gauge transformation $\Omega_{gl}$:
\eqb
\label{gatugsu2}
\tilde{\Omega}=\Omega(\tau) Z(\tau)\Omega_{gl}\,,
\eqe
where $\Omega(\tau)\equiv\exp[\mp i\pi\frac{\tau}{\beta}\lambda_3]$, 
$Z(\tau)=\left(2\Theta(\tau-\frac{\beta}{2})-1\right)\UM_2$, and 
$\Omega_{gl}=\exp[-i\frac{\pi}{4}\lambda_2]$. $\Theta$ denotes the Heavyside step function:
\eqb
\label{Hevayside}
\Theta(x)=\left\{\begin{array}{c}
0\,,\ \ \ \ (x<0)\,,\\ 
\frac{1}{2}\,,\ \ \ \ (x=0)\,,\\ 
1\,,\ \ \ \ \ (x>0)\,.
\end{array}\right.\,.
\eqe
Applying $\tilde\Omega$ 
to $a_\mu=a^{bg}_\mu+\delta a_\mu$, 
where $\delta a_\mu$ is a {\sl periodic} fluctuation in winding gauge, we have
\eab
\label{trafiwtoug}
a_\mu&\rightarrow &\tilde{\Omega}^\dagger(a^{bg}_\mu+\delta a_\mu)\tilde{\Omega}+
\frac{i}{e}\pd_\mu \tilde{\Omega}^\dagger\tilde{\Omega}\nonumber\\ 
&=&\Omega_{gl}^\dagger\left(\Omega^\dagger(a^{bg}_\mu+\delta a_\mu)\Omega+
\frac{i}{e}\left((\pd_\mu\Omega^\dagger)\Omega+(\pd_\mu Z)Z\right)\right)\Omega_{gl}\nonumber\\ 
&=&\Omega_{gl}^\dagger\left(\Omega^\dagger\delta a_\mu\Omega+
\frac{2i}{e}\delta(\tau-\frac{\beta}{2})Z\right)\Omega_{gl}\nonumber\\ 
&=&(\Omega\Omega_{gl})^\dagger\delta a_\mu\Omega\Omega_{gl}\,.
\eae
Since $\Omega\Omega_{gl}(0)=-\Omega\Omega_{gl}(\beta)$ 
we conclude that if the fluctuation $\delta a_\mu$ 
is periodic in winding gauge then it is also periodic in unitary
gauge. It thus is irrelevant whether we integrate out the 
fluctuations $\delta a_\mu$ in winding or unitary gauge in a loop expansion 
of thermodynamical quantities: Hosotani's mechanism \cite{Hosotani1983} 
does not take place. What changes though under 
the gauge transformation $\tilde{\Omega}$ is the Polyakov loop evaluated on 
the background configuration $a^{bg}_\mu$:
\eqb
\label{Polch}
{\bf P}[a^{bg}_\mu]=-\UM_2 \rightarrow {\bf P}[a^{bg}_\mu=0]=\UM_2\,.
\eqe
We conclude that the theory has two equivalent ground states and 
that the global electric $Z_2$ symmetry is dynamically broken. Thus we have shown that 
the elecric phase is {\sl deconfining}.\vspace{0.1cm}\\   

\noindent\underline{SU(3) case:}\vspace{0.1cm}\\   
Let us now compute the Polyakov loop on the configuration 
$a^{bg}_\mu$ of Eq.\,(\ref{puregaugesu3}). We have
\eab
\label{Polsu3}
{\bf P}[a^{bg}_\mu]&=&\exp\left[\pm i\frac{\pi}{3}\tilde{\lambda}_3\right]
\exp\left[\pm i\frac{\pi}{3}\bar{\lambda}_3\right]\exp\left[\pm i\frac{\pi}{3}\lambda_3\right]\nonumber\\ 
&=&\exp\left[i\frac{\pi}{3}(\pm\tilde{\lambda}_3\pm\bar{\lambda}_3\pm\lambda_3)\right]\,.
\eae
The $+$ or $-$ sign can be chosen independently for 
each SU(2) algebra. The following combinations are possible:
\eab
\label{insu2com}
\pm\left(+\tilde{\lambda}_3+\bar{\lambda}_3+\lambda_3\right)&=&\pm 2\,\bar{\lambda}_3\,,\nonumber\\ 
\pm\left(-\tilde{\lambda}_3-\bar{\lambda}_3+\lambda_3\right)&=&\pm 2\,\tilde{\lambda}_3\,,\nonumber\\ 
\pm\left(-\tilde{\lambda}_3+\bar{\lambda}_3+\lambda_3\right)&=&\pm 2\,\lambda_3\,,\nonumber\\ 
\pm\left(+\tilde{\lambda}_3-\bar{\lambda}_3+\lambda_3\right)&=&0\,.
\eae
The corresponding values of the Polyakov loop are
\eab
\label{Polcosu3}
{\bf P}^\pm_1&=&\left(\begin{array}{ccc}\exp[\pm\frac{2\pi i}{3}]&0&0\\ 
0&1&0\\ 
0&0&\exp[\mp\frac{2\pi i}{3}]\end{array}\right)\,,\ \ \ \ \ 
{\bf P}^\pm_2=\left(\begin{array}{ccc}1&0&0\\ 
0&\exp[\pm\frac{2\pi i}{3}]&0\\ 
0&0&\exp[\mp\frac{2\pi i}{3}]\end{array}\right)\,,\nonumber\\  
{\bf P}^\pm_3&=&\left(\begin{array}{ccc}\exp[\pm\frac{2\pi i}{3}]&0&0\\ 
0&\exp[\mp\frac{2\pi i}{3}]&0\\ 
0&0&1\end{array}\right)\,,\ \ \ \ \ \ {\bf P}_4=\UM_3.
\eae
${\bf P}_4$ is a trivial representation of the center group. 
The set ${\bf P}^\pm_1, {\bf P}^\pm_2,{\bf P}^\pm_3$ closes 
under multiplication with the center elements 
${\bf P}=\exp[\pm\frac{2\pi i}{3}]\UM_3,\UM_3$. It is a six 
dimensional, reducible representation of the center group. 
The two three dimensional irreducible representations, which collapse on one another, are spanned by 
\eqb
\label{irrcenter}
\frac{1}{3}\UM_3\left({\bf P}^\pm_1+{\bf P}^\mp_2+{\bf P}^\pm_3\right)\,,\ \ \ 
\frac{1}{3}\exp[\mp\frac{2\pi i}{3}]\UM_3\left({\bf P}^\pm_1+{\bf P}^\mp_2+{\bf P}^\pm_3\right)\,.
\eqe
We conclude that the ground state has a 
$Z_3$ degeneracy: The electric $Z_3$ symmetry is dynamically broken and thus 
we have discussed a {\sl deconfining} phase.

What about a gauge rotation to unitary gauge $a_\mu^{bg}=0$ and 
$\phi=|\phi|\lambda_3$ or $\phi=|\phi|\bar{\lambda}_3$ or $\phi=|\phi|\tilde{\lambda}_3$? 
Such a gauge transformation $\tilde{\Omega}$ is given as
\eqb
\label{gatugsu3}
\tilde{\Omega}=\left\{
\begin{array}{c}\exp[\mp i\frac{\pi}{\beta}\tau\lambda_3]
\exp[-i\frac{\pi}{4}\lambda_2]\,,\ \ \ \ \ \ \ \ \ \ \ \ \ (0\le\tau<\frac{\beta}{3})\nonumber\\ 
\exp[\mp i\pi\frac{\pi}{\beta}(\tau-\frac{\beta}{3})\bar{\lambda}_3]
\exp[-i\frac{\pi}{4}\bar{\lambda}_2]\,,\ \ \ \ \ (\frac{\beta}{3}\le\tau<\frac{2\beta}{3})\nonumber\\ 
\exp[\mp i\frac{\pi}{\beta}(\tau-\frac{2\beta}{3})\tilde{\lambda}_3]
\exp[-i\frac{\pi}{4}\tilde{\lambda}_2]\,,\ \ \ \ \ (\frac{2\beta}{3}\le\tau<\beta)\,.
\end{array}\right.
\eqe
By construction $\tilde{\Omega}$ is periodic, at $\tau=\beta$ it jumps back 
to its value at $\tau=0$, and thus a 
fluctuation $\delta a_\mu$, which is periodic in winding gauge, is 
also periodic in unitary gauge. 

A comment concerning SU($N$) theories with $N\ge 4$ is in order. We only discuss the case when $N$ is even. 
Since at any $\tau$ the maximal number of {\sl independent} SU(2) subgroups contributing with 
calorons of topological charge one to the 
macroscopic field $\phi$ is $N/2$ the dynamical 
gauge-symmetry breaking is not as maximal as 
it is for SU(2) and SU(3). For example, SU(4) breaks only down to SU(2)$^2\times$U(1). 
The question is whether there is a single critical temperature $T_{c,E}$ where 
the unbroken nonabelian subgroups get broken to U(1) factors 
by the condensation of color magnetic monopoles into macroscopic 
adjoint Higgs fields and pure gauges, and where the abelian 
magnetic monopoles with respect to the remaining U(1) factors 
condense as soon as they are created, or whether this is 
a stepwise process. The lattice seems to 
favor the former situation \cite{LuciniTeperWenger2005} which, 
however, does not appear natural to the author.

\subsection{Excitations}

Now that the derivation of a macroscopic ground state 
for the electric phase is completed we are in a position to discuss 
the properties of its on-shell excitations and the role of 
residual quantum fluctuations. The temperature dependent 
mass is a measure for the strength of coupling between gauge modes and 
the nontrivial ground state. The evolution of this 
coupling with temperature is 
a manifestation of the thermodynamical selfconsistency 
of the separation into topological configurations 
and gauge modes with trivial topology. As we 
shall see, this evolution represents a decoupling between 
high and low temperature physics and, for temperatures 
sufficiently above the critical 
temperature $T_{c,E}$, indicates the conservation of 
magnetic charge associated with the 
isolated and screened BPS monopoles that are liberated by 
the dissociation of calorons with a large holonomy.         

\subsubsection{Mass spectrum of thermal quasiparticles}

We first discuss some general aspects of the mass spectrum 
and then specialize to the cases SU(2) and SU(3). We refer to 
gauge modes which acquire a quasiparticle mass on tree level by the adjoint Higgs mechanism  
as tree-level heavy (TLH) and to those which remain 
massless as tree-level massless (TLM). 

In unitary gauge the mass spectrum calculates as 
\eqb
\label{massspectrum}
m_a^2=-2e^2\,\mbox{tr}\,[\phi,t^{a}][\phi,t^{a}]\,,
\eqe
where $t^a$ are the group generators normalized 
as tr\,$t^a t^b=\frac{1}{2}\delta^{ab}$. For SU($N$) off-diagonal generators 
can be chosen as
\eab
\label{TLgen}
t^{IJ}_{rs}&=&1/2\,(\delta_r^I\delta_s^J+\delta_s^I\delta_r^J)\,,\ \ \ 
{\bf t}^{IJ}_{rs}=-i/2\,(\delta_r^I\delta_s^J-\delta_s^I\delta_r^J)\,,\nonumber\\ 
(I&=&1,\cdots,N;\,J>I;\,r,s=1,\cdots,N)\,. 
\eae
This yields
\eqb
\label{TLHmasses}
m_{IJ}^2={\bf m}_{IJ}^2=e^2(\phi_I-\phi_J)^2\,
\eqe
where $\phi_I,\phi_J$ denote the diagonal elements of $\phi$ in unitary gauge.\vspace{0.1cm}\\    
\noindent\underline{SU(2) case:}\vspace{0.1cm}\\  
In this case $\phi$ breaks the gauge symmetry dynamically down to U(1). 
Thus we have one TLM mode and two TLH modes whose degenerate 
masses are, according to Eqs.\,(\ref{backsu2}) and (\ref{TLHmasses}), 
given as 
\eqb
\label{n2masses}
m_{12}^2={\bf m}_{12}^2=4\,e^2|\phi|^2=4\,e^2 \frac{\Lambda_E^3}{2\pi T}\,.
\eqe
\noindent\underline{SU(3) case:}\vspace{0.1cm}\\  
Here $\phi$ is diagonal in either one of the three SU(2) subgroups, and the gauge symmetry is 
dynamically broken to U(1)$^2$. We have
\eab
\label{n3masses}
m_{12}^2&=&{\bf m}_{12}^2=4\,e^2\,\frac{\Lambda_E^3}{2\pi T}\,,\nonumber\\ 
m_{13}^2&=&{\bf m}_{13}^2=m_{23}^2={\bf m}_{23}^2=e^2 \,\frac{\Lambda_E^3}{2\pi T}\,,
\ \ \ \ \ \mbox{or}\nonumber\\ 
m_{13}^2&=&{\bf m}_{13}^2=4\,e^2\,\frac{\Lambda_E^3}{2\pi T}\,,\nonumber\\ 
m_{12}^2&=&{\bf m}_{12}^2=m_{23}^2={\bf m}_{23}^2=e^2 \,\frac{\Lambda_E^3}{2\pi T}\,,
\ \ \ \ \ \mbox{or}\nonumber\\ 
m_{23}^2&=&{\bf m}_{23}^2=4\,e^2\,\frac{\Lambda_E^3}{2\pi T}\,,\nonumber\\ 
m_{12}^2&=&{\bf m}_{12}^2=m_{13}^2={\bf m}_{13}^2=e^2 \,\frac{\Lambda_E^3}{2\pi T}\,.
\eae
For $T$ sufficiently far above $T_{c,E}$ we will 
see in Sec.\,\ref{eveffgc} that $e$ is practically 
independent of $T$. As a consequence, the TLH masses 
die off according to the power law in 
Eqs.\,(\ref{n2masses}) and (\ref{n3masses}). Moreover, the energy density and the pressure of the 
ground state are only linear in $T$, $P^{gs}=-\rho^{gs}=-4\pi\,\La_E^3\,T$. 
We will show in Sec.\,\ref{Radcor} that the one-loop 
result for thermodynamical quantities is reliable on the $0.1$\% level 
throughout the electric phase. Thus the Stefan-Boltzmann 
limit $P=\frac{\pi^2}{15}T^4$ and $\rho=\frac{\pi^2}{5}T^4$ (SU(2)) 
or $P=\frac{8\pi^2}{45}T^4$ and $\rho=\frac{8\pi^2}{15}T^4$ (SU(3)) 
is reached very quickly apart from a factor arising from the extra polarizations of TLH modes. This 
result is in agreement with early 
lattice simulations using the differential method.

\subsubsection{Thermodynamical selfconsistency}

In this section we provide a conceptual basis for the notion of 
thermodynamical selfconsistency. 

As a result of the existence of a nontrivial macroscopic 
ground state, which is built of interacting calorons of 
topological charge-one, the ground-state physics and the 
properties of the excitations are temperature 
dependent. We have discussed in Sec.\,\ref{macgs} 
how the temperature dependence of the ground-state pressure 
and its energy density arises due to caloron 
interactions. These interactions are 
encoded in a pure-gauge configuration 
on the macroscopic level. For topologically 
trivial fluctuations $\delta a_\mu$ the two following properties are induced 
by the ground-state physics. First, in unitary gauge off-Cartan fluctuations 
are massive in a temperature dependent way, compare with 
Eqs.\,(\ref{n2masses}) and (\ref{n3masses}). Notice that the 
quasiparticle masses are related to the ground-state 
pressure by their respective dependences on temperature 
if the temperature dependence of the 
effective gauge coupling $e$ is known. Second, there are 
constraints for the maximal off-shellness of the fluctuations 
$\delta a_\mu$ arising from the compositeness of 
the ground-state field $\phi$. Namely, a fluctuation 
$\delta a_\mu$, which was generated in a thermal equilibrium situation 
by the ground state, is not capable of 
destroying this ground state: In a physical gauge vacuum fluctuations or scattering 
processes with momenta or momentum transfers larger 
than the (temperature dependent) 
scale $|\phi|$ are forbidden.  

Thermodynamical quantities such as the pressure, 
the energy density, or the
entropy density are interrelated by Legendre transformations. 
These transformations can be derived from the partition function 
of the fundamental theory where the temperature dependence only occurs in 
an explicit way through the Boltzmann weight. A reformulation of the 
theory into a spatially coarse-grained Lagrangian, where certain parameters are 
temperature dependent (implicit temperature dependences) 
by virtue of a separation into a ground state and (very weakly interacting) excitations, 
see Eq.\,(\ref{actiontotal}), must not alter 
the Legendre transformations between thermodynamical quantities. 
For this to be true in the effective theory 
one needs to impose that in each transformation law the total 
derivatives with respect to temperature involves the explicit 
temperature dependences only. That is, derivatives with respect to implicit 
temperature dependences ought to cancel in a given Legendre 
transformation. 

A particular and essential Legendre transformation maps the total pressure onto the total 
energy density as
\eqb
\label{rhoPree}
\rho=T\frac{dP}{dT}-P\,.
\eqe
If the effective theory has temperature dependent quasiparticle fluctuations of mass 
$m_a=c_a\,m$, where $c_a$ are dimensionless constants,  
and a ground-state pressure $P^{gs}$, which can be regarded a function of $m$, 
then the condition of thermodynamical selfconsistency 
is expressed as \cite{Gorenstein1995}
\eqb
\label{TSCm}
\pd_m P=0\,.
\eqe
Eq.\,(\ref{TSCm}) assures that 
in Eq.\,(\ref{rhoPree}) only derivatives with respect to the 
explicit temperature dependence of $P$ contribute since 
$\frac{dP}{dT}=\pd_T P+\pd_m P\frac{d\,m}{d\,T}$. In $\pd_m P=0$ the 
derivative of the pressure contribution arising from fluctuations cancels against that 
arising from the ground state. Eq.\,(\ref{TSCm}) governs 
the temperature evolution of the effective coupling $e$ 
at any loop order that $P$ is expanded in.

The higher the loop order the more 
complicated the implementation of Eq.\,(\ref{TSCm}). 
In Sec.\,\ref{eveffgc} we perform an analysis on the 
one-loop level (noninteracting quasiparticles) 
which is more than sufficient 
for many practical purposes. A discussion of Eq.\,(\ref{TSCm}) on 
the two-loop level is given in Sec.\,\ref{loopexp}.

\subsubsection{Compositeness constraint and pressure at one loop}

We work in a 
physical gauge where TLM modes are transverse 
(two polarizations, Coulomb gauge with respect to the 
unbroken gauge group), and where TLH modes have three polarizations and do not 
interact with the pure-gauge configuration of the ground state (unitary gauge), 
for details see Sec.\,\ref{Radcor}. 
This is a physical gauge fixing which needs no introduction of additional fields 
since the Coulomb ghosts decouple from the dynamics. 

\begin{figure}
\begin{center}
\leavevmode
\leavevmode
\vspace{2.5cm}
\includegraphics{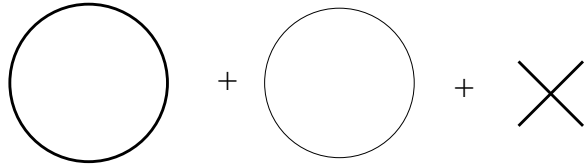}
\end{center}
\caption{Diagrams contributing to the pressure when radiative corrections are 
ignored. A thick line corresponds to TLH modes and a 
thin one to TLM modes. The cross depicts the ground-state contribution 
that arises from interacting calorons.  
\label{diagpress}}      
\end{figure}

On the one-loop level, see Fig.\,\ref{diagpress}, there are no interactions 
between the fluctuations $\delta a_\mu$. 
The only relevant compositeness 
constraint, related to the maximal off-shellness of a 
quantum fluctuation created by the ground state, thus is
\eab
\label{comconnoi}
|p^2-m_a^2|&\le & |\phi|^2\ \ \ \ (\mbox{Minkowskian signature})\nonumber\\ 
p^2+m_a^2&\le & |\phi|^2\ \ \ \ (\mbox{Euclidean signature})\,.
\eae
\noindent\underline{SU(2) case:}\vspace{0.1cm}\\ 
Before we discuss the thermal contribution to the 
one-loop pressure let us investigate what the 
constraints (\ref{comconnoi}) imply for the one-loop 
quantum correction $-\Delta V_E$. Setting $m_a\equiv 0$ in (\ref{comconnoi}) and considering 
two polarizations for TLM and 
three polarizations for TLH modes, 
an upper bound on $|\Delta V_E|$ can be obtained as
\eqb
\label{1loopvacbubble}
|\Delta V_E|<\frac{1}{\pi^2}\int_0^{|\phi|}dp\,
 p^3\log\left(\frac{p}{|\phi|}\right)=\frac{|\phi|^4}{16\pi^2}\,.
\eqe
Thus we have
\eqb
\label{ratioVE}
\left|\frac{\Delta V_E}{V_E}\right|<
\frac{1}{32\pi^2}\left(\frac{|\phi|}{\La_E}\right)^6=
\frac{\lambda_E^{-3}}{32\pi^2}\,.
\eqe
Since $\lambda_E>13.867$ when omitting $\Delta V_E$ 
in the one-loop evolution of $e$, see Sec.\,\ref{eveffgc}, this omission 
is justified: One-loop quantum corrections 
to the pressure are suppressed as compared to the tree-level 
result by a factor less than $2\times 10^{-6}$. 

Omitting the tiny quantum part, the one-loop expression for 
the pressure, including the ground-state contribution, reads:
\eqb
\label{treelPsu2}
P(\lambda_E)=-\La_E^4\left\{\frac{2\lambda_E^4}{(2\pi)^6}\left[2\bar{P}(0)+
6\bar{P}(2a)\right]+2\lambda_E\right\}\, 
\eqe
where $\lambda_E\equiv\frac{2\pi T}{\La_E}$ and 
\eqb
\label{aofela}
a\equiv\frac{m}{2T}=e\frac{|\phi|}{T}=e\sqrt{\frac{\La_E^3}{2\pi T^3}}=2\pi\,e\lambda_E^{-3/2}\,.
\eqe
The (negative) function $\bar{P}(a)$ is defined as
\eqb
\label{P(y)}
\bar{P}(a)\equiv\int_0^\infty dx \,x^2\,\log[1-\exp(-\sqrt{x^2+a^2})]\,.
\eqe
\noindent\underline{SU(3) case:}\vspace{0.1cm}\\ 
Here a similar estimate for the quantum contribution to the 
one-loop pressure holds as in Eq.\,(\ref{1loopvacbubble}), 
and thus, again, we only have to consider the thermal part. 
It reads   
\eqb
\label{treelPsu3}
P(\lambda_E)=-\La_E^4\left\{\frac{2\lambda_E^4}{(2\pi)^6}\left[4\bar{P}(0)+3\left(
4\bar{P}(a)+2\bar{P}(2a)\right)\right]+
2\lambda_E\right\}\,
\eqe
with the same definitions as in the SU(2) case.

\subsubsection{One-loop evolution of effective gauge coupling\label{eveffgc}}

\begin{figure}
\begin{center}
\leavevmode
\leavevmode
\vspace{4.5cm}
\includegraphics{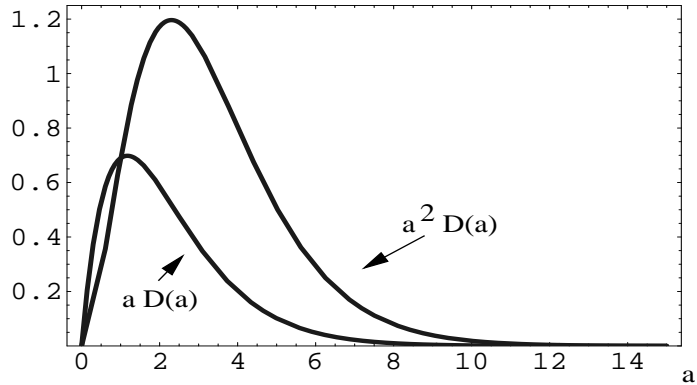}
\end{center}
\caption{The functions $a\,D(a)$ and $a^2\,D(a)$. \label{DAf}}      
\end{figure}

Let us now implement the condition (\ref{TSCm}). We have
\eqb
\label{TSCa}
\pd_m P=0\ \ \Leftrightarrow\ \ \pd_{(aT)} P=0\,.
\eqe
\noindent\underline{SU(2) case:}\vspace{0.1cm}\\ 
Substituing Eq.\,(\ref{treelPsu2}) 
into Eq.\,(\ref{TSCa}) yields the following evolution 
equation
\eqb
\label{eeq}
\pd_a \lambda_E=-\frac{24\,\lambda_E^4\,a}{(2\pi)^6}\frac{D(2a)}{1+\frac{24\,\lambda_E^3\,a^2}{(2\pi)^6}\,D(2a)}\,,
\eqe
where the function $D(a)$, see Fig.\,\ref{DAf}, 
is defined as
\eqb
\label{DA}
D(a)\equiv \int_0^{\infty} dx\,
\frac{x^2}{\sqrt{x^2+a^2}}\frac{1}{\exp(\sqrt{x^2+a^2})-1}\,.
\eqe
\noindent\underline{SU(3) case:}\vspace{0.1cm}\\ 
The evolution equation is obtained by substituting Eq.\,(\ref{treelPsu3}) into Eq.\,(\ref{TSCa}):
\eqb
\label{eeq3}
\pd_a \lambda_E=-\frac{12\,\lambda_E^4\,a}{(2\pi)^6} 
\frac{D(a)+2\,D(2a)}{1+\frac{12\,\lambda_E^3\,a^2}{(2\pi)^6}\,\left(D(a)+2\,D(2a)\right)}\,.
\eqe
Each of the Eqs.\,(\ref{eeq}) and (\ref{eeq3}) has two fixed points, 
one at $a=0$ and one at $a=\infty$, see Fig.\,\ref{DAf}. The points $\lambda_{P,E}\equiv\lambda_E(a\to 0)$ and 
$\lambda_{c,E}\equiv\lambda_E(a=\infty)$ are 
associated with the highest and the lowest temperatures, respectively, which are attainable 
in the electric phase. 

Above $\lambda_{P,E}$ the 
ground state would be trivial (topological fluctuations are absent) and thus 
no tree-level quasiparticle mass were generated by a caloron induced Higgs mechanism. 
This re-introduces the problem of the diverging loop expansion as it is encountered 
in thermal perturbation theory \cite{Linde1980} and 
thus makes the thermalized Yang-Mills theory inconsistent 
as an interacting field theory. We conclude that $\lambda_{P,E}$ marks the 
point in temperature where the field theoretic implementation 
of the gauge principle breaks down. For a physics model, whose gauge group is a product of 
SU(2) and/or SU(3) groups, we expect that 
$\lambda_{P,E}\sim \frac{2\pi M_P}{\La_E}$ where 
$M_P=1.22\times\,10^{19}\,$GeV is the Planck mass. 

As we shall see below, the point $\lambda_{c,E}$, where each 
tree-level quasiparticle mass diverges, marks a transition 
to a (pre-confining) phase with condensed magnetic monopoles, dynamically broken dual 
gauge symmetries U(1)$_D$ (SU(2)) and U(1)$_D^2$ (SU(3)), and isolated but instable 
center-vortex loops: The magnetic phase.   

Notice that the right-hand sides of Eqs.\,(\ref{eeq}) and (\ref{eeq3}) 
are negative definite. As a consequence, the solutions $\lambda_E(a)$ 
to Eqs.\,(\ref{eeq}) and (\ref{eeq3}) can be inverted, and, according to Eq.\,(\ref{aofela}), 
one obtains the evolution of the effective gauge coupling $e$ with temperature as 
\eqb
\label{coupP}
e(\lambda_E)=\frac{1}{2\pi}a(\lambda_E)\lambda_E^{3/2}\,.
\eqe
Inspecting the right-hand of Eqs.\,(\ref{eeq}) and (\ref{eeq3}) 
in the vicinity of the point $\lambda_{c,E}$, it follows 
with Eq.\,(\ref{coupP}) that $e$ diverges logarithmically at $\lambda_{c,E}$:
\eqb
\label{divloglae}
e(\lambda_E)\sim -\log(\lambda_E-\lambda_{c,E})\,.
\eqe
\begin{figure}
\begin{center}
\leavevmode
\leavevmode
\vspace{4.8cm}
\includegraphics{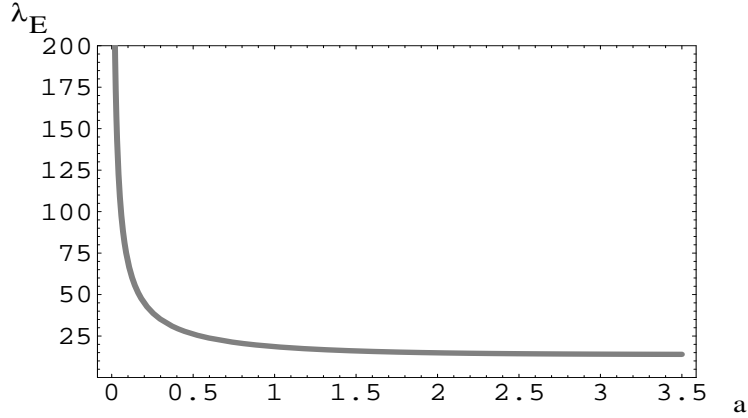}
\end{center}
\caption{The solution $\lambda_E(a)$ to Eq.\,(\protect\ref{eeq}) subject to the initial 
condition $\lambda_{P,E}=10^7$.\label{laofa}}      
\end{figure}
In Fig.\,\ref{laofa} a solution to Eq.\,(\ref{eeq}) subject to 
the initial condition $\lambda_{P,E}\equiv\lambda_E(a=0)=10^7$ is shown. 
We have noticed numerically that the low-temperature behavior of $\lambda_E(a)$ 
is practically independent of the value $\lambda_{P,E}$ as long as $\lambda_{P,E}$ is 
sufficiently large. Let us show this analytically. For $a$ sufficiently smaller than unity we may 
expand the right-hand side of Eq.\,(\ref{eeq}) only taking the 
linear term in $a$ into account. In this regime, the 
inverse of the solution $\lambda_E(a)$ is of the following form
\eqb
\label{anasolv}
a\propto \lambda_E^{-3/2}\sqrt{1-\left(\frac{\lambda_E}{\lambda_{E,P}}\right)^3}\,.
\eqe
If $\lambda_E$ is sufficiently smaller than $\lambda_{P,E}$ then this can be approximated as
\eqb
\label{anasolapp}
a\propto\lambda_E^{-3/2}\,.
\eqe
Thus the dependence in Eq.\,(\ref{anasolapp}) 
is an {\sl attractor} of the downward-in-temperature evolution 
as long as $a$ remains sufficiently small: If we are only interested 
in the behavior of the theory not too far above $\lambda_{c,E}$ 
then it is {\sl irrelevant} what the value of 
$\lambda_{P,E}$ is as long as 
it is sizably larger than $\lambda_{c,E}$. 
This result is reminiscent of the 
ultraviolet-infrared perturbative decoupling property of the renormalizable, 
underlying theory. Notice that the dependence of $a$ on $\lambda_E$ in 
Eq.\,(\ref{anasolapp}) is canceled by the explcicit dependence of 
$e$ on $\lambda_E$ in Eq.\,(\ref{coupP}). Thus a plateau value $e(\lambda_E)=\mbox{const}$ 
is reached rapidly. 

\begin{figure}
\begin{center}
\leavevmode
\leavevmode
\vspace{5.5cm}
\includegraphics{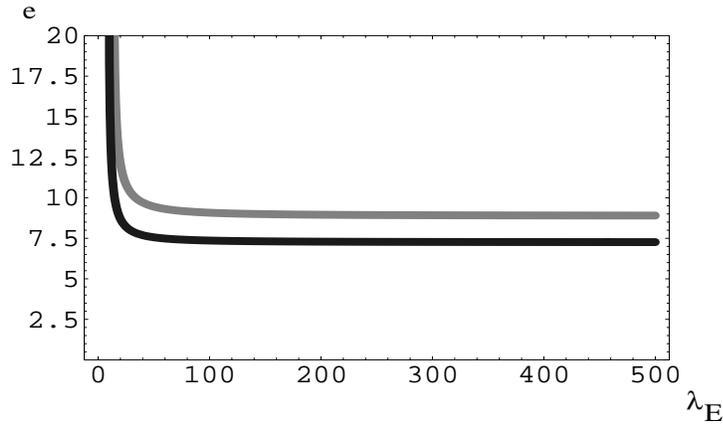}
\end{center}
\caption{The temperature evolution of the gauge 
coupling $e$ in the electric phase for 
SU(2) (grey line) and SU(3) (black line). The gauge coupling 
diverges logarithmically, $e\propto -\log(\lambda_{E}-\lambda_{c,E})$, at 
$\lambda_{c,E}=13.867$ (SU(2)) and $\lambda_{c,E}=9.475$ SU(3). 
The respective plateau values are $e=8.89$ and $e=7.26$.\label{eoflam}}      
\end{figure}

In Fig.\,\ref{eoflam} the temperature dependence of $e$ 
for SU(2) and SU(3), subject to the initial condition $\lambda_{P,E}=10^7$, is shown 
for $\lambda_E\le 500$. 
Before we interpret our results a remark on the 
interpretation of the effective gauge coupling constant $e$ is in order. 
Since $e$ determines the 
strength of the interaction between topologically trivial gauge 
field fluctuations $\delta a_\mu$ 
and the {\sl macroscopic} manifestation $\phi$ of interacting calorons in the ground state 
$e$ is {\sl not} equal to the perturbatively generated 
gauge coupling constant $\bar{g}$ of the fundamental Yang-Mills theory. 
From Fig.\,\ref{eoflam} we see that the effective gauge coupling $e$ 
evolves to values larger than unity. Naively, one would 
conclude that the theory is strongly coupled and 
that the one-loop evolution of $e$ contradicts 
itself. This, however, is an incorrect conclusion. Because of 
compositeness constraints and the infrared regularization provided by 
TLH masses higher loop orders turn out to be very small 
as compared to the one-loop result for the pressure, 
see Sec.\,\ref{loopexp}. 

We interpret the fact that the 
gauge coupling constant $e$ is constant for $\lambda_E$ sizably 
larger than $\lambda_{c,E}$ as another indication for the 
existence of spatially isolated and conserved 
magnetic charges in the system. These 
charges are screened by calorons with a 
small holonomy, and thus $e$ measures 
the {\sl effective} magnetic charge of a BPS monopole which 
is given as $g=\frac{4\pi}{e}$. 
The plateau values for $e$ are $e\sim 8.89$ (SU(2)) and $e\sim 7.26$ (SU(3)). 
At $\lambda_{c,E}=13.867$ (SU(2)) or $\lambda_{c,E}=9.475$ (SU(3)) 
both the core size $\sim \frac{\beta}{e}$ of a screened BPS 
monopole and its mass $\sim \frac{4\pi}{e}\,u_{\tiny\mbox{max}}=\frac{4\pi^2}{e\beta}$  
{\sl vanish}, see Sec.\,\ref{BPSM}. (Notice that monopoles are very massive at high temperatures.) 
Thus monopoles are not well separated anymore at $\lambda_{c,E}$ because it is extremely easy to move them. Local magnetic 
charge conservation is violated since 
in a typical volume $\beta^3$ the number 
of monopoles no longer is conserved. The increasing mobility of monopoles and the 
increasing violation of local 
charge conservation can be seen 
in the evolution of $e$ for 
$\lambda_E\searrow\lambda_{c,E}$ 
where an increase of $e$ as 
compared to the plateau value is observed, 
see Fig.\,\ref{eoflam} and Eq.\,(\ref{divloglae}). 

For $\lambda_E\searrow\lambda_{c,E}$ TLH modes decouple (their masses diverge) and 
thus the (small) interaction between TLM modes 
dies off. This is the macroscopic 
manifestation of the fact that the magnetic 
charge of (dynamical and screened) BPS monopoles vanishes at $\lambda_{c,E}$ 
making them unavailable as `scattering centers' 
for TLM modes. This is, indeed, seen as a result 
of a two-loop calculation of the SU(2) 
pressure in the electric phase \cite{HerbstHofmannRohrer2004,SHG2006}, 
see Sec.\,\ref{loopexp}.            

\subsection{Radiative corrections for SU(2)\label{Radcor}}

\subsubsection{Electric screening mass for TLM modes\label{elscmass}}

In this section we investigate for SU(2) the one-loop contribution 
to the electric screening mass for the TLM mode. 

\begin{figure}
\begin{center}
\leavevmode
\leavevmode
\vspace{3.5cm}
\includegraphics{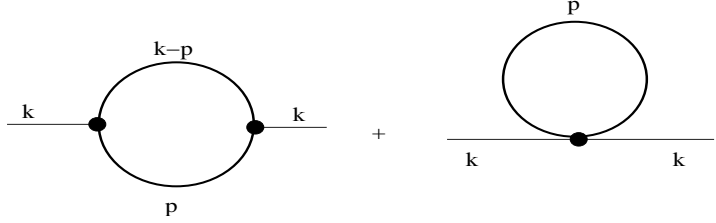}
\end{center}
\caption{Diagrams for the TLM polarization tensor at one loop. 
Thick and thin lines denote TLH and TLM propagation, respectively.\label{Fig3e}}      
\end{figure}

In unitary gauge the analytically continued propagator of 
a free TLH mode $D^{\tiny\mbox{TLH},IJ,0}_{\mu\nu,ab}(k,T)$ 
is that of a massive vector boson \cite{LandsmanWeert1987}
\eab
\label{TLHprop} 
D^{\tiny\mbox{TLH},IJ,0}_{\mu\nu,ab}(k,T)&=&-\delta_{ab}
\left(g_{\mu\nu}-\frac{k_\mu k_\nu}{4\,(e|\phi|)^2}\right)\times\nonumber\\
&&\left[\frac{i}{k^2-4\,(e|\phi|)^2}+2\pi\delta(k^2-4\,(e|\phi|)^2)n_B(|k_0|/T)\right]\,,
\eae
where $n_B(x)=\frac{1}{\e^x-1}$ is the Bose distribution and 
$k_0=\pm\sqrt{\vec{k}^2+4\,(e|\phi|)^2}$. The electric (or Debye) 
screening mass $m_D$ is related \cite{Linde1980} to the 00-component of the polarization 
tensor $\Pi_{\mu\nu}(k)$ in the limit $k_0=0,\vec{k}\to 0$. 
$\Pi_{\mu\nu}(k)$ is calculated according to 
the diagrams in Fig.\,\ref{Fig3e}.
The vertices for the interactions of TLH and TLM modes are the usual ones. In addition to 
the compositeness constraint (\ref{comconnoi}) we have constraints  associated with 
the maximally allowed momentum transfer in a four vertex. Let the three independent momenta be $p_1, p_2$, 
and $p_3$. Then we have \cite{HofmannSept2006}:  
\eab
\label{comcon4vert}
|(p_1+p_2)^2|&\le&|\phi|^2\,,\ \ \ (s\ \mbox{channel})\ \ \ \ \ \ \ \ \
|(p_3-p_1)^2|\le|\phi|^2\,,\ \ \ (t\ \mbox{channel})\nonumber\\ 
|(p_2-p_3)^2|&\le&|\phi|^2\,,\ \ \ (u\ \mbox{channel})\,.
\eae
One can easily see that these 
constraints together with the constraint (\ref{comconnoi}) 
do not allow for any (neither from quantum nor from thermal propagation of the intermediate states) 
contribution of the local (or tadpole) diagrams in Fig.\,\ref{Fig3e} in the 
limit $k_0=0,\vec{k}\to 0$ and for $e_{\tiny\mbox{plateau}}=8.89$ (SU(2)). A calculation of the nonlocal 
diagram reveals that only thermal intermediate 
states contribute in the limit $k_0=0$ and for $e=13.867$. 
Notice that in perturbation theory the nonlocal diagram does not contribute 
to $\Pi_{00}(k_0=0,\vec{k})$. The result for $\Pi_{00}(k_0=0,\vec{k})$ reads\footnote{The author 
would like to thank Ulrich Herbst for performing this calculation.}
\eab
\label{Pi00limit}
&&\Pi_{00}(k_0=0,\vec{k})=\nonumber\\ 
&&\frac{ie^2}{2\pi}\left\{\frac{1}{2}\left(\frac{12}{|\vec{k}|}+
4\frac{|\vec{k}|}{4(e|\phi|)^2}+\frac{|\vec{k}|^3}{(4(e|\phi|)^2)^2}\right)\times\right.\nonumber\\ 
&&\left.\int_{\frac{|\vec{k}|}{2}}^\infty d|\vec{p}|\,|\vec{p}|\sqrt{|\vec{p}|^2+4(e|\phi|)^2} 
\left[n_B\left(\frac{\sqrt{|\vec{p}|^2+4(e|\phi|)^2}}{T}\right)\right]^2-\right.\nonumber\\ 
&&\left.\left(4|\vec{k}|+\frac{|\vec{k}|^3}{4(e|\phi|)^2}\right)
\int_{\frac{|\vec{k}|}{2}}^\infty d|\vec{p}|\,\frac{|\vec{p}|}{\sqrt{|\vec{p}|^2+4(e|\phi|)^2}}
\left[n_B\left(\frac{\sqrt{|\vec{p}|^2+4(e|\phi|)^2}}{T}\right)\right]^2\right\}\,.\nonumber\\ 
\eae
Thus $\Pi_{00}(k_0=0,\vec{k})$ is purely 
{\sl imaginary} and {\sl diverges} for $|\vec{k}|\to 0$: The electric screening 
mass $m_D$, which is the positive square root of $\Pi_{00}(k_0=0,\vec{k}\to 0)$, 
has an infinite and positive real part at {\sl finite} coupling $e$: Static 
electric fields of long wavelength are completely screened 
by calorons. For $e\to\infty$ the Boltzmann factors 
in the integrals in Eq.\,(\ref{Pi00limit}) make $\Pi_{00}(k_0=0,\vec{k})$ 
vanish at {\sl any} momentum $\vec{k}$.

\subsubsection{Two-loop result for the SU(2) pressure\label{loopexp}}

\begin{figure}
\begin{center}
\leavevmode
\leavevmode
\vspace{3.5cm}
\includegraphics{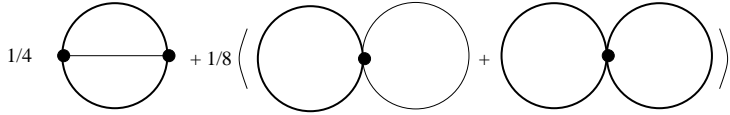}
\end{center}
\caption{Nonvanishing two-loop diagrams contributing to the pressure. Thick lines denote propagators of TLH modes, 
thin lines those of TLM modes.\label{looppress}}      
\end{figure}

The nonvanishing two-loop diagrams contributing to the SU(2) pressure 
in the electric phase \cite{LandsmanWeert1987} are shown in Fig.\,\ref{looppress}. 
Because of the strong screening of near-to-static electric 
modes, compare with Sec.\,\ref{elscmass}, the 
TLM propagator in Coulomb gauge can safely be approximated as
\eqb
\label{TLMprop}
D^{\tiny\mbox{TLM},0}_{\mu\nu,ab}(k,T)=-\delta_{ab}\left\{P^T_{\mu\nu}
\left(\frac{i}{k^2}+2\pi\delta(k^2)n_B(|k_0|/T)\right)-i\frac{u_\mu u_\nu}{\vec{k}^2}\right\}\,
\eqe
where 
\eab
\label{PT}
P^T_{00}&=&P^T_{0i}=P^T_{i0}=0\,,\nonumber\\ 
P^T_{ij}&=&\delta_{ij}-\frac{k_ik_j}{\vec{k}^2}\,,
\eae
$k_0=\pm |\vec{k}|$, and $u_\mu=\delta_{0\mu}$. 

The result of a calculation of the diagrams in 
Fig.\,\ref{looppress} was published in \cite{HerbstHofmannRohrer2004}. 
We do only outline this (lengthy) calculation here. 

\begin{figure}
\begin{center}
\leavevmode
\leavevmode
\vspace{10.0cm}
\includegraphics{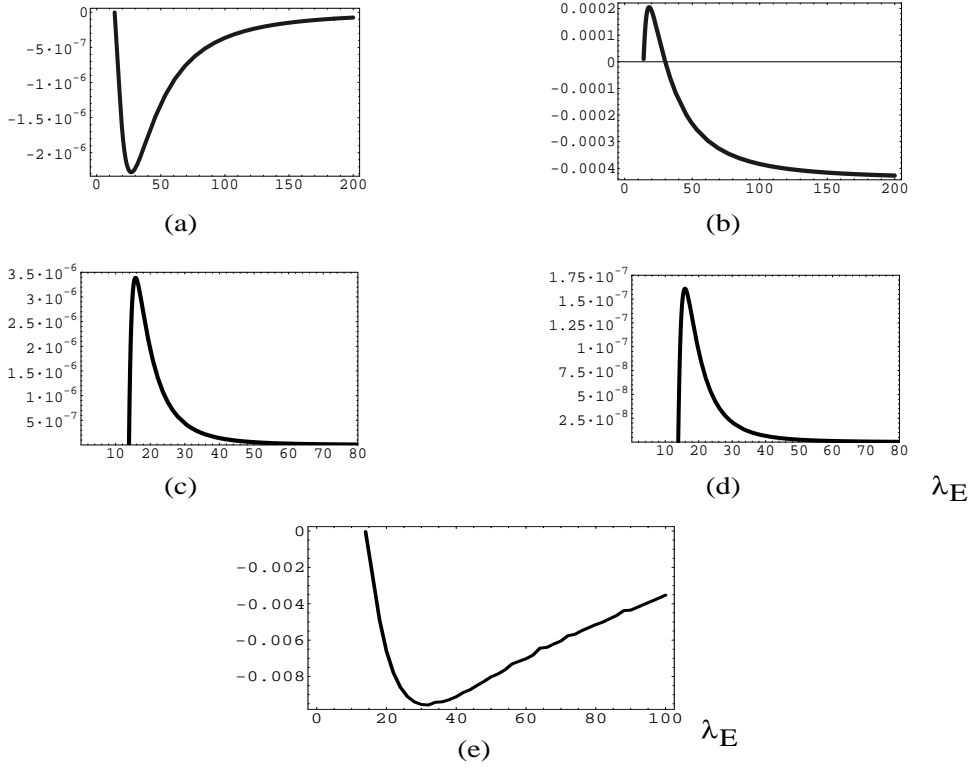}
\end{center}
\caption{Ratios (a) $\frac{1}{8 P_{1-loop}}\Delta P^{HH}_{tt}$ , 
(b) $\frac{1}{8 P_{1-loop}}\Delta P^{HM}_{tt}$, (c) $>\frac{1}{8 P_{1-loop}}\Delta P^{HM}_{tv}$, 
(d) $>\frac{1}{8 P_{1-loop}}\Delta P^{HM}_{tc}$, and (e) $\frac{1}{4 P_{1-loop}}\left(\Delta P^{HHM}_{ttv}+
\Delta P^{HHM}_{ttc}\right)$ as functions 
of temperature. 
\label{Fig7b}}      
\end{figure}

The following nomenclature is useful. Each diagram can be split into contributions 
arising from the vacuum ($v$), the Coulomb ($c$) ($\propto u_\mu u_\nu$ in Eq.\,(\ref{TLMprop})), and the thermal ($t$) parts of the 
involved propagators. Moreover, if a TLH propagator 
contributes to a given diagram then this situation is abreviated by 
$H$, in the other case by $M$. With $e$ being larger than the one-loop plateau value $e_{\tiny\mbox{plateau}}=8.89$ 
the compositeness constraint in Eq.\,(\ref{comconnoi}) allows for the five following two-loop radiative 
corrections to the pressure only:
\eqb
\label{deltacorrpre}
\frac{1}{8}\Delta P^{HH}_{tt}\,,\ \ 
\frac{1}{8}\Delta P^{HM}_{tt}\,,\ \ 
\frac{1}{8}\Delta P^{HM}_{tv}\,,\ \ 
\frac{1}{8}\Delta P^{HM}_{tc}\,,\ \ 
\frac{1}{4}\left(\Delta P^{HHM}_{ttv}+
\Delta P^{HHM}_{ttc}\right)\,.
\eqe
The ratio of the two-loop corrections and the one-loop result 
(excluding the ground-state contribution) is plotted in Fig.\,\ref{Fig7b}. 
For $\frac{1}{8 P_{1-loop}}\Delta P^{HM}_{tv}$ the plot represents only an 
upper bound for the modulus of the correction, all other plots 
are exact results. The dominating 
correction $\frac{1}{4 P_{1-loop}}\Delta P^{HHM}_{ttv}$ 
arises from the nonlocal diagram in 
Fig.\,\ref{looppress}. 
It is negative. The dip is microscopically 
explained by the increasing effect of TLM modes scattering off of decreasingly massive 
magnetic monopoles close to the phase 
transition at $\lambda_{c,E}=13.867$. The effect 
of this scattering is suppressed for $\lambda_E\gg \lambda_{c,E}$ 
since then the monopoles are dilute and massive scattering centers. (Recall that their 
mass is $\propto \frac{T}{e}$ after screening.)
Notice, that the presence of massive but dilute 
scattering centers causes the contribution 
$\frac{1}{8 P_{1-loop}}\Delta P^{HM}_{tt}$ to remain finite but small 
for asymptotically high temperatures.     

A comment concerning thermodynamical selfconsistency on the two-loop level is in order. 
Recall that on the one-loop level 
we have obtained an evolution equation from the requirement of thermal 
selfconsistency $\pd_m P=0$. This gave a functional relation between temperature and mass  
which could be inverted for all temperatures in the electric phase. 
After the relation Eq.\,(\ref{coupP}) between 
coupling constant $e$ and mass $a$ was exploited we obtained 
a functional dependence of the effective gauge coupling constant 
$e$ on temperature. Equivalently, we could have 
demanded $\pd_e P=0$ since $e$ is the only variable parameter (apart from the scale $\La_E$) 
of the effective theory in the electric phase. This would {\sl directly} have 
generated an evolution equation for temperature 
as a function of $e$. Because each diagram comes with a prefactor $e^2$ 
and due to the compositeness constraints 
radiative corrections $\Delta P$ to the 
pressure have a separate dependence on $a$ and $e$, 
\eqb
\label{radpressure}
\Delta P=T^4 \Delta\tilde{P}(e,a,\lambda_E)\,,
\eqe
where $\Delta\tilde{P}$ is a dimensionless function of its dimensionless arguments. 
To implement thermodynamical selfconsistency by demanding $\pd_m P=0$ one has to 
express the explicitly appearing $e$ in Eq.\,(\ref{radpressure}) 
in terms of $a$ by means of Eq.\,(\ref{coupP}) and distinguish temperature 
dependences arising from a simple rescaling and those 
arising from the $T$ dependent ground-state physics. For SU(2) we have 
\eqb
\label{e(a)}
\frac{m^2}{|\phi|^2}\equiv e^2(a,\lambda_E)=T^2\,\times\frac{a^2}{|\phi|^2}=
\frac{\lambda_E^2}{4\pi^2}\times a^2\lambda_E\,.
\eqe
The first factor on the right-hand sides of Eq.\,(\ref{e(a)}) arises 
from rescaling, so only the second factor needs to be differentiated: 
\eqb
\label{e(a)D}
\pd_a e^2(a,\lambda_E)=\frac{\lambda_E^2}{4\pi^2}\times\left(2a\lambda_E+a^2\pd_a\lambda_E\right)\,.
\eqe
After solving $\pd_m P=0$ for the term $\pd_a\lambda_E$ 
we obtain a modified right-hand side of the 
evolution equation Eq.\,(\ref{eeq}). The two-loop evolution of $e$ 
is work in progress \cite{RohrerHofmann2005}. The 
investigation of the screening masses and of the two-loop 
pressure for SU(3) is left for future work. The latter 
will be of a similar magnitude as in the SU(2) case.

\section{The magnetic phase\label{MP}}

The electric phase is terminated at the temperature $T_{c,E}$ by the 
condensation of massless magnetic monopoles. These condensates can be described macroscopically 
by quantum mechanically and 
statistically inert complex scalar fields and pure gauges. The latter describe 
the interactions between monopoles which induce a 
negative pressure by the generation of 
isolated but collapsing magnetic flux loops (center-vortex loops). 
At $T_{c,E}$ one can not distinguish between the unbroken U(1) and the 
dual gauge group U(1)$_D$ (SU(2)) or the unbroken U(1)$^2$ and the 
dual gauge group U(1)$^2_D$ (SU(3)): An exact 
electric-magnetic duality occurs. For SU(2) 
the point $T_{c,E}$ is stabilized by a dip of the energy density. For 
SU(3) a similar stabilization takes place but with a much larger slope of the 
energy density: The phase transition appears to be 
weakly first order. It turns out that the electric 
$Z_2$ (SU(2)) or $Z_3$ (SU(3)) 
degeneracy of the ground state as it was observed in the electric phase 
is lifted: A unique 
ground state characterizes 
the magnetic phase. On the other hand, the expectation of the 
Polyakov loop, though strongly suppressed as compared to that 
in the electric phase, does not vanish entirely 
in the magnetic phase. This is a manifestation 
of the fact that the ground state in the 
magnetic phase does allow for the propagation 
of massive dual gauge modes 
despite the confinement of fundamental, fermionic, and 
heavy test charges by the monopole condensates. 
In that sense, the magnetic phase is only preconfining.  
 
The critical 
behavior in the vicinity of $T_{c,E}$ is 
investigated and compared with results that seem to be 
related to the Yang-Mills theory by universality arguments. 

The monopole condensates are characterized by infinite 
correlation lengths. In a thermodynamical simulation on a finite-size 
lattice performed in the magnetic phase not much can be learned about thermodynamical 
quantities such as the pressure or the energy density which are sensitive to the infrared behavior of the 
theory.

\subsection{Prerequisites}

\subsubsection{The BPS monopole\label{BPSM}}

Here we provide some facts about the 't Hooft-Polyakov 
monopole \cite{'tHooft1974,Polayakov1974} in the BPS limit \cite{PrasadSommerfield1975} 
since we will need them in 
Sec.\,\ref{defphismag}. We consider an SU(2) gauge model 
with the Lagrangian of Eq.\,(\ref{actiontotal}) 
with the modification that the potential is absent. 
The BPS monopole is a 
static, particle-like solution to the equations of motion of this model 
saturating the BPS bound on 
the mass $M$. When centered at $\vec{x}=0$ it is given as
\eqb
\label{BPSmonopole}
\phi^a=\hat{x}_a|\phi|\,F(\rho)\,,\ \ \ \ a^a_4=0\,,\ \ \ \ \ \ \ a_i^a=\frac{{\epsilon}_{aij}}{er}\hat{x_j}\,G(\rho)\,,
\eqe
where $r\equiv |\vec{x}|$, $\hat{x}_i\equiv\frac{x_i}{r}$, $|\phi|\equiv\sqrt{\phi^a\phi^a}(r\to\infty)$, and 
$\rho=e|\phi|r$. The antimonopole solution is obtained by letting 
$\hat{x}\to -\hat{x}$ in Eq.\,(\ref{BPSmonopole}). The functions 
$F$ and $G$ can be determined analytically. They are given as \cite{PrasadSommerfield1975}
\eqb
\label{FG}
F(\rho)=\coth\rho-\frac{1}{\rho}\,,\ \ \ \ \ \ \ \ \ \ \ \ G(\rho)=1-\frac{\rho}{\sinh\rho}\,.
\eqe
The mass $M$ of a BPS monopole or antimonopole calculates as
\eqb
\label{BPSmass}
M=\frac{4\pi}{e}|\phi|\,.
\eqe
(Notice that in Eq.\,(\ref{BPSmass}) $|\phi|$ is replaced by $u\sim u_{\tiny\mbox{max}}=\frac{\pi}{\beta}$ 
for a screened BPS monopole generated by the dissociation of a large-holonomy caloron, compare 
with Eq.\,(\ref{Vmu}).) The magnetic charge $g$ is obtained by integrating the 't Hooft tensor
\eqb
\label{tHoofttensor}
{\cal F}_{\mu\nu}=\pd_\mu(\hat{\phi}^a a_\nu^a)-\pd_\nu(\hat{\phi}^a
a_\mu^a)-\frac{1}{e}\epsilon^{abc}\hat{\phi}^a\pd_\mu\hat{\phi}^b\pd_\nu\hat{\phi}^c\,,\ \ \ \ \ (\hat{\phi^a}=
\frac{\phi^a}{\sqrt{\phi^b\phi^b}})\,,
\eqe
over a two-sphere $S_2$ with infinite radius which is centered at $\vec{x}=0$. 
It is given as
\eqb
\label{chargeg}
g=\int_{S_2,R=\infty}\,d\Sigma_{\mu\nu}\,{\cal F}_{\mu\nu}=\pm\frac{4\pi}{e}\,.
\eqe
In Eq.\,(\ref{chargeg}) $d\Sigma_{\mu\nu}$ denotes the differential surface element.

\begin{figure}
\begin{center}
\leavevmode
\leavevmode
\vspace{4.9cm}
\includegraphics{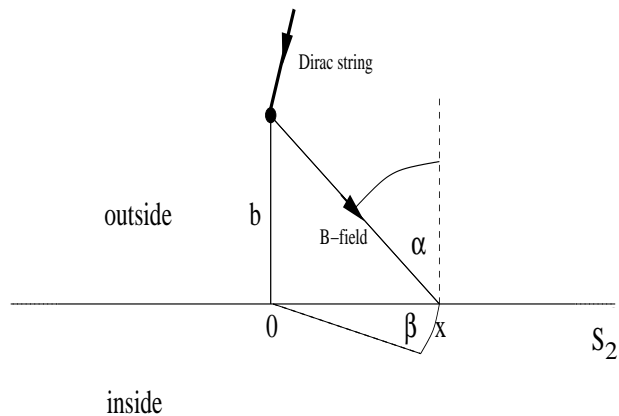}
\end{center}
\caption{A BPS monopole in unitary gauge outside of an $S_2$ with 
infinite radius. The Dirac string does not pierce the surface of the $S_2$. 
\label{Fig30}}      
\end{figure}

In unitary gauge, where the color orientation of $\phi^a$ is `combed' into a fixed 
direction in adjoint color space, the magnetic field 
$B_i=\mp \frac{\hat{x}_i}{er^2}$ associated with Eq.\,(\ref{tHoofttensor}) 
is accompanied by a Dirac string 
along this direction in position space: 
If the monopole lies inside an $S_2$ 
then the magnetic flux through this $S_2$ of the Dirac string 
precisely cancels that of the hedge-hog magnetic field. If a monopole or antimonopole lies 
outside of an $S_2$ with infinite radius an finite distance $b$ away from its surface 
and the monopole's or the antimonopole's Dirac string does not pierce the surface then the magnetic flux $F$
through the surface, see Fig.\,\ref{Fig30}, calculates as
\eab
\label{magfluxs2}
F&=&\int_{\tiny\mbox{plane}}d\Sigma_{\mu\nu}\,{\cal F}_{\mu\nu}\nonumber\\ 
&=&\pm\frac{1}{e}\int_0^{2\pi}d\beta\int_{-\infty}^{\infty}dx\,|x|\cos\alpha(x,b)\frac{1}{x^2+b^2}\nonumber\\ 
&=&\pm\frac{4\pi}{e}\,b\int_{0}^{\infty}dx\,\frac{|x|}{(x^2+b^2)^{3/2}}\nonumber\\ 
&=&\pm\frac{4\pi}{e}\,.
\eae
We conclude that static monopoles or antimonopoles lying inside an $S_2$ of 
{\sl infinite} radius do not contribute to the flux through $S_2$ 
while they may contribute to the flux when situated 
outside of this $S_2$. 

\subsubsection{Derivation of the phases of 
macroscopic complex scalar fields\label{defphismag}}

There is one species of magnetic monopoles in the SU(2) case 
while there are two independent species 
of magnetic monopoles for SU(3). 
\vspace{0.1cm}\\ 
\noindent\underline{SU(2) case:}\vspace{0.1cm}\\ 
In unitary gauge we consider an isolated system of a monopole and an antimonopole, which both are 
at rest and do not interact, outside of an $S_2$ with {\sl infinite} radius. 
We characterize their Dirac strings by 
unit vectors $\hat{x}_m$ and $\hat{x}_a$ which point away 
from the core of the monopole and the antimonopole, respectively. 
Let $\Pl$ be the plane perpendicular to $S_2$ such that the intersection line 
$L=\Pl\cap S_2$ coincides with the intersection line of $S_2$ with 
the plane spanned by $\hat{x}_m$ and $\hat{x}_a$. (The case where $\hat{x}_m$ and $\hat{x}_a$ lie 
in $S_2$ is inessential for what follows.) Whether or not this system contributes 
to the magnetic flux through $S_2$ depends on the angle 
$\delta=\angle(\hat{x}_m,\hat{x}_a)$ and on the angle 
$\gamma$ which the projection of $\hat{x}_m$ onto $\Pl$ 
forms with $L$. 

A magnetic flux through $S_2$ is generated if and only 
if either $\hat{x}_m$ or $\hat{x}_a$ {\sl alone} pierces $S_2$. 
For a given angle $\delta$ the angle $\gamma$ 
is uniformly distributed. In the absence of a heat bath 
the probability of measuring a flux $\frac{4\pi}{e}$ or 
a flux $-\frac{4\pi}{e}$ through $S_2$ thus is given as 
$\frac{\delta}{2\pi}$. We conclude that for a given angle $\delta$ the average plus or minus 
flux through $S_2$ reads
\eqb
\label{avflux}
\bar{F}_{\pm}=\pm\frac{\delta}{2\pi}\frac{4\pi}{e}=\pm\frac{2\delta}{e}\,,\ \ \ \ \ \ (0\le\delta\le\pi)\,. 
\eqe
Notice that $\delta$ and $2\pi-\delta$ generate the same average flux $\bar{F}_{\pm}$, 
thus the restriction $0\le\delta\le\pi$ in Eq.\,(\ref{avflux}). 

So far we have discussed the flux through $S_2$ which is
generated by an isolated monopole-antimonopole system with no interactions. 
To derive the phase of the macroscopic complex field $\varphi$ describing the Bose 
condensate of such systems we couple the system to the heat bath, 
project onto zero-momentum states (condensate) of the monopole-antimonopole 
system such that each constituent does not carry momentum 
and perform the massless limit $e\to\infty$ which takes place for $T\le T_{c,E}$, compare 
with Eq.\,(\ref{divloglae}). (The rare case of a zero-momentum 
state with opposite and finite momenta of the constituents generates a 
closed and instable magnetic flux line, see Fig.\,\ref{Fig0}. 
This situation takes place if a large number of large-holonomy 
calorons dissociate into monopole-antimonopole pairs 
almost simultaneously inside a small 
spatial volume. On the macroscopic level, the thermal average over these flux loops, 
which collapse as soon as they are created, will later be described by a pure gauge configuration.) 

The monopole and antimonopole 
are generated by the dissociation of a large-holonomy caloron. According to 
Eq.\,(\ref{massesMonLL}) the sum of 
monopole and antimonopole mass, $M_{m+a}$, is, after screening, given as
\eqb
\label{massesma}
M_{m+a}=\frac{8\pi^2}{e\beta}\,.
\eqe
The thermally averaged flux of the zero-momentum system at 
finite coupling $e$ is obtained as
\eab
\label{avfluxsys}
\bar{F}_{\pm,\tiny\mbox{th}}(\delta)&=&\,4\pi\,\int d^3p\,\delta^{(3)}(\vec{p})\, n_B(\beta |E(\vec{p})|)\,\bar{F}_{\pm}\nonumber\\ 
&=&\pm\frac{8\pi\,\delta}{e}\int d^3p\,\frac{\delta^{(3)}(\vec{p})}{\exp\left[\beta\sqrt{M_{m+a}^2+\vec{p}^2}\right]-1}\,.
\eae
After setting $\vec{p}=0$ in $\left(\exp\left[\beta\sqrt{M_{a+b}^2+\vec{p}^2}\right]-1\right)$ and by
appealing to Eq.\,(\ref{massesma}), the expansion of this term reads
\eqb
\label{expexpon}
\lim_{\vec{p}\to 0}\left(\exp\left[\beta\sqrt{M_{m+a}^2+\vec{p}^2}\right]-1\right)=\frac{8\pi^2}{e}\left(1+\frac{1}{2}\frac{8\pi^2}{e}+
\frac{1}{6}\left(\frac{8\pi^2}{e}\right)^2+\cdots\right)\,.
\eqe
Appealing to Eq.\,(\ref{expexpon}), the limit $e\to\infty$ can 
now safely be performed in Eq.\,(\ref{avfluxsys}). We have
\eqb
\label{avfluxsys,einf}
\lim_{e\to\infty} \bar{F}_{\pm,\tiny\mbox{th}}(\delta)=\pm\frac{\delta}{\pi}\,,\ \ \ \ \ \ (0\le\delta\le\pi)\,.
\eqe
The right-hand side of Eq.\,(\ref{avfluxsys,einf}) defines the 
argument of the complex and periodic 
function $f$ with 
\eqb
\label{fmag}
f(\frac{\delta}{\pi})\equiv C\frac{\varphi}{|\varphi|}(\frac{\delta}{\pi})
\eqe
where $C$ is an undetermined (complex) constant. (Recall, that $\delta$ is an angle). Since $f$'s argument 
was obtained by a projection onto zero spatial 
momentum the only admissible nontrivial periodic dependence is that on the 
Euclidean time $\tau$. Without restriction of generality 
we can thus set $\frac{\delta}{\pi}=\frac{\tau}{\beta}$. 
Since $f$ is periodic, $f(\tau=0)=f(\tau=\beta)$, it can be expanded into a 
Fourier series:
\eqb
\label{FourSer}
f(\frac{\tau}{\beta})=\sum_{n=-\infty}^{n=\infty} f_n \exp\left[2\pi i n\frac{\tau}{\beta}\right]\,
\eqe
where $f_n$ are (complex) constants. According to Eq.\,(\ref{fmag}) $f\bar{f}=|C|^2$ 
is constant in $\tau$. Thus the only possibility in Eq.\,(\ref{FourSer}) is $f_n=C\,\delta_{mn}$ {\sl or} 
$f_n=\bar{C}\,\delta_{(-m)n}$ for a 
fixed value of $m$. Moreover, only $m=1$ is allowed since the 
physical situation generating the continuous parameter $\delta$ 
does not repeat itself for $0\le\delta\le\frac{\pi}{m}$, 
$\frac{\pi}{m}\le\delta\le\frac{2\pi}{m},\cdots$ if $m>1$ because no 
higher-charge monopoles exist. (Recall that only calorons with a 
large-holonomy and topological charge unity are allowed to contribute to 
the thermodynamics in the electric phase.) We conclude that the equation of motion satisfied by $f$ is:
\eqb
\label{eomfm}
\pd_\tau^2 f(\frac{\tau}{\beta})+\left(\frac{2\pi}{\beta}\right)^2\,f(\frac{\tau}{\beta})=0\,.
\eqe

\subsubsection{Derivation of the modulus of 
macroscopic complex scalar fields\label{defphismagmod}}

What about $\varphi$'s modulus? The reasoning is completely analoguous to that for the derivation of 
$\phi$'s modulus. First, since $\varphi$ is composed of massless, noninteracting 
monopoles being at rest its $\tau$ dependence ought to be BPS saturated.  
Second, for $\varphi$ to have the phase $\exp[\pm 2\pi i\frac{\tau}{\beta}]$ 
the right-hand side of its BPS equation ought 
to be {\sl linear} in $\varphi$. Third, for the existence of smooth 
deformations $\beta\to\beta+\delta\beta$, subject to a 
perturbative expansion in $\frac{\delta\beta}{\beta}$ away from a 
phase boundary, the right-hand side of the BPS equation must 
be analytic. Fourth, we assume that a scale $\La_M$ is given externally. 
Fifth, no explicit dependence on $\beta$ may appear on the 
right-hand side of the BPS equation since at zero momentum the temperature dependence of the 
mass of a monopole cancels against the factor $\beta$ in the 
Boltzmann weight. The only possibility for the BPS equation satisfying these conditions is
\eqb
\label{BPSvarphi}
\pd_\tau \varphi=\pm i\frac{\La_M^3\,\varphi}{|\varphi|^2}=\pm i\frac{\La_M^3}{\bar{\varphi}}\,.
\eqe
From Eq.\,(\ref{BPSvarphi}) it follows that $\varphi$'s modulus 
is given as 
\eqb
\label{BPSvarphimod}
|\varphi|=\sqrt{\frac{\La_M^3\beta}{2\pi}}\,. 
\eqe
The right-hand side of Eq.\,(\ref{BPSvarphi}) defines the square root of the 
potential $V_M$. In the absence of interactions 
between (screened) monopoles the effective theory for $\varphi$ thus reads
\eqb    \label{actionvarphi}
S_{\varphi} = \int_0^\beta d\tau \int d^3x 
\left( \frac{1}{2}\,\overline{\partial_\tau \varphi} \partial_\tau \varphi + 
\frac{1}{2}\,\frac{\La_M^6}{\bar{\varphi}\varphi}  \right)  \,.
\eqe
\noindent\underline{SU(3) case:}\vspace{0.1cm}\\ 
Here we have two independent monopole species which 
do not interact. The situation for each species is 
completely analogous to the SU(2) case. The two macroscopic 
fields $\varphi_1$ and $\varphi_2$ both satisfy the 
BPS equation (\ref{BPSvarphi}) and are given as
\eqb
\label{varphi12}
\varphi_1(\tau)=\varphi_2(\tau)=\sqrt{\frac{\La_M^3\beta}{2\pi}}\,\exp\left[\pm 2\pi i\frac{\tau}{\beta}\right]\,.
\eqe
In the absence of interactions 
between monopoles the effective theory for $\varphi_1,\varphi_2$ reads
\eqb    
\label{actionvarphi12}
S_{\varphi} = \int_0^\beta d\tau \int d^3x 
\left( \frac{1}{2}\,\overline{\partial_\tau \varphi_1} \partial_\tau \varphi_1 + 
\frac{1}{2}\,\overline{\partial_\tau \varphi_2} \partial_\tau \varphi_2 +
\frac{1}{2}\,\frac{\La_M^6}{\bar{\varphi_1}\varphi_1} + \frac{1}{2}\,
\frac{\La_M^6}{\bar{\varphi_2}\varphi_2}  \right)  \,.
\eqe

\subsection{A macroscopic ground state\label{gsmag}}

In close analogy to the electric phase we now derive the full macroscopic 
ground-state dynamics. First, we establish the quantum mechanical and statistical 
inertness of the fields $\varphi$ (SU(2)) and $\varphi_1,\varphi_2$ 
(SU(3)). Subsequently, we solve the equations of motion for the 
topologically trivial dual gauge field in these backgrounds to 
obtain pure-gauge configurations describing, on a macroscopic level, the interaction between monopoles in the 
ground state. It will turn out that the Polyakov loops when evaluated on 
these pure-gauge configurations are trivial: The electric $Z_2$ (SU(2)) and $Z_3$ (SU(3)) 
degeneracies observed in the electric phase no longer exist. Thus the magnetic 
phase confines fundamental test charges. 

The ratios of the mass-squared of potential $\varphi$-field fluctuations with 
$|\varphi|^2$ and $T^2$ are
\eqb
\label{qsflphiM}
\frac{\pd^2_{|\varphi|} V_{M}(\varphi)}{|\varphi|^2}=6\,\lambda_M^3\,,\ 
\ \ \ \ \ \ \ \frac{\pd^2_{|\varphi|} V_{M}(\varphi)}{T^2}=24\pi^2\,,
\eqe
where $\lambda_M\equiv\frac{2\pi T}{\La_M}$. We will show below 
that $\lambda_M\ge 7.075$ (SU(2)) and $\lambda_M\ge 6.467$ (SU(3)). 
Thus $\varphi$-field (SU(2)) and $\varphi_1,\varphi_2$-field (SU(3)) 
fluctuations neither exist on-shell nor off-shell. 

\subsubsection{Pure-gauge configurations\label{pgmag}}

\noindent\underline{SU(2) case:}\vspace{0.1cm}\\ 
The topologically trivial sector is coupled to $\varphi$ in a minimal fashion, and 
the following effective action arises
\eqb
\label{effactM2}
S=\int_0^{\beta}
d\tau\int d^3x\,\left[\frac{1}{4}\,
G^D_{\mu\nu}G^D_{\mu\nu}+
\frac{1}{2}\overline{{\cal D}_{\mu}\varphi}
{\cal D}_{\mu}\varphi+\frac{1}{2}\frac{\La_M^6}{\bar{\varphi}\varphi}\right]\,,
\eqe
where $G^D_{\mu\nu}=\pd_\mu a^D_\nu-\pd_\nu a^D_\mu$ denotes the Abelian field strength 
of the dual gauge field $a^D_\mu$ and ${\cal D}_{\mu}\equiv \pd_\mu+ig\,a^D_\mu$ 
denotes the covariant derivative involving the magnetic gauge coupling $g$. 

Since the field $\varphi$ does not fluctuate it is a background 
to the macroscopic gauge-field equations of motion which follows from Eq.\,(\ref{effactM2}): 
\eqb
\label{eomdualG2}
\pd_\mu G^D_{\mu\nu}=ig\left[\overline{{\cal D}_{\nu}\varphi}\varphi-\bar{\varphi}
{\cal D}_{\nu}\varphi\right]\,.
\eqe
A pure-gauge solution to Eq.\,(\ref{eomdualG2}) with $D_\mu\varphi\equiv0$ 
is given as
\eqb
\label{pgsolM2}
a^{D,bg}_{\mu}=\pm\delta_{\mu 4}\frac{2\pi}{g\beta}\,.
\eqe
On $\varphi$ and on $a^{D,bg}_{\mu}$ only the potential does not vanish in 
Eq.\,(\ref{effactM2}): Interactions 
between magnetic monopoles create a nonvanishing energy density $\rho^{gs}$ 
and pressure $P^{gs}$ where 
\eqb
\label{Pmag2}
\rho^{gs}=\pi\,\La_M^3\,T=-P^{gs}\,.
\eqe
We shall see in Sec.\,\ref{ANOBPS}, compare with Eq.\,(\ref{Pvort}), that the negative 
ground-state pressure in Eq.\,(\ref{Pmag2}) originates from center-vortex 
loops which collapse as soon as they are created. 

\noindent\underline{SU(3) case:}\vspace{0.1cm}\\ 
The situation is the same as for SU(2) except that 
we have gauge dynamics subject to U(1)$_D^2$ and not only 
U(1)$_D$. The effective action reads
\eab
\label{effactM3}
S&=&\int_0^{\beta}
d\tau\int d^3x\,\left[\frac{1}{4}\,
G^D_{\mu\nu,1}G^D_{\mu\nu,1}+\frac{1}{4}\,
G^D_{\mu\nu,2}G^D_{\mu\nu,2}+\right.\nonumber\\ 
&&\left.\frac{1}{2}\overline{{\cal D}_{\mu,1}\varphi_1}
{\cal D}_{\mu,1}\varphi_1+\frac{1}{2}\overline{{\cal D}_{\mu,2}\varphi_2}
{\cal D}_{\mu,2}\varphi_2+\frac{1}{2}\frac{\La_M^6}{\bar{\varphi_1}\varphi_1}
+\frac{1}{2}\frac{\La_M^6}{\bar{\varphi_2}\varphi_2}\right]\,.
\eae
The Abelian field strengths $G^D_{\mu\nu,1}, G^D_{\mu\nu,2}$ 
and the covariant derivatives ${\cal D}_{\mu,1},{\cal D}_{\mu,2}$ 
are defined as for the SU(2) case with the replacements $a^D_\mu\to a^D_{\mu,1},a^D_\mu\to a^D_{\mu,2}$, 
respectively. (The magnetic gauge coupling $g$ is universal since both species of 
monopoles couple with the same strength to their respective gauge field.) 

The equations of motion for the fields $a^D_{\mu,1},a^D_{\mu,2}$ in the background 
of the fields $\varphi_1,\varphi_2$ are   
\eqb
\label{eomdualG3}
\pd_\mu G^D_{\mu\nu,1}=ig\left[\overline{{\cal D}_{\nu,1}\varphi_1}\varphi-\bar{\varphi_1}
{\cal D}_{\nu,1}\varphi_1\right]\,,\ \ 
\pd_\mu G^D_{\mu\nu,2}=ig\left[\overline{{\cal D}_{\nu,2}\varphi_2}\varphi-\bar{\varphi_2}
{\cal D}_{\nu,2}\varphi_2\right]\,.
\eqe
Pure-gauge solutions to these equations with 
${\cal D}_{\nu,1}\varphi_1={\cal D}_{\nu,2}\varphi_2=0$ are given as
\eqb
\label{pgsolM3}
a^{D,bg}_{\mu,1}=\pm\delta_{\mu 4}\frac{2\pi}{g\beta}\,,\ \ \ \ \ \ \ 
a^{D,bg}_{\mu,2}=\pm\delta_{\mu 4}\frac{2\pi}{g\beta}\,.
\eqe
On $\varphi_1,\varphi_2$ and on $a^{D,bg}_{\mu,1},a^{D,bg}_{\mu,2}$ 
only the potentials do not vanish in Eq.\,(\ref{effactM3}): Interactions 
between magnetic monopoles create a nonvanishing energy density $\rho^{gs}$ 
and pressure $P^{gs}$ where 
\eqb
\label{Pmag3}
\rho^{gs}=2\pi\,\La_M^3\,T=-P^{gs}\,.
\eqe
Again, the negative ground-state pressure in Eq.\,(\ref{Pmag3}) originates from center-vortex 
loops which collapse as soon as they are created. 

\subsubsection{Polyakov loop and rotation to unitary gauge\label{PUGM}}

\noindent\underline{SU(2) case:}\vspace{0.1cm}\\ 
The Polyakov loop ${\bf P}\equiv \exp\left[ig\int_0^\beta d\tau\, a^D_4\right]$, when 
evaluated on the pure-gauge configuration in Eq.\,(\ref{pgsolM2}), reads
\eqb
\label{pGPol2}
{\bf P}=\exp[\pm ig\int_0^\beta d\tau\,\frac{2\pi}{g\beta} ]=1\,.
\eqe
A gauge rotation $a_\mu^{D,bg}\to a_\mu^{D,bg}+\frac{i}{g}\left(\pd_\mu \Omega^\dagger\right)\Omega$ 
to unitary gauge $\varphi=|\varphi|, a_\mu^{D,bg}=0$ is mediated by the U(1) group element 
$\Omega=\exp\left[\pm 2\pi i\frac{\tau}{\beta}\right]$: The Polyakov loop {\bf P} is invariant 
under this gauge transformation. We conclude that the electric 
$Z_2$ ground-state degeneracy, which  was observed in the electric 
phase, no longer exists in the magnetic phase: The ground state confines 
fundamentally charged, heavy and fermionic test charges.\vspace{0.1cm}\\   
\noindent\underline{SU(3) case:}\vspace{0.1cm}\\ 
Here the Polyakov loop ${\bf P}$ is a product of the Polyakov 
loops ${\bf P}_1$ and ${\bf P}_2$ computed on the respective pure-gauge 
configurations $a^{D,bg}_{\mu,1}$ and $a^{D,bg}_{\mu,1}$ in Eq.\,(\ref{pgsolM3}). We have
\eqb
\label{pGPol3}
{\bf P}={\bf P}_1{\bf P}_2=\exp[\pm 2ig\int_0^\beta d\tau\,\frac{2\pi}{g\beta} ]=1\,\ \ \ \ \mbox{or}\ \ \ \ \ 
{\bf P}={\bf P}_1{\bf P}_2=\exp[0]=1\,.
\eqe
Gauge rotations to unitary gauge 
$\varphi_1=|\varphi_1|=\varphi_2, a_{\mu,1}^{D,bg}=a_{\mu,2}^{D,bg}=0$ are 
mediated by the U(1) group elements 
$\Omega_1=\exp\left[\pm 2\pi i\frac{\tau}{\beta}\right]=\Omega_2$. 
Again, ${\bf P}$ is invariant under these gauge rotations. We conclude that the electric 
$Z_3$ ground-state degeneracy does not exist in the magnetic phase: Fundamentally charged, heavy and fermionic 
test charges are confined by the monopole condensates.

\subsection{Excitations\label{ExcM}}

\subsubsection{Mass spectrum of thermal quasiparticles\label{MSTQPM}}

The dual Abelian Higgs mechanism generates tree-level
quasiparticle masses $m$ and $m_1,m_2$ for the fluctuations $\delta a^D_{\mu}$ (SU(2)) 
and $\delta a^D_{\mu,1},\delta a^D_{\mu,2}$ (SU(3)), respectively. We have 
\eqb
\label{massspecM}
m=g|\varphi|=m_1=m_2=a\,T\,,\ \ \ \ \ \ \ (a\equiv 2\pi\,g\,\lambda_M^{-3/2})\,
\eqe
where $\lambda_M\equiv\frac{2\pi T}{\La_M}$. 

\subsubsection{Thermodynamical selfconsistency and evolution equation\label{TSCevM}}

Due to the absence of interactions between the dual gauge fields 
the thermodynamics of the magnetic phase is exact on the one-loop level. 
Again, a magnetic modification of the compositeness condition Eq.\,(\ref{comconnoi}) applies. 

Let us first compare the contribution $\Delta V_M$ of 
quantum fluctuations to the pressure arising from dual gauge modes with the tree-level 
result $-1/2\,V_M=-\pi\,\La_M^3 T$ (SU(2)) and $-1/2\,V_M=-2\pi\,\La_M^3 T$ (SU(3)). 
In both cases we have
\eqb
\label{deltatreeM}
\frac{\Delta V_M}{V_M}=\frac{\lambda_M^{-3}}{24\pi^2}\,.
\eqe
Considering that $\lambda_M\ge 7.075$ (SU(2)) and $\lambda_M\ge 6.467$ (SU(3)) this 
is smaller than $1.2\times 10^{-5}$ and $1.6\times 10^{-5}$, respectively. 
Thus the quantum contribution to the 
one-loop pressure can safely be neglected. 

\begin{figure}
\begin{center}
\leavevmode
\leavevmode
\vspace{5.0cm}
\includegraphics{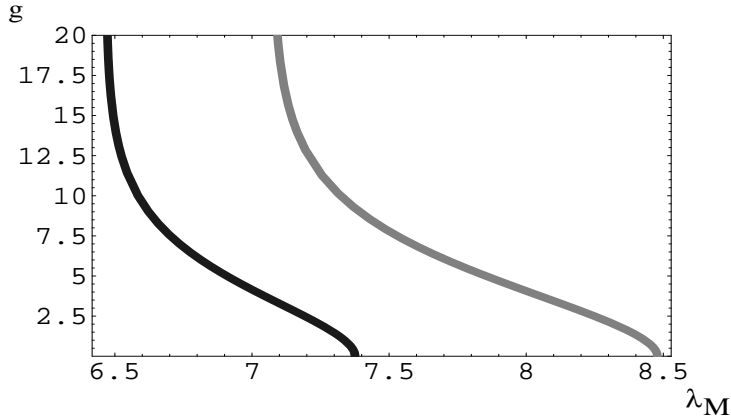}
\end{center}
\caption{The evolution of the effective gauge coupling $g$ in the magnetic phase for 
SU(2) (thick grey line) and SU(3) (thick black line). At 
$\lambda_{c,M}=7.075$ (SU(2)) and $\lambda_{c,M}=6.467$ (SU(3)) $g$  diverges logarithmically, 
$g\sim -\log(\lambda_{M}-\lambda_{c,M})$. \label{gevol}}      
\end{figure}

For SU(2) the thermal contribution to the pressure reads
\eqb
\label{Pre2M}
P(\lambda_M)=-\La_M^4\left[\frac{6\lambda_M^4}{(2\pi)^6}\bar{P}(a)+\frac{\lambda_M}{2}\right]\, 
\eqe
where the (negative) function $\bar{P}(a)$ is defined in Eq.\,(\ref{P(y)}). The SU(3) 
pressure is just twice the SU(2) pressure. From the condition $\pd_a P=0$ of 
thermodynamical selfconsistency the following evolution equation arises for both SU(2) and SU(3):
\eqb
\label{evol23M}
\pd_a\lambda_M=-\frac{12\lambda_M^4\,a}{(2\pi)^6}\,\frac{D(a)}{1+\frac{12\lambda_M^3\,a^2}{(2\pi)^6}\,D(a)}
\eqe
where the (positive) function $D(a)$ is defined in Eq.\,(\ref{DA}). 
In analogy to the electric phase the $\lambda_M$ dependence of the gauge coupling constants $g$ 
is obtained by inverting the solutions to Eq.\,(\ref{evol23M}) 
and by subsequently using the relation between $g$, $\lambda_M$ and $a$ in 
Eq.\,(\ref{massspecM}). The temperature evolution of $g$ is 
shown in Fig.\,\ref{gevol}.

\subsubsection{Interpretation of results\label{intM}}

The magnetic gauge coupling $g$ increases continuously from $g=0$ at the electric-magnetic 
phase boundary ($T=T_{c,E}$) until it diverges logarithmically at $T_{c,M}$. 
A variation of the magnetic 
coupling with temperature is not in contradiction with magnetic charge conservation 
since no {\sl isolated} magnetic charges appear 
in the magnetic phase: Magnetic monopoles either are condensed 
or they conspire to form instable magnetic flux loops (center-vortex loops), 
see Fig.\,\ref{Fig0}. From Fig.\,\ref{gevol} one can see that the 
magnetic phase is more narrow for SU(3) than it is for SU(2). Despite the fact 
that the ground-state degeneracies with respect to the electric $Z_2$ symmetry (SU(2)) and 
the electric $Z_3$ symmetry (SU(3)) are absent in the magnetic phase 
the fully averaged Polyakov loop (including the massive dual gauge-mode excitations) 
does not vanish at $T_{c,E}$, see Sec.\,\ref{polyaloop}. The expectation of the Polyakov loops only 
vanishes at $T_{c,M}$ where {\sl all} dual gauge modes are decoupled because of a 
diverging mass. If we take the mass of the dual 
gauge modes to be the order parameter for the dynamical 
breaking of U(1)$_D$ (SU(2)) and U(1)$_D^2$ (SU(3)) 
then the electric-magnetic transition is second order with 
mean-field exponents in both cases. The best one 
can do to relate this order-parameter to electric 
$Z_2$ or $Z_3$ restoration is to look for the point 
where its exponent is least sensitive to the length $\Delta\lambda_M$ of the 
fitting interval. The expectation of the 't Hooft loop, which is a dual order parameter for complete 
confinement and which is nonzero if center-vortices are condensed, 
vanishes inside the magnetic phase where center-vortex loops 
are isolated and instable defects. At $T_{c,M}$ the expectation of the 't Hooft loop 
jumps to a finite value. The associated transition 
is, however, neither second nor first order 
but of the Hagedorn type see Sec.\,\ref{Hagedorn}. Because of the 
infinite correlation length (massless condensed magnetic monopoles) 
a finite-size lattice simulation of the order parameter as well as infrared sensitive 
thermodynamical quantities such as the pressure is problematic.

\subsection{Polyakov loop in the electric and the magnetic phase\label{polyaloop}}

In this section we show, on a qualitative level, that the expectation of the 
Polyakov loop $\la{\bf P}\ra$, which is an order parameter 
for the confinement-deconfinement transition, is finite both in the electric 
phase and the magnetic phase. Deep in the magnetic phase $\la{\bf P}\ra$ is, however, 
strongly suppressed as compared to its value at $T_{c,E}$. 

In each phase the Polyakov-loop expectation $\la{\bf P}\ra$
can be computed in unitary(-Coulomb) gauge. Let us first discuss the electric phase. 
The sector $S_{f,E}$ in the effective action 
(\ref{actiontotal}), which involves fluctuating fields, is given 
as
\eqb
\label{SfE}
S_{f,E}=\int_0^{\beta}d\tau\,d^3x\left[\frac14\,G^a_{\mu\nu}G^a_{\mu\nu}+
\frac12\sum_a m_a^2\,(\delta a^a_\mu)^2\right]\,.
\eqe
Here $a=1,2,3$ and $m_1=m_2>0, m_3=0$ for SU(2) and $a=1,\cdots,8$ 
and $m_1=m_2=\frac{1}{2}m_3=\frac{1}{2}m_4=\frac{1}{2}m_5=\frac{1}{2}m_6>0, 
m_7=m_8=0$ for SU(3). Since in unitary gauge the Polyakov loop is 
unity in the ground state no direct ground-state contribution arises 
in the expectation $\la{\bf P}\ra$: The associated factor in the numerator 
cancels that in the denominator. We have
\eqb
\label{Ployeff}
\la{\bf P}\ra=Z_{f,E}^{-1}\times
\int \left[d\delta a_\mu\right]\,\exp\left[ie\int_0^{\beta} d\tau\,\delta a_4\right] \exp[-S_{f,E}]\,,
\eqe
where $\left[d\delta a_\mu\right]$ denotes the path-integral measure and 
$Z_{f,E}\equiv \int \left[d\delta a_\mu\right] \exp[-S_{f,E}]$. 

Since the fluctuations $\delta a_\mu$ 
are periodic in $\tau$ they can be decomposed into a Matsubara sum:
\eqb
\label{fourierb}
\delta a_\mu(\tau,{\vec x})=\sum_{n=-\infty}^{n=\infty} 
\exp\left[2\pi in \frac{\tau}{\beta}\right]\,\delta\bar{a}_{\mu,n}({\vec x})\,.
\eqe
Modes with $n\not=0$ in Eq.\,(\ref{fourierb}) render the Polyakov-loop 
phase in Eq.\,(\ref{Ployeff}) to be unity and are 
action-suppressed. Zero modes ($n=0$) contribute  
to $\la{\bf P}\ra$ sizably 
if they are not action-suppressed. This is the case if and only if 
both of the following conditions are met: (i) as compared to $T$ some or all masses 
$m_a$ in Eq.\,(\ref{SfE}) are small and (ii) 
$\pd_i\delta\bar{a}_{\mu,0}({\vec x})$ is small compared with $T^2$ and the field configuration 
is still localized in space. Here $\pd_i$ denotes a spatial derivative. In the electric 
phase there are massless modes, the conditions (i) and (ii) can be satisfied, and 
thus a finite Polyakov-loop expectation emerges.  

A similar consideration can be performed for the magnetic phase. 
Deep inside this phase condition (i) is badly violated since the mass of {\sl all} dual 
gauge modes is much larger than $T$ by virtue of Fig.\,\ref{gevol} 
and the mass formula in Eq.\,(\ref{massspecM}). Thus $\la{\bf P}\ra$, though nonvanishing, is strongly 
suppressed deep inside the magnetic phase as compared 
to its value at $T_{c,E}$. For quantitative results the average $\la{\bf P}\ra$ can be performed 
on a lattice or analytically by using the respective 
effective theory for the electric and the magnetic phase.

\subsection{Critical behavior at the electric-magnetic transition\label{critexp}}

The electric-magnetic transition, which goes with the dynamical breakdown 
of U(1)$_D$ (SU(2)) and U(1)$^2_D$ (SU(3)), is second 
order in both cases. The difference is that 
in the SU(3) case the magnetic phase is more narrow than for SU(2), 
see Fig.\,\ref{gevol}, and that the peak-value 
of the specific heat per volume is much larger for SU(3) than it is for SU(2), 
see Fig.\,\ref{SH}. In addition, the entropy density, which measures 
the mobility of dual gauge modes and which is used to extract an 
apparent latent heat on the lattice \cite{Brown1988}, 
drops much more rapidly in the magnetic phase for SU(3) as compared to the SU(2) case. 
This is a plausible explanation for the apparent first-order nature of the confinement-deconfinement 
transition observed in lattice simulations \cite{LuciniTeperWenger2003,LuciniTeperWenger2005}.

The order parameter for the electric-magnetic transition is the 
mass $\propto a\lambda_M$ of the dual gauge bosons. (The monopole mass vanishes 
like an inverse {\sl logarithm} on the electric side of the transition, and thus it is not an order parameter.) 
The following model applies to the 
behavior of $a\,\lambda_M$ close to the critical temperature $\lambda_{c}=9.24$ (SU(2)) 
and $\lambda_{c}=6.81$ (SU(3)):
\eqb
\label{modeldata}
a\,\lambda_M (\lambda_M)=K\,|\lambda_M-\lambda_{c}|^\nu\,,
\eqe
where $K$ and $\nu$ are constants, and $\lambda_{c}$ is the critical 
temperature $T_{c,E}$ in units of $\frac{\Lambda_M}{2\pi}$. By demanding continuity of 
the pressure across the electric-magnetic transition, see Sec.\,\ref{MCC}, 
we derive $\lambda_{c}=8.478$ (SU(2)) and $\lambda_{c}=7.376$ (SU(3)). 
 
The magnetic phase is not 
entirely confining (dual gauge bosons propagate although fundamental, heavy, 
and fermionic test charges are confined), and thus the expectation of the Polyakov 
loop $\la{\bf P}\ra$ is not exactly zero. The best one can do in order 
to compare the behavior of $a\,\lambda_M$ to the behavior of $\la{\bf P}\ra$ inferred 
from universality arguments \cite{SvetitskyYaffe1982-1,SvetitskyYaffe1982-2} 
is to look for the point where the fitted value of $\nu$ 
in Eq.\,(\ref{modeldata}) is least sensitive to a variation of the length $\Delta\lambda_M$ 
of the fitting interval. 

\begin{figure}
\begin{center}
\leavevmode
\leavevmode
\vspace{4.5cm}
\includegraphics{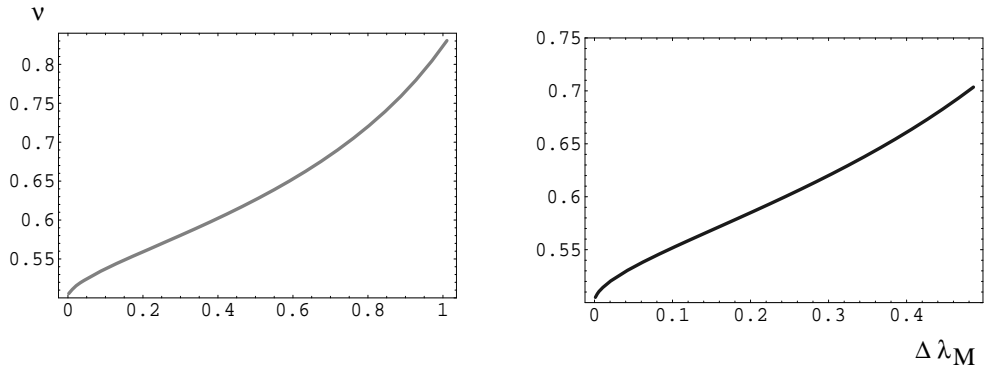}
\end{center}
\caption{The critical exponent $\nu$ for the mass of the dual gauge modes and for 
the magnetic-electric transition as a function of the length $\Delta\lambda_M$ 
of the fitting interval. The left panel is the SU(2) and the 
right panel the SU(3) result. \label{expfit}}      
\end{figure}

To perform the fit to the model in Eq.\,(\ref{modeldata}) we 
have used Mathematica's NonlinearFit function which is contained in the 
statistics package. The function $a(\lambda_M)$, subject to the 
initial conditions $a(\lambda_{c}=8.478)=0$ (SU(2)) and $a(\lambda_{c}=7.376)=0$ 
(SU(3)), was generated by an inversion of the corresponding numerical 
solutions to Eq.\,(\ref{evol23M}). (A step-size $\delta a=5\times 10^{-9}$ 
was used in the Runge-Kutta algorithm\footnote{The author would like to thank 
Jochen Rohrer for performing the numerical calculation.}.) In Fig.\,\ref{expfit} the 
fitted exponent $\nu$ is shown as a function of $\Delta\lambda_M$. 
Two things are important to observe. First, the magnetic-electric 
transition is second order with the mean-field 
exponent $\nu=0.5$ for both SU(2) and SU(3). Second, for each case there 
is a point $\Delta\lambda^*_M$ of least sensitivity for $\nu$ under 
variations in $\Delta\lambda_M$. For SU(2) we have $\Delta\lambda^*_M=0.28 \pm 0.03$ and 
$\nu(\Delta\lambda^*_M)=0.576\pm 0.008$, and 
for SU(3) $\Delta\lambda^*_M=0.16 \pm 0.02$ and 
$\nu(\Delta\lambda^*_M)=0.572 \pm 0.006$. 
By universality we expect $\nu(\Delta\lambda^*_M)$ for SU(2) to 
be close to the exponent $\nu_{\tiny\mbox{IM}}$ for the corresponding order 
parameter of a 3D Ising model \cite{SvetitskyYaffe1982-1,SvetitskyYaffe1982-2}. 
One has $\nu_{\tiny\mbox{IM}}\sim 0.63$. The SU(2) exponent $\nu(\Delta\lambda^*_M)$ only 
deviates by about 8.5\% from $\nu_{\tiny\mbox{IM}}$.

\section{The center phase\label{CVCM}} 

\subsection{Prerequisites} 

\subsubsection{ANO vortex in the BPS limit\label{ANOBPS}}

Just like the isolated defects in the electric phase 
are screened BPS monopoles, the isolated defects in 
the magnetic phase are screened and closed 
magnetic flux lines (vortex-loops). In Fig.\,\ref{Fig0}, see also \cite{Olejnik1997}, we have given 
a figurative interpretation of these flux 
lines: They are composed of magnetic monopoles and 
antimonopoles which move into opposite directions. Thus there is a net magnetic 
current in the vortex core. A vortex core can 
be viewed as locations in space where U(1)$_D$ (SU(2)) or one of the factors in 
U(1)$^2_D$ (SU(3)) are restored. Hence the picture of 
isolated monopoles, contributing to the magnetic current, applies to the vortex core. 

We have explained in Sec.\,\ref{intro} why the magnetic flux carried by a vortex-loop 
is independent of the state of motion of a 
particular monopole contributing 
to the vortex: The amount of flux carried by a vortex solely is a function of the charge 
of each BPS monopole contributing to it. Because 
large-holonomy calorons of topological charge one and 
thus monopoles with one unit of magnetic charge only are 
thermodynamically excited Abrikosov-Nielsen-Olesen (ANO) vortices are 
center vortices in the magnetic phase of an SU(2) or an SU(3)
Yang-Mills theory. 

A mesoscopic description of a static ANO vortex in the BPS limit 
is given by the action Eq.\,(\ref{effactM2}) when omitting 
the potential for $\varphi$. (This potential measures the 
energy density of the {\sl macroscopic} ground state and thus must 
be subtracted when discussing the typical energy of a 
solitonic configuration on a mesoscopic level.)  
The following cylindrically symmetric (with axis along the $x_3$ 
direction) and static ansatz for the gauge field $a^D_\mu$ is made to describe the vortex
\cite{NielsenOlesen1973}:
\eqb
\label{ansatzaM}
a^D_4=0\,,\ \ \ \ \ \ \ \ a^D_i=\epsilon_{ijk}\hat{r}_j \,e_k A(r)\,,
\eqe
where $\hat{r}$ is a radial unit vector in the $x_1x_2$ plane, and $\vec e$ is a 
unit vector along the $x_3$ direction. Writing 
$\varphi=|\varphi|(r)\exp[i\theta]$, the equations 
of motion for $|\varphi|(r)$ and $A(r)$ read
\eab
-\frac{1}{r}\frac{d}{dr}\left(r\frac{d}{dr}|\varphi|\right)+
\left(\frac{1}{r}-g\,A\right)^2|\varphi|&=&0\,,\label{phir}\\ 
-\frac{d}{dr}\left(\frac{1}{r}\frac{d}{dr}(rA)\right)+
g\,|\varphi|^2\left(g\,A-\frac{1}{r}\right)&=&0\,.\label{Ar}
\eae
We keep in mind that $|\varphi|(r\to\infty)=\sqrt{\frac{\La_M^3\beta}{2\pi}}$, 
see Eq.\,(\ref{BPSvarphimod}). Let us first search for BPS saturated solutions to the system 
(\ref{phir},\ref{Ar}). The question is under what 
condition a solution to the first-order equation
\eqb
\label{fovarphi}
\frac{d}{dr}|\varphi|=\left(\frac{1}{r}-g\,A\right)|\varphi|
\eqe
also solves Eq.\,(\ref{phir}). Substituting Eq.\,(\ref{fovarphi}) 
into Eq.\,(\ref{phir}), we observe that
\eqb
\label{foA}
A=-r\frac{d}{dr}\,A\,.
\eqe
The solution to Eq.\,(\ref{foA}) is 
\eqb
\label{solfoA}
A(r)=\frac{\mbox{const}}{r}\,. 
\eqe
Substituting Eq.\,(\ref{solfoA}) into Eq.\,(\ref{Ar}), 
the constant in Eq.\,(\ref{solfoA}) is determined to be 
$\frac{1}{g}$. Eq.\,(\ref{fovarphi}) is solved for $r>0$ by 
\eqb
\label{fovarphisol}
|\varphi|(r)\equiv\sqrt{\frac{\La_M^3\beta}{2\pi}}\,.
\eqe
We have found an analytical solution to the system 
(\ref{phir},\ref{Ar}) for $r>0$ which has one unit of 
magnetic flux $F_v(r)\equiv\frac{2\pi}{g}=\oint_C dz_\mu\,a^D_{\mu}$, where $C$ 
is a circular curve of radius $r$ in the $x_1 x_2$ plane 
centered at $x_1=x_2=0$, and which has a vanishing vortex core. 
The energy density $\rho_v(r)$, when evaluated 
on the configuration $A(r)=\frac{1}{g\,r}$, $|\varphi|(r)\equiv\sqrt{\frac{\La_M^3\beta}{2\pi}}$, 
reduces to that of the magnetic field $H(r)=\frac{1}{2\pi r}\frac{d}{dr}\,F_v(r)$: 
(By Stoke's theorem the magnetic field $H$ must be proportional to a 
two-dimensional $\delta$-function at $r=0$. Thus the energy per unit 
vortex length diverges on the configuration (\ref{solfoA}) and (\ref{fovarphisol}).) 
\eqb
\label{endensmag}
\rho_v(r)=\frac12\,H^2(r)\equiv 0\,,\ \ \ \ \ \ (r>0)\,.
\eqe
The (isotropic in the $x_1x_2$ plane) pressure 
$P_v(r)$ outside the vortex core is given as 
\eqb
\label{Pvort}
P_v(r)=-\frac12\,\frac{\La_M^3\beta}{2\pi}\,\frac{1}{g^2\,r^2}\,,\ \ \ \ \ \ \ \ (r>0)\,.
\eqe
Eq.\,(\ref{Pvort}) is the mesoscopic reason for the macroscopic 
results in Eqs.\,(\ref{Pmag2}) and (\ref{Pmag3}). Because of the {\sl negative} pressure in 
Eq.\,(\ref{Pvort}) vortex-{\sl loops} start to collapse as soon as 
they are created at finite coupling $g$. (The pressure is more negative inside 
than outside of the vortex-loop.) Notice that in the limit $g\to\infty$ we have 
$P_v(r)\to 0$: For temperatures below $T_{c,M}$ 
vortex-loops do exist as particle-like excitations. 

\subsubsection{Leaving the BPS limit}

Let us now discuss how the solutions in Eqs.\,(\ref{fovarphisol}) and 
(\ref{solfoA}) are deformed when the BPS limit is left at 
finite coupling $g$. In this case only approximate analytical 
solutions to the second-order system (\ref{phir}) and (\ref{Ar}) 
are known \cite{NielsenOlesen1973}. Assuming $|\varphi|$ to be constant, 
which is viable sufficiently far away from the core around $r=0$, the 
solution to Eq.\,(\ref{Ar}) reads 
\eqb
\label{arnoBPS}
A(r)=\frac{1}{gr}-|\varphi|\,K_1(g|\varphi|r)\to \frac{1}{gr}-|\varphi|\sqrt{\frac{\pi}{2g|\varphi|r}}\,
\exp[-g|\varphi|r]\,,\ \ \ \ (r\to\infty)\,,
\eqe
where $K_1$ is a modified Bessel function. (Notice that the $1/r$ divergence at $r=0$ 
of the solution in Eq.\,(\ref{solfoA}) is absent in the configuration in Eq.\,(\ref{arnoBPS}).) 
Now $|\varphi|$ is not constant inside the vortex 
core but smoothly approaches zero for $r\to 0$. So there 
is a gradient contribution from $|\varphi|$ to the energy per unit length $\frac{E_v}{2\pi R}$  
along the vortex where $R$ denote the typical radius 
of a vortex loop. Let us first calculate the magnetic energy per unit length $\frac{E_{m,v}}{2\pi R}$. 
One has \cite{NielsenOlesen1973}
\eqb
\label{magenANO}
\frac{E_{m,v}}{2\pi R}=\frac12\,\int_0^\infty dr\,2\pi r H^2(r)=
\pi\,|\varphi|^2\int_0^\infty dy\,K^2_0(y)\,y=\frac{\pi}{2}\,|\varphi|^2\,,
\eqe
where $|\varphi|$ is given in Eq.\,(\ref{fovarphisol}). The gradient contribution 
$\frac{E_{\varphi,v}}{2\pi R}$ is comparable to $\frac{E_{m,v}}{2\pi R}$. 
Thus the typical energy $E_v$ of the vortex loop is obtained by 
multiplying $\frac{E_{m,v}}{2\pi R}+\frac{E_{\varphi,v}}{2\pi R}$ with 
the typical circumference $2\pi R\sim \frac{1}{g|\varphi|}$ of the loop. We have
\eqb
\label{EANOvortex}
E_v\sim 2\,\frac{\pi}{2}\,|\varphi|^2\times \frac{1}{g|\varphi|}=
\pi\,\frac{|\varphi|}{g}\,.
\eqe
From Eq.\,(\ref{EANOvortex}) we conclude that vortex loops 
become massless in the limit $g\to\infty$. 
For $r\gg \frac{1}{g|\varphi|}$ the (isotropic in the $x_1x_2$ plane) pressure 
$P_v(r)$ of the vortex configuration is still given by Eq.\,(\ref{Pvort}): 
For finite coupling $g$ vortex loops collapse as soon as they are created.

\subsection{Derivation of the phase of a macroscopic complex scalar field}

\begin{figure}
\begin{center}
\leavevmode
\leavevmode
\vspace{5.2cm}
\includegraphics{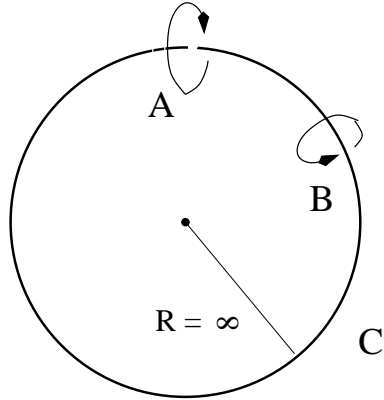}
\end{center}
\caption{Two center-vortex loops of opposite flux being 
pierced by an $S_1$ of infinite radius.\label{Fig35}}      
\end{figure}

We consider a pair of center-vortex loops that are pierced by a circular 
contour $C$ of infinite radius, see Fig.\,\ref{Fig35}. The total 
center flux $F_{\pm,0}$ through the minimal surface spanned by 
$C$ is  
\eqb
\label{centerfluxC}
F_{\pm,0}=\left\{\begin{array}{c}\pm\frac{2\pi}{g}\\ 
0\end{array}\right.
\eqe
depending on whether at finite coupling $g$ the loop $A$ ($B$) 
collapses to nothing well before the loop $B$ ($A$) or whether this roughly happens 
at the same time. (In any case, vortex loops which start out {\sl without} getting pierced 
by $C$ do not contribute a center flux through $C$.)

Unlike in the case of a pair of a 
monopole and an antimonopole discussed in 
Sec.\,\ref{defphismag} the flux $F_{\pm,0}$ takes {\sl discrete} values. 
In analogy to the derivation 
of a macroscopic monopole condensate we may investigate the thermally averaged 
flux of the vortex-antivortex (spin-0) system in the limit where 
there is no spatial momentum of this system and where 
$g\to\infty$: 
\eab
\label{avfluxsysV}
\lim_{g\to\infty}\,F_{\pm,0;\tiny\mbox{th}}(\delta)&=&
\,4\pi\,\int d^3p\,\delta^{(3)}(\vec{p})\, n_B(\beta |2\,E_v(\vec{p})|)\,F_{\pm,0}\nonumber\\ 
&=&0,\pm\frac{2}{\beta|\varphi|}=0,\pm\frac{\lambda^{3/2}_{c,M}}{\pi}\,.
\eae
The phase of a macroscopic complex field $\Phi$ is defined as
\eqb
\label{phasePhi}
\Gamma\frac{\Phi}{|\Phi|}(x)\equiv \lim_{g\to\infty}\la \exp[ig\oint_{{C}(x)}dz_\mu\, 
(a^D)^\mu]\ra
\eqe
where $\Gamma$ is a complex constant, and $(a^D)^\mu$ denotes 
the gauge field of a center vortex. The expectation on the right-hand side of Eq.\,(\ref{phasePhi}) 
is proportional to the expectation of 't Hooft's 
loop operator \cite{'tHooft1978}. (Green functions of this 
operator change their phase by $-1$ (SU(2)) and $\exp[\pm \frac{2\pi i}{3}]$ (SU(3)) 
under an exchange of the order of any two of their arguments: 
A feature which is familiar from Green functions of fermionic fields.) 
The possible values of $\Phi$'s phase are parametrized by the average center 
flux $\lim_{g\to\infty}\,F_{\pm,0;\tiny\mbox{th}}(\delta)$ in Eq.\,(\ref{avfluxsysV}).   

\begin{figure}
\begin{center}
\leavevmode
\leavevmode
\vspace{4.5cm}
\includegraphics{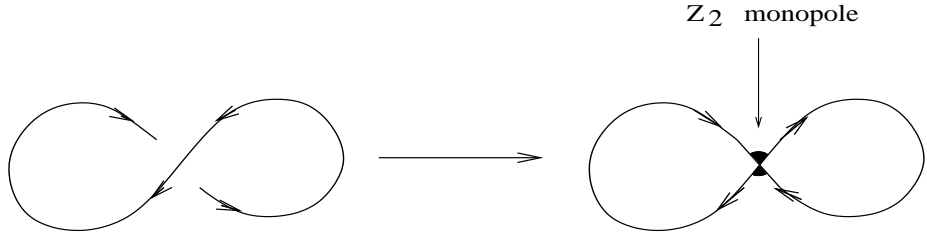}
\end{center}
\caption{The creation of an isolated $Z_2$ monopole by self-intersection 
of a center-vortex loop. \label{intersect}}      
\end{figure}

According to Eq.\,(\ref{avfluxsysV}) the condensate $\Phi$ of center-vortex loops 
is determined by discrete parameter values which can be 
normalized as $\hat{\tau}=\pm 1,0$. Recall that at $\lambda_{c,M}$ 
vortex loops start to be stable excitations since 
their pressure $P_v(r)$ is zero outside the (infinitely thin) 
vortex core. Once the field $\Phi$ 
acquires a nonzero modulus its phase is observed to jump locally 
in space. (Each jump corresponds to a stable vortex loop travelling in from infinity and getting 
pierced by ${\cal C}$.) Thus we are led to interpret jumps in 
$\Phi$'s phase as creation processes for (fermionic) 
particles. (There are two degenerate 
polarizations of these particles: The two possible directions of center 
flux in a given vortex-loop. By travelling in from infinity the 
vortex loop makes $\Phi$'s phase jump twice: A created unit of 
flux is associated with a forward jump (piercing by $C$) while a backward jump corresponds to 
minus this flux (no piercing by $C$, center-vortex loop lies inside $C$).) If a single 
center-vortex loop is created with sufficiently large momentum then a part of 
its kinetic energy can be converted 
into the mass of self-intersection points by subsequent twisting. Self-intersection points are 
$Z_2$ magnetic monopoles, each contributing $\sim \Lambda_C$ 
to the mass of the state where $\Lambda_C$ is a mass scale. Twisting does 
not alter the fact that only two possible polarizations (spin-1/2 fermions) occur. 
(The magnetic flux is reversed by a $Z_2$ monopole \cite{Reinhardt2001}, see Fig.\,\ref{intersect}.)
We conclude that the mass spectrum $m_n$ of 
fermionic excitations is equidistant: 
\eqb
\label{massspectferm}
m_n\sim n\,\Lambda_C\,,\ \ \ \ \ (n=0,1,2,\cdots)\,.
\eqe
The process of fermion creation violates the spatial homogeneity 
of the system and thus thermal equilibrium. Fermion creation, that is, the process of 
sucking in stable center-vortex loops from infinity, can only go on so long as 
the energy density provided by the field $\Phi$ 
is finite. Thus the field 
$\Phi$ must eventually relax to one of the possible zero-energy 
minima of its potential. This phenomenon is 
generally known as tachyonic pre- and re-heating \cite{nonequ}.  

\noindent\underline{SU(2) case:}\vspace{0.1cm}\\ 
The symmetry, which is dynamically broken by center-vortex condensation, 
is a local magnetic $Z_2$. After a relaxation of 
$\Phi$ to zero energy density the ground state must exhibit 
the associated $Z_2$ degeneracy. We conclude that for 
SU(2) the parameters $\hat{\tau}=\pm 1$ 
must be identified: They describe the same 
minimum of $\Phi$'s potential. The parameter value 
$\hat{\tau}=0$ corresponds to the other degenerate minimum. 
The center flux carried by a given 
flux line is associated with the differences in $\Phi$'s phase in each 
minimum of $\Phi$'s potential.\vspace{0.1cm}\\   
\noindent\underline{SU(3) case:}\vspace{0.1cm}\\ 
For SU(3) the dynamically broken symmetry is a local magnetic $Z_3$. 
As a consequence, each of the three possible values 
$\hat{\tau}=\pm 1,0$ describes 
one of the three possible, distinct minima of $\Phi$'s potential. 
Before relaxation to zero energy density local jumps of $\Phi$'s phase 
generate two distinct species of flux loops: Each associated with the three differences 
in $\Phi$'s phase modulo three. (A short jump between two neighbouring unit roots is 
equivalent to a long jump into the opposite direction involving 
the third unit root as a brief stop-over.) 

\subsection{The potential of the macroscopic complex scalar field}

At $T_{c,M}$ dual gauge modes decouple. Moreover, a 
condensate of (Cooper-like) pairs of single center-vortex--center-antivortex loops confines fundamental electric and fermionic 
test charges. This happens 
because each condensed center-vortex loop represents an 
electric dipole. A condensate of such dipoles squeezes an externally applied 
electric field into a flux tube: Oppositely charged test particles are subject to 
a linear potential at large distances. Thus the center phase 
is confining both test charges and {\sl all} gauge modes: 
There is complete confinement. (The Polyakov loop 
expectation is zero below $T_{c,M}$, the 't Hooft loop expectation $\Phi$, which is the 
dual order parameter for confinement, jumps to a finite value.) 

The effective action for 
the center phase thus is only a functional of $\Phi$ and $\bar{\Phi}$. Moreover, 
thermal equilibrium (that is, periodicity in Euclidean time) 
is no longer applicable to constrain $\Phi$'s potential. According to our discussion in 
the last section $\Phi$'s potential $V_C$ must 
be (i) invariant under center jumps only (invariance under a larger 
(continuous or discontinuous) symmetry is forbidden), (ii) it must allow for 
fermion creation by center jumps, and (iii) center-degenerate minima of 
zero energy density have to occur. Moreover, (iv) we insist on the occurrence 
of a single mass scale $\La_C$ only. From (i) we conclude that $V_C$ can not 
be a function of $\bar{\Phi}\Phi$ alone.     

\noindent\underline{SU(2) case:}\vspace{0.1cm}\\ 
The only potential $V_C$ satisfying (i),(ii), (iii), and (iv) is given by
\eqb
\label{2potC}
V_C=\overline{v_C}\,v_C\equiv\overline{\left(\frac{\Lambda_C^3}{\Phi}-\La_C\,\Phi\right)}\,
\left(\frac{\Lambda_C^3}{\Phi}-\La_C\,\Phi\right)\,.
\eqe
The zero-energy minima of $V_M$ 
are at $\Phi=\pm \Lambda_C$. It is easy to show that 
adding or subtracting powers $(\Phi^{-1})^{2l+1}$ or 
$\Phi^{2k+1}$ in $v_C$, where $k,l=0,1,2,3,\cdots$, 
generates additional minima and thus destroys 
the center degeneracy of the ground state after relaxation 
and/or violates the demand for zero energy-density at a finite value of $\Phi$ 
in these minima (requirement (iii)). \vspace{0.1cm}\\   
\noindent\underline{SU(3) case:}\vspace{0.1cm}\\ 
The only potential $V_C$ satisfying (i),(ii), (iii), and (iv) is given by
\eqb
\label{3potC}
V_C=\overline{v_C}\,v_C\equiv\overline{\left(\frac{\Lambda_C^3}{\Phi}-\Phi^2\right)}\,
\left(\frac{\Lambda_C^3}{\Phi}-\Phi^2\right)\,.
\eqe
The zero-energy minima of $V_C$ 
are at $\Phi=\Lambda_C\exp\left[\pm\frac{2\pi i}{3}\right]$ and $\Phi=\Lambda_C$. 
Again, adding or subtracting powers $({\Phi}^{-1})^{3l+1}$ or $(\Phi)^{3k-1}$ 
in $v_C$, where $l=0,1,2,3,\cdots$ and $k=1,2,3,4,\cdots$, violates requirement (iii).

\begin{figure}
\begin{center}
\leavevmode
\leavevmode
\vspace{5.5cm}
\includegraphics{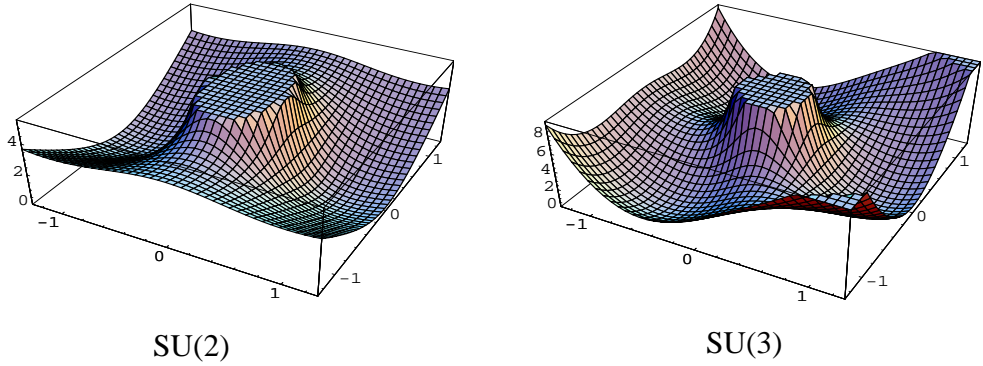}
\end{center}
\caption{The potential $V_C=\overline{v_C(\Phi)}v_C(\Phi)$ for the center-vortex 
condensate $\Phi$. Notice the regions of negative tangential curvature 
inbetween the minima.\label{Fig37}}      
\end{figure}

In Fig.\,\ref{Fig37} plots of the potentials in Eq.\,(\ref{2potC}) and Eq.\,(\ref{3potC}) are shown. 
The ridges of negative tangential curvature are classically forbidden: 
The field $\Phi$ tunnels through these ridges, and a 
phase change, which is determined by an element of the center $Z_2$ (SU(2)) or $Z_3$ (SU(3)), 
occurs locally in space. This is the afore-mentioned generation of one unit of center flux. 

\subsection{Thermodynamics close to the Hagedorn transition \label{Hagedorn}}

The action describing the process of relaxation of $\Phi$ 
to one of $V_M$'s minima is  
\eqb    \label{actionPhi}
S = \int d^4x 
\left(\frac{1}{2}\,\overline{\partial_\mu \Phi} \partial^\mu \Phi - \frac12\, V_C \right) \,.
\eqe
In contrast to the electric and magnetic phases the action $S$ in Eq.\,(\ref{actionPhi}) 
does not determine a classical, macroscopic ground state if $\Phi$ is not 
in one of $V_C$'s minima. Though tunneling processes occur 
in real time they can be described by a Euclidean simulation, 
WKB-like approximations are thinkable. Alternatively, the computation of fermion creation 
rates can be performed on a finite-temperature 
lattice based on the theory (\ref{actionPhi}). An interesting object to be measured
is $\Phi$'s two-point correlator $\Pi(x)\equiv\la\bar{\Phi}(x)\Phi(0)\ra$. 
Projecting onto a given spatial momentum $\vec{p}$ at a given temperature, the strength of intermediate tachyonic 
modes can be extracted by a Fourier analysis 
of the $\tau$ dependence in $\int d^3x\,\exp[i\vec{p}\cdot \vec{x}]\,\Pi(\tau,\vec{x})$. This gives a measure for the 
density of states $\rho_n$ for fermions of 
mass $\sim n\,\Lambda_C$ and spatial momentum $\vec{p}$ \cite{HofmannSchefflerStamatescu2005}. 

Let us estimate $\rho_n$ for SU(2). 
The multiplicity of fermion states with $n$ self-intersections is given by twice the number $L_n$
of bubble diagrams with $n$ vertices in a scalar $\lambda \phi^4$ theory. 
In \cite{BenderWu1969} the minimal number of such diagrams $L_{n,min}$ was 
estimated to be $L_{n,min}=n!3^{-n}$. Using Stirling's formula 
this can be approximated as 
\eab
\label{Stirling}
L_{n,min}&\sim&\frac{1}{3}\sqrt{2\pi n}\,\left(3\e\right)^{-n}\,n^{n}\nonumber\\ 
&=&\frac{\sqrt{2\pi}}{3}\,\exp\left[n\Big(\log n-(\log3+1)\Big)+\frac12\log n\right]\nonumber\\ 
&\sim&\frac{\sqrt{2\pi}}{3}\,\exp\left[n\log n\right]\,
\eae
for $n\gg 1$. So the number $F_n$ of fermion states with mass $m_n\sim n\,\La_C$ is 
bounded from below roughly by
\eqb
\label{Fn}
F_n\sim 2\times \frac{\sqrt{2\pi}}{3}\,\exp\left[n\log n\right]=
\frac{\sqrt{8\pi}}{3}\,\exp\left[n\log n\right]\,.
\eqe
We now estimate the density $\rho_{n,0}$ of fermion states at 
rest $\vec{p}=0$ (or $\tilde{\rho}(E=n\La_C)$) 
by differentiating $F_n$ with respect to 
$n$ and dividing the result by the level-spacing 
$\delta m_n=\Lambda_C$. We have
\eab
\label{statdes}
\rho_{n,0}&>&\frac{\sqrt{8\pi}}{3\La_C}\,\exp[n\log n]\Big(\log n+1\Big)\,\ \ \ \ \ \mbox{or}\nonumber\\ 
\tilde{\rho}(E)&>&\frac{\sqrt{8\pi}}{3\La_C}\exp[\frac{E}{\La_C}\log\frac{E}{\La_C}]
\Big(\log\frac{E}{\La_C}+1\Big)\,.
\eae
Eq.\,(\ref{statdes}) tells us that the density of static fermion 
states is more than exponentially increasing with energy $E$. The partition function 
$Z_{\Phi}$ thus is estimated as
\eab
\label{partfunctionphi}
Z_{\Phi}&>&\int_{E^*}^\infty dE\,\tilde{\rho}(E)\,n_F(\beta E)\nonumber\\ 
&>&\frac{\sqrt{8\pi}}{3\La_C}\,\int_{E^*}^\infty dE\,\exp\left[\frac{E}{\La_C}\right]\,
\exp[-\beta E]\,,
\eae
where $E^*\gg \Lambda_C$ is the energy where we start to trust 
our approximations. Thus $Z_{\Phi}$ diverges at some 
temperature $T_H<\Lambda_C$. (Strictly speaking, $T_H=0$ 
according to Eq.\,(\ref{statdes}). This is an artefact of 
our assumption that all states are infinitely narrow. 
There are, however, finite widths for higher-charge states ($n>1$) since contact interactions 
exist between vortex lines and intersection points. 
Moreover, in the real world higher-charge states are even broader 
due to their decay and their mutual annihilation into charge-one and 
charge-zero states. This happens because an SU(2) theory, which is not 
confining at the present temperature of the Universe, 
mixes with the theory under consideration 
at large temperatures and thus couples its massless gauge mode -- the photon -- to the 
$Z_2$ charges of the latter. The larger $n$ the 
broader the associated state and the less reliable our 
assumption of infinitely narrow states.) This is the 
celebrated Hagedorn transition. (At $T_H$ the entropy 
diverges and the system condenses self-intersecting center-vortex 
loops into a new ground state: The monopole condensate of 
the magnetic phase. The process of monopole condensation from below violates the spatial 
homogeneity of the system: $Z_2$ charges loose their 
mass by dense packing only. Thus the Hagedorn transition is 
genuinely nonthermal.)  
 
Once $\Phi$ has is settled into $V_C$'s 
minima $\Phi_{\tiny\mbox{min}}$ there are no 
quantum fluctuations $\delta\Phi$. Writing $\Phi=|\Phi|\exp\left[i\frac{\theta}{\La_c}\right]$, 
this is a consequence of the following fact:  
\eqb
\label{minimacur}
\left.\frac{\pd^2_{\theta} V_C(\Phi)}{|\Phi|^2}\right|_{\Phi_{\tiny\mbox{min}}}=
\left.\frac{\pd^2_{|\Phi|} V_C(\Phi)}
{|\Phi|^2}\right|_{\Phi_{\tiny\mbox{min}}}
=\left\{\begin{array}{c}8\,\ \ \ \ \ (\mbox{SU(2)})\\ 
18\,\ \ \ \ \ (\mbox{SU(3)})\end{array}\right.\,.
\eqe
Thus radial {\sl and} tangential 
fluctuations around $\Phi_{\tiny\mbox{min}}$ would have a mass $m_{\delta\Phi}$ which is 
sizably larger than the compositeness scale $|\Phi_{\tiny\mbox{min}}|$ 
for both SU(2) and SU(3). Since $|p^2_{\delta\Phi}+m^2_{\delta\Phi}|\le 
|\Phi_{\tiny\mbox{min}}|^2$ for any allowed Euclidean momentum 
$p^2_{\delta\Phi}>0$ this means that the fluctuations $\delta\Phi$ 
are absent: After relaxation the ground state of the 
center phase has a vanishing pressure and a vanishing energy density.

\section{Matching the pressure\label{MCC}}

The mass scales $\La_E$ and $\La_M$, which determine the modulus of the 
adjoint Higgs field $\phi$ and the moduli of the 
monopole condensates $\varphi$ (SU(2)) and $\varphi_1,\varphi_2$ (SU(3)), respectively, are 
related. This derives from the fact that across a second-order 
transition the pressure is continuous. We have
\eab
\label{laElaM}
\Lambda_M&=&\left(4+\frac{\lambda_{c,E}^3}{720\pi^2}\right)^{1/3}\La_E\,,\ \ \ \ \ \ \ (\mbox{SU(2)})\,,\nonumber\\ 
\Lambda_M&=&\left(2+\frac{\lambda_{c,E}^3}{720\pi^2}\right)^{1/3}\La_E\,,\ \ \ \ \ \ \ (\mbox{SU(3)})\,.
\eae
Across the magnetic-center transition the pressure is not 
continuous. We may, however, estimate the scale 
$\Lambda_C$ by assuming thermal equilibrium at the 
{\sl onset} of this transition. (This assumption underlies 
Eq.\,(\ref{avfluxsysV}).) This gives
\eqb
\label{laMlaC}
\Lambda_M\sim 2^{1/3}\,\La_C\,,\ \ \ \ (\mbox{SU(2)})\,\ \ \ \ \ \mbox{and}\ \ \ \ \ 
\Lambda_M\sim \La_C\,,\ \ \ \ (\mbox{SU(3)})\,.
\eqe

\section{Thermodynamical quantities\label{PEEN}}

\subsection{Results}

In this section we present our numerical results for one-loop 
temperature evolutions of thermodynamical quantities throughout the 
electric and magnetic phase.\vspace{0.1cm}\\  
\noindent\underline{SU(2) case:}\vspace{0.1cm}\\ 
\noindent In the electric phase the ratio of pressure $P$ and $T^4$ is given as
\eqb
\label{pressureEP2}
\frac{P}{T^4}=-\frac{(2\pi)^4}{\lambda_E^4}\left[\frac{2\lambda_E^4}{(2\pi)^6}\left(2\bar{P}(0)+
6\bar{P}(2a)\right)+2\lambda_E\right]\, 
\eqe
where the function $\bar{P}(a)$ and the dimensionless mass parameter 
$a$ are defined in Eq.\,(\ref{P(y)}) and 
Eq.\,(\ref{aofela}), respectively. In 
the magnetic phase we have
\eqb
\label{pressureMP2}
\frac{P}{T^4}=-\frac{(2\pi)^4}{\lambda_M^4}
\left[\frac{6\lambda_M^4}{(2\pi)^6}\bar{P}(a)+\frac{\lambda_M}{2}\right]\,
\eqe
where $a$ is defined in Eq.\,(\ref{massspecM}). 

\noindent In the electric phase the ratio of energy density $\rho$ 
and $T^4$ is given as
\eqb
\label{rhoEP2}
\frac{\rho}{T^4}=\frac{(2\pi)^4}{\lambda_E^4}
\left[\frac{2\lambda_E^4}{(2\pi)^6}\left(2\bar{\rho}(0)+
6\bar{\rho}(2a)\right)+2\lambda_E\right]\, 
\eqe
where the function $\bar{\rho}(a)$ is defined as
\eqb
\label{rhobar}
\bar{\rho}(a)\equiv \int_{0}^{\infty}dx\,x^2 \frac{\sqrt{x^2+a^2}}{\exp(\sqrt{x^2+a^2})-1}\,.
\eqe
In the magnetic phase we have
\eqb
\label{rhoMP2}
\frac{\rho}{T^4}=\frac{(2\pi)^4}{\lambda_M^4}
\left[\frac{6\lambda_M^4}{(2\pi)^6}\bar{\rho}(a)+\frac{\lambda_M}{2}\right]\,.
\eqe
The ratio of entropy density and $T^3$ is given as
\eqb
\label{ent}
\frac{s}{T^3}=\frac{1}{T^4}\left(\rho+P\right)\,.
\eqe
Because the ground-state contributions cancel in $\frac{s}{T^3}$ this quantity 
is not as infrared sensitive as, e. g., $\frac{\rho}{T^4}$ or $\frac{P}{T^4}$: 
Lattice simulations are in a position to correctly measure the 
entropy density at low temperatures. 
\vspace{0.1cm}\\  
\noindent\underline{SU(3) case:}\vspace{0.1cm}\\ 
In the electric phase we have
\eqb
\label{pressureEP3}
\frac{P}{T^4}=-\frac{(2\pi)^4}{\lambda_E^4}
\left[\frac{2\lambda_E^4}{(2\pi)^6}\left(4\bar{P}(0)+
3(4\,\bar{P}(a)+2\,\bar{P}(2a))\right)+
2\lambda_E\right]\,
\eqe
and
\eqb
\label{rhoEP3}
\frac{\rho}{T^4}=\frac{(2\pi)^4}{\lambda_E^4}
\left[\frac{2\lambda_E^4}{(2\pi)^6}\left(4\bar{\rho}(0)+
3(4\,\bar{\rho}(a)+2\,\bar{\rho}(2a))\right)+
2\lambda_E\right]\,.
\eqe
In the magnetic phase we have
\eqb
\label{pressureMP3}
\frac{P}{T^4}=-\frac{(2\pi)^4}{\lambda_M^4}
\left[\frac{12\lambda_M^4}{(2\pi)^6}
\bar{P}(a)+\lambda_M\right]\,
\eqe
and
\eqb
\label{rhoMP3}
\frac{\rho}{T^4}=\frac{(2\pi)^4}{\lambda_M^4}
\left[\frac{12\lambda_M^4}{(2\pi)^6}
\bar{\rho}(a)+\lambda_M\right]\,.
\eqe
The ratio of entropy density and $T^3$ is given in Eq.\,(\ref{ent}) where 
now the SU(3)-expressions for $P$ and $\rho$ have to be used.

\begin{figure}
\begin{center}
\leavevmode
\leavevmode
\vspace{4.5cm}
\includegraphics{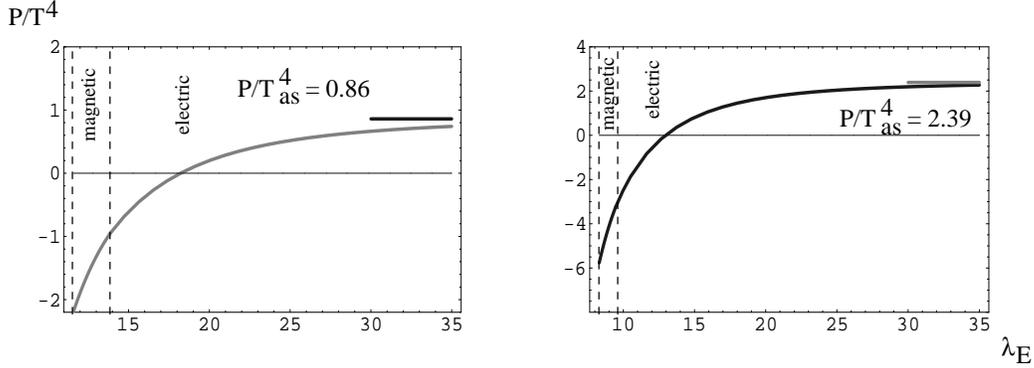}
\end{center}
\caption{\protect{$\frac{P}{T^4}$ as a function of temperature for SU(2) (left panel) and SU(3) (right panel). 
The horizontal lines indicate the respective asymptotic values, the dashed vertical lines are the phase boundaries.  
\label{pressure}}}      
\end{figure}

\begin{figure}
\begin{center}
\leavevmode
\leavevmode
\vspace{4.5cm}
\includegraphics{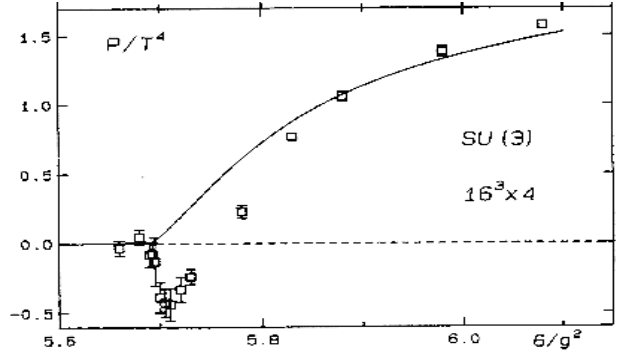}
\end{center}
\caption{\protect{$\frac{P}{T^4}$ as a function of temperature for SU(3) as obtained on 
a ($16^3\times 4$)-lattice using the differential method with a universal two-loop 
perturbative $\beta$ function \protect\cite{Brown1988,Deng1988} and using the integral 
method (solid line) \protect\cite{EngelsFingberg1990}. The figure is taken 
from \protect\cite{EngelsFingberg1990}.\label{pressureLat}}}      
\end{figure}

The result for $\frac{P}{T^4}$ is plotted in Fig.\,\ref{pressure} as 
a function of temperature throughout the electric and 
magnetic phase, Fig.\,\ref{pressureLat} depicts SU(3)-lattice results. 
Notice that the pressure is negative 
in the electric phase close to $\lambda_{E,c}$ and 
even more so in the magnetic phase where the ground state 
strongly dominates the thermodynamics of infrared sensitive quantities. Notice also the 
negative pressure in Fig.\,\ref{pressureLat} obtained close to the phase transition 
when the differential method is used in the lattice simulation. 
(For a discussion of differential versus integral 
method see Sec.\,\ref{DVI}.) We conclude that the finite-size 
constraints of realistic lattices have a severe effect on the obtained values of the 
pressure shortly above the first confining transition 
and even more so below this transition. 

Let us now discuss the behavior close to $T_{c,M}$ where the thermodynamical relation
\eqb
\label{pdt}
dP=S\,dT\,
\eqe
starts to be  violated. Eq.\,(\ref{pdt}) implies that 
in a homogeneous, thermalized system the pressure needs to be a strictly 
monotonic increasing function of temperature since $S$ is never negative. 
Crossing the point $T_{c,M}$ from above, 
the pressure jumps from a negative to a positive value. (On the magnetic side of the phase 
boundary the ground state strongly dominates the excitations, on the 
center side the vortex-condensate has zero pressure while fermionic 
excitations give a positive contribution.) There are two ways 
of seeing that thermal equilibrium must break down close to 
$T_{c,M}$. First, spatial homogeneity starts to be badly violated by 
discontinuous and local 
phase changes of the field $\Phi$ as soon as the system starts to 
condense center-vortex loops. The derivation of Eq.\,(\ref{pdt}), 
however, relies on thermal equilibrium and thus on spatial homogeneity. 
Second, one may assume thermal equilibrium and then lead this assumption to a contradiction.  
In thermal equilibrium the spectral density $\rho(E)$ in the center phase is more than  
exponentially increasing with energy $E$, see Eq.\,(\ref{statdes}). Thus the 
partition function diverges at $T=T_{H}<\Lambda_C$: A 
homogeneous system would need an infinite amount of energy per volume 
to increase its temperature beyond $T_{H}$. But this is a contradiction to the fact that 
a magnetic phase exists for $T>T_{H}$. (In an extended thermalized 
system the transition from the center phase to the magnetic phase is accomplished 
by an increase of the overall energy density: The excitation 
of very massive dual gauge modes, though very unlikely, is furnished energetically 
by the large energy residing in the system. If 
the considered system is of a small spatial extent, such as the interaction 
vertex in a scattering process, then the total energy of the system, e.g., 
the center-of-mass energy being deposited into a vertex, needs to be larger than the 
very large mass of the dual gauge modes on the magnetic 
side of the phase boundary.) Thus thermal equilibrium breaks down 
for $T\sim T_H$. 

\begin{figure}
\begin{center}
\leavevmode
\leavevmode
\vspace{5.5cm}
\includegraphics{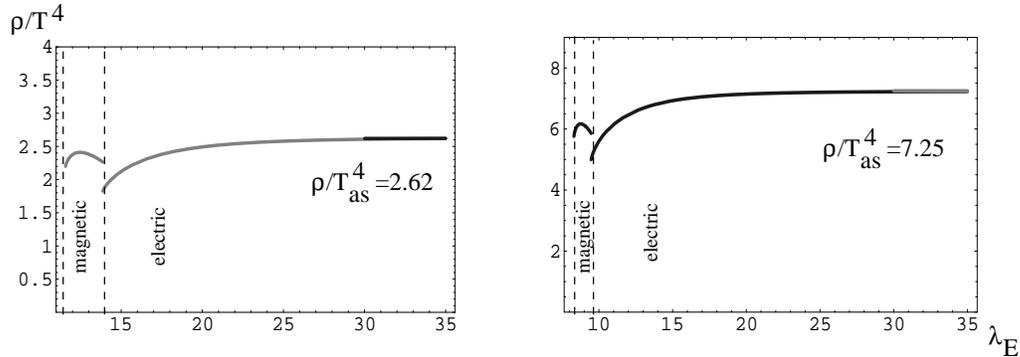}
\end{center}
\caption{$\frac{\rho}{T^4}$ as a function of temperature for SU(2) (left panel) and SU(3) (right panel). 
The horizontal lines indicate the respective asymptotic values, the dashed 
vertical lines are the phase boundaries.\label{rho}}      
\end{figure}

\begin{figure}
\begin{center}
\leavevmode
\leavevmode
\vspace{6cm}
\includegraphics{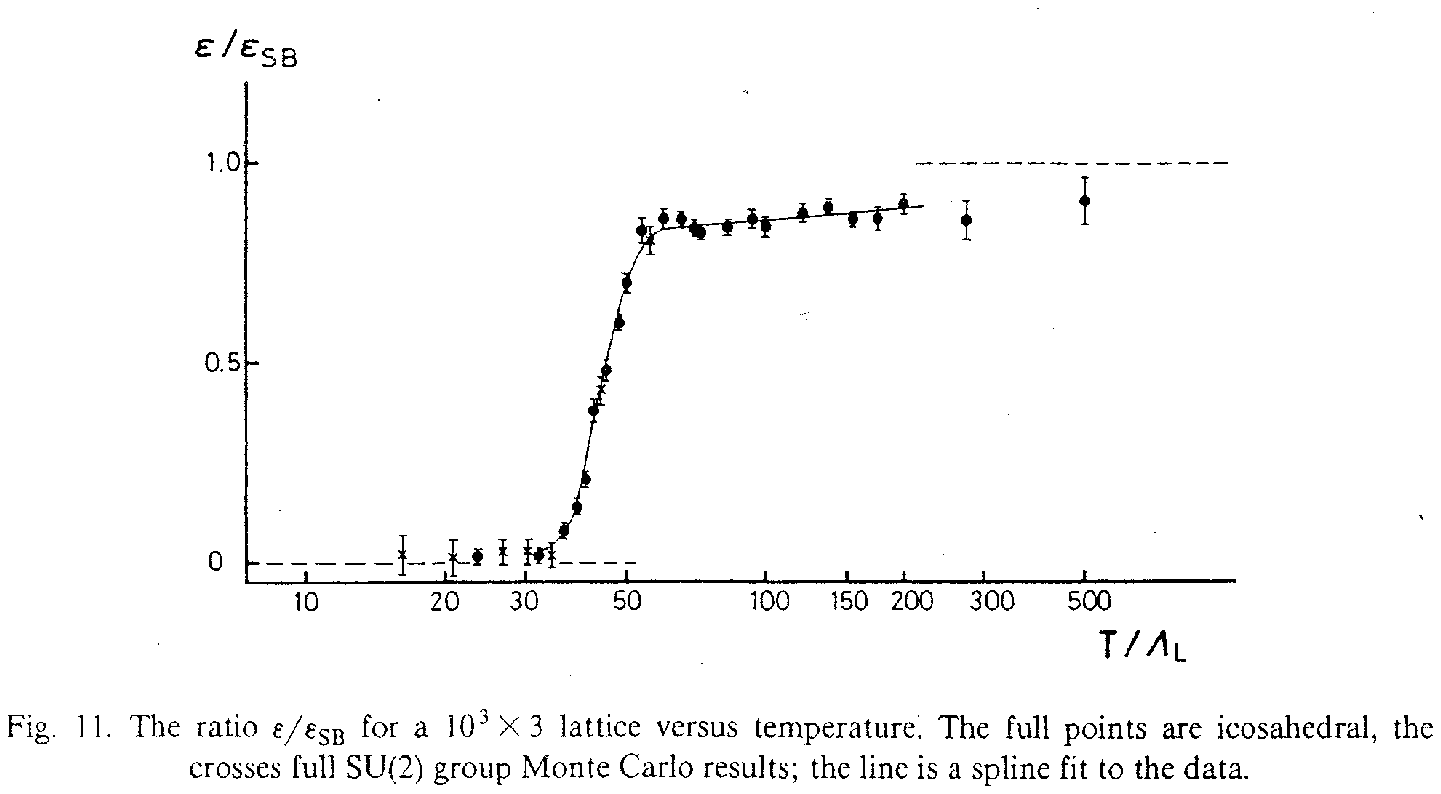}
\end{center}
\caption{$\frac{\rho}{T^4}$ as obtained from the SU(2)-lattice simulation in \protect\cite{EnKaSaMo1982}.\label{En1}}      
\end{figure}

The result for $\frac{\rho}{T^4}$ as 
a function of temperature throughout the electric and 
magnetic phase is shown in Fig.\,\ref{rho}. Fig.\,\ref{En1} depicts an 
SU(2)-lattice result \cite{EnKaSaMo1982}. Notice the (small) discontinuities at 
$\lambda_{c,E}$. Their occurrence is explained by the fact that by crossing the 
electric-magnetic phase boundary the system needs to generate an 
extra polarization for each dual gauge mode compared to the two polarizations 
of a TLM mode. Extra polarizations are extra fluctuating 
degrees of freedom which increase the energy 
density on the magnetic side of the phase boundary. 
The situation is somewhat peculiar: On the one hand, the order parameter for the dynamical breaking 
of the dual gauge groups U(1)$_D$ (SU(2)) and U(1)$_D^2$ (SU(3)) 
is continuous. On the other hand, there is a small amount of 
latent heat being released across the {\sl magnetic-electric} transition. (That is, by heating up 
the system starting in the magnetic phase.) As we shall see in 
Sec.\,\ref{Apps}, this is the reason for a dynamical stabilization of 
the temperature of the cosmic microwave background 
against gravitational expansion. (Thus we may look forward to 
enjoy the privilege of the photon's masslessness 
for another sizable fraction of today's age of the Universe \cite{GH2005}.) Again, 
the energy density is dominated by the 
ground-state contribution in the electric phase close to the electric-magnetic transition 
and even more so in the magnetic
phase. Notice also that $\rho=-P$ at the point, where the system starts 
to condense center-vortex loops, that the magnetic phase is 
narrower for SU(3) than it is for SU(2), and that $\frac{\rho}{T^4}$ 
dips in a much steeper way at the electric-magnetic transition for SU(3) than for SU(2). 

\begin{figure}
\begin{center}
\leavevmode
\leavevmode
\vspace{5.5cm}
\includegraphics{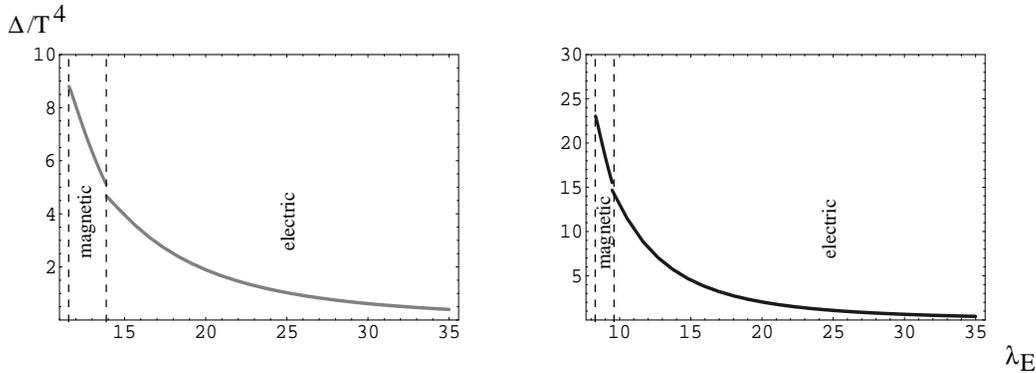}
\end{center}
\caption{The interaction measure $\frac{\Delta}{T^4}$ as a function of temperature 
for SU(2) (left panel) and SU(3) (right panel). The asymptotic value in both cases is $\frac{\Delta}{T^4}=0$, the 
dashed vertical lines are the phase boundaries.\label{IM}}      
\end{figure}

\begin{figure}
\begin{center}
\leavevmode
\leavevmode
\vspace{7.5cm}
\includegraphics{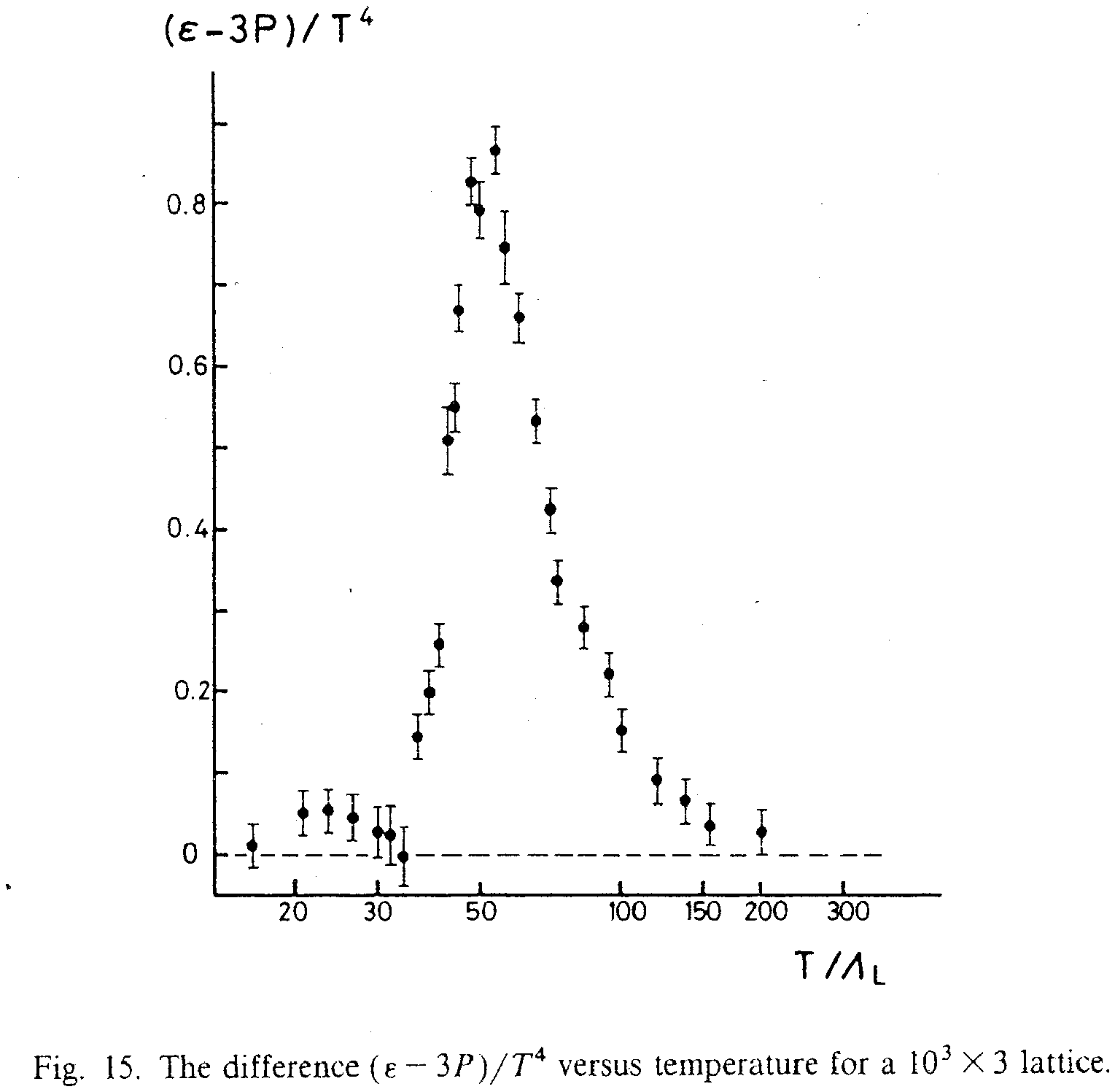}
\end{center}
\caption{$\frac{\Delta}{T^4}$ as obtained from the SU(2)-lattice 
simulation in \protect\cite{EnKaSaMo1982}.\label{En3}}      
\end{figure}

\begin{figure}
\begin{center}
\leavevmode
\leavevmode
\vspace{12.8cm}
\includegraphics{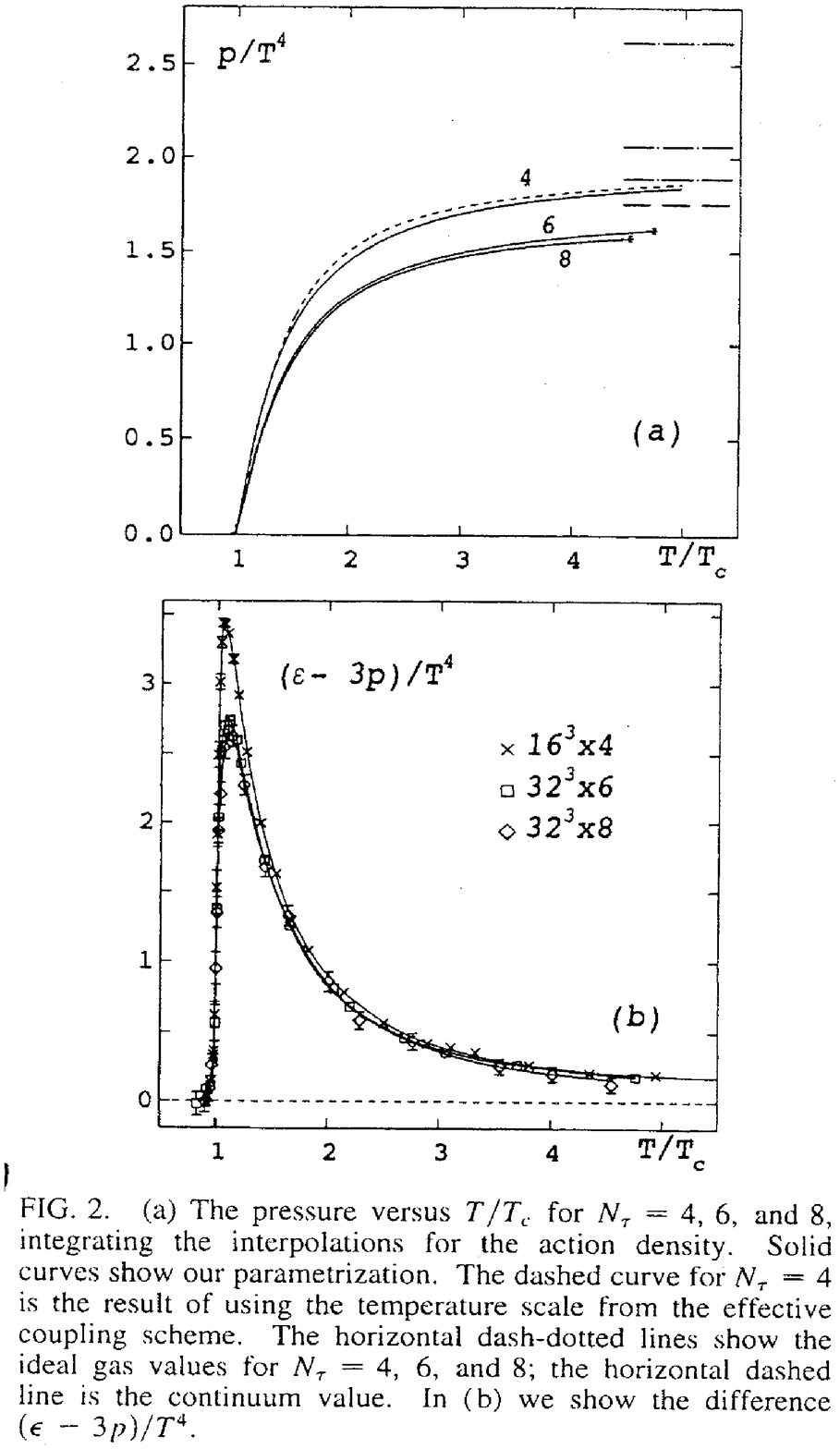}
\end{center}
\caption{$\frac{P}{T^4}$ and $\frac{\Delta}{T^4}$ as obtained from the SU(3)-lattice 
simulation in \protect\cite{Bielfeld1996}.\label{B1}}      
\end{figure}

The result for the interaction measure 
$\frac{\Delta}{T^4}\equiv\frac{\left(\rho-3P\right)}{T^4}$ is shown in 
Fig.\,\ref{IM}. Figs.\,\ref{En3} and \ref{B1} are lattice results. 
Notice the rapid approach to the free-gas limit in Fig.\,\ref{IM} and 
the large values of $\frac{\Delta}{T^4}$ in the magnetic phase. Interestingly, there is a small 
bump to the left of the phase boundary in Fig.\,\ref{En3}.  

\begin{figure}
\begin{center}
\leavevmode
\leavevmode
\vspace{5.5cm}
\includegraphics{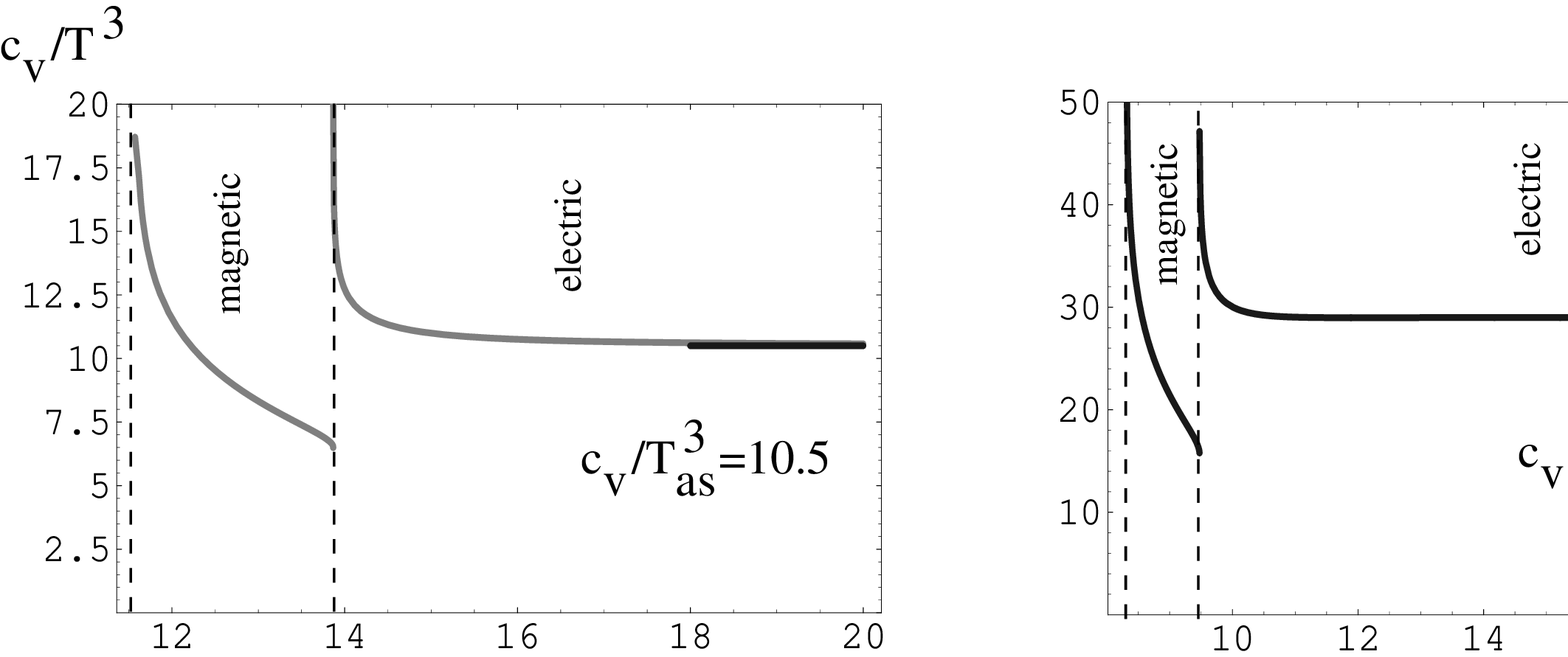}
\end{center}
\caption{$\frac{c_V}{T^3}$ as a function of temperature for SU(2) (left panel) and SU(3) (right panel). 
The horizontal lines signal the respective asymptotic values, the dashed 
vertical lines are the phase boundaries.\label{SH}}      
\end{figure}

\begin{figure}
\begin{center}
\leavevmode
\leavevmode
\vspace{6.0cm}
\includegraphics{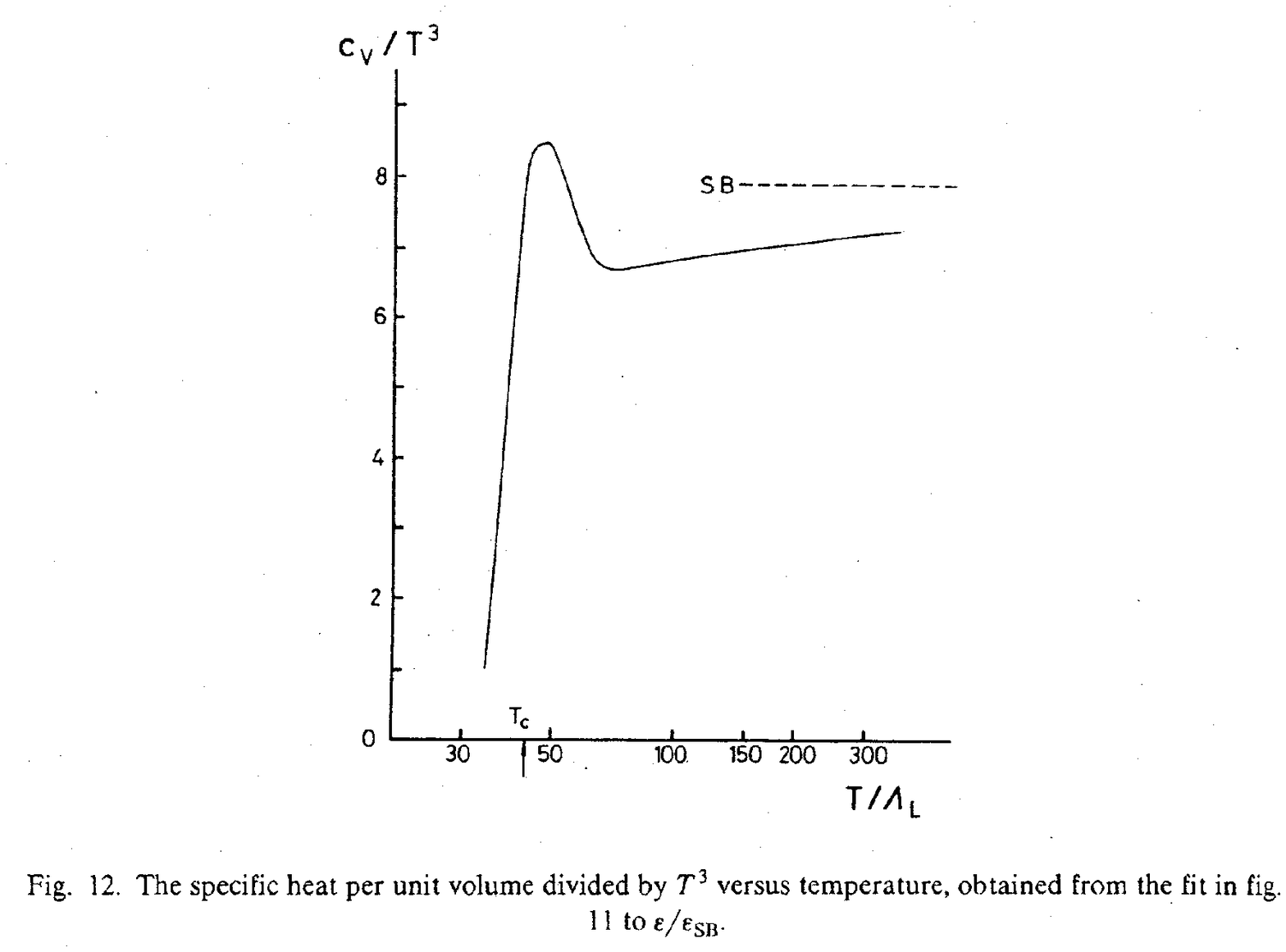}
\end{center}
\caption{$\frac{c_V}{T^3}$ as obtained from the SU(2)-lattice simulation in \protect\cite{EnKaSaMo1982}.\label{En2}}      
\end{figure}

The result for the ratio of the specific heat per unit volume $c_V\equiv \frac{d\rho}{dT}$ and 
$T^3$ is shown in Fig.\,\ref{SH}, Fig.\,\ref{En2} is an SU(2)-lattice result \cite{EnKaSaMo1982}. 
The quantity $\frac{c_V}{T^3}$ peaks both at the electric-magnetic and 
the magnetic-center transition. The finite peak at the former phase boundary is 
in agreement with the electric-magnetic transition being second-order. Moreover, 
we have $\left.\frac{c_V}{T^3}\right|_{T_{c,E};\tiny\mbox{SU(3)}}\sim 3\,
\left.\frac{c_V}{T^3}\right|_{T_{c,E};\tiny\mbox{SU(2)}}$. This 
explains why lattice simulations prefer to identify the confining transition 
as weakly first order for SU(3), see \cite{LuciniTeperWenger2003,LuciniTeperWenger2005} 
and references therein. Now, 
$3\not=\infty$ but in the vicinity of $T_{c,E}$ lattice results 
for infrared sensitive quantities, such as $\frac{c_V}{T^3}$, are 
not reliable anyway.  

The result for $\frac{S}{T^3}$ is shown in Fig.\,\ref{ST3}. 
Fig.\,\ref{ST3lat} depicts a lattice result for SU(3) 
obtained with the differential method \cite{Brown1988}. 
The entropy density $S$ is a measure for the `mobility' of gauge modes. Notice 
the jump of $S/T^3$ which, again, is explained by the additional 
polarization of the dual gauge mode in the magnetic phase. Notice also that $\frac{S}{T^3}$ vanishes at 
the point $T_{c,M}$ where the system condenses center vortices. At this point dual gauge modes are 
infinitely heavy: The thermodynamics is completely determined by the ground state. The numerical 
agreement between the lattice result (d) (largest lattice) in Fig.\,\ref{ST3lat} and the SU(3)-result 
in Fig.\,\ref{ST3} is striking. The two data points to the left of the jump 
in (d) indicate that an ambiguity exists for the value of $\frac{S}{T^3}$ 
very close to the transition. 
The jump itself corresponds to the large slope of $\frac{S}{T^3}$ on the electric side of the phase 
boundary in Fig.\,\ref{ST3}. The observed agreement is explained by the small sensitivity of 
the quantity $\frac{S}{T^3}$ on the ground-state physics making the finite-volume 
lattice simulation reliable close to the electric magnetic transition. 

The data files needed to generate the plots in Figs.\,\ref{pressure}, \ref{rho}, \ref{IM}, \ref{SH}, 
and \ref{ST3} are provided by the author upon request. 

\subsection{Comparison with the lattice\label{complat}}

\subsubsection{Specific observations}

\noindent\underline{SU(2) case:}\vspace{0.1cm}\\ 
The results of an early lattice measurements of the energy density $\rho$ 
and the interaction measure $\Delta\equiv \rho-3P$ in a pure SU(2) gauge theory 
were reported in \cite{EnKaSaMo1982}. In that work the critical temperature $T_c$ for 
the deconfinement transition was determined from the critical behavior of the 
Polyakov-loop expectation and the peak position 
of the specific heat using a Wilson action. The function $\Delta(T)$ was extracted by multiplying the 
lattice $\beta$ function with the difference of 
plaquette expectations at finite and zero temperature (symmetric lattice). 
This assures that $\Delta$ vanishes for $T\to 0$. What is subtracted in \cite{EnKaSaMo1982} 
at finite $T$ is, however, {\sl not} the value $\Delta(T=0)$ since the plaquette expectation on the symmetric lattice 
is multiplied with the {\sl finite}-$T$ value of the $\beta$ function. Apart from this 
approximation, the use of a perturbative $\beta$ function was assumed for all 
temperatures. The simulation was carried out on a (rather small) 
$(10^3\times 3)$-lattice. 

Let us compare our results with those of \cite{EnKaSaMo1982}. 
\begin{figure}
\begin{center}
\leavevmode
\leavevmode
\vspace{5.0cm}
\includegraphics{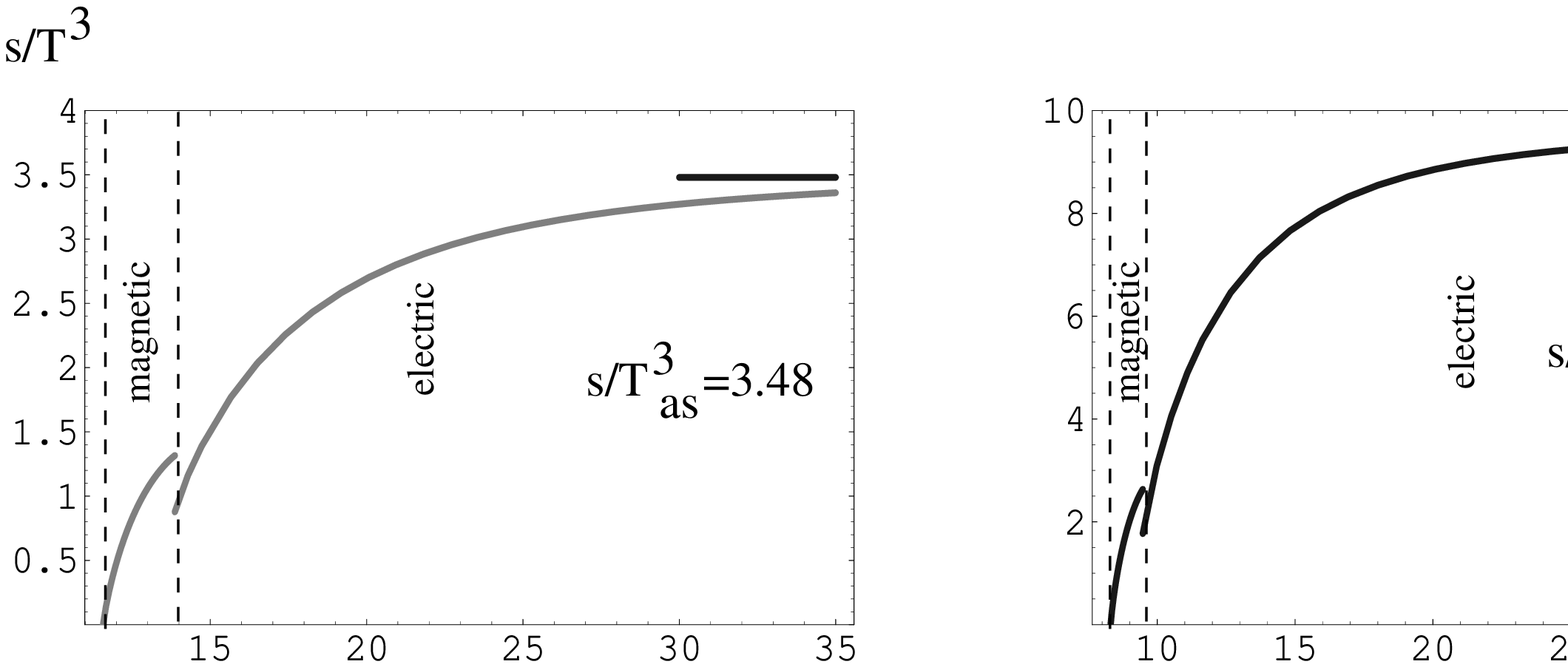}
\end{center}
\caption{$\frac{S}{T^3}$ as a function of temperature for SU(2) (left panel) and SU(3) (right panel). 
The horizontal lines signal the respective asymptotic values. \label{ST3}}      
\end{figure}
\begin{figure}
\begin{center}
\leavevmode
\leavevmode
\vspace{5.0cm}
\includegraphics{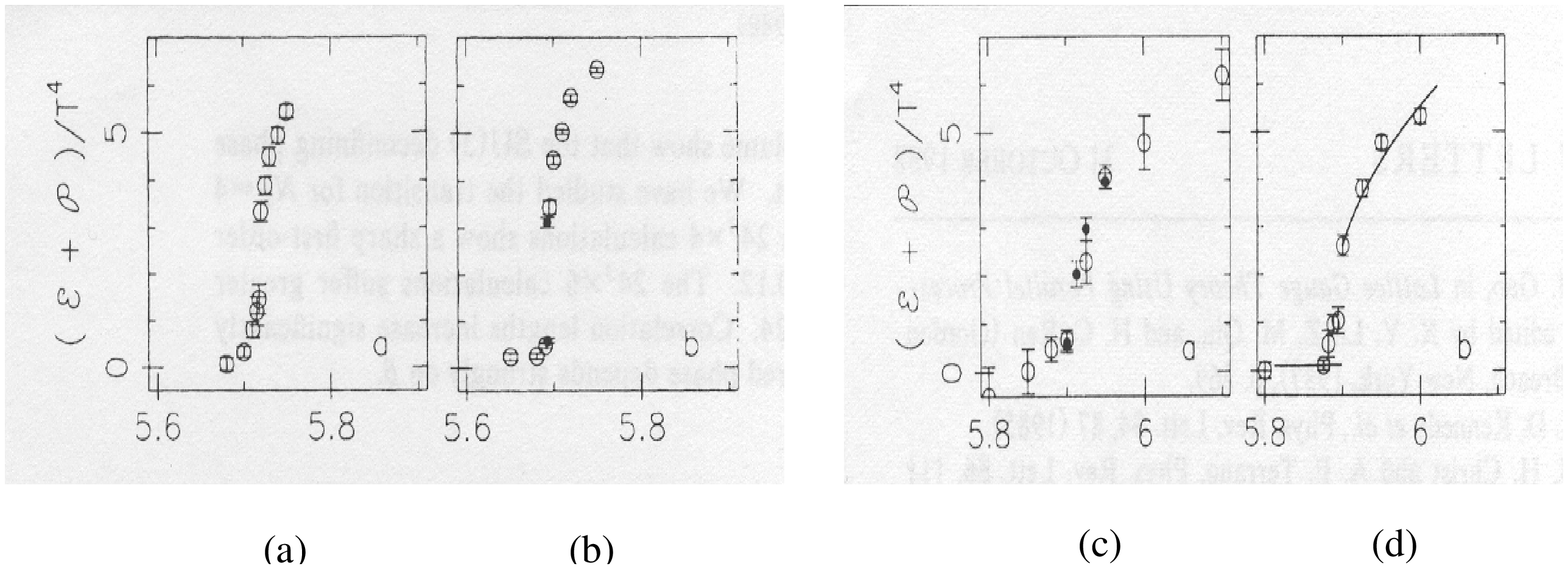}
\end{center}
\caption{$\frac{S}{T^3}$ as a function of $\beta$ 
obtained in SU(3) lattice gauge theory using the differential method and a 
perturbative beta function \protect\cite{Brown1988}. The simulations were performed on (a) $16^3\times 4$, (b) 
($24^3\times 4$)-, (c) ($16^3\times 6$)- (open circles) and ($20^3\times 6$)- (closed circles), and 
(d) ($24^3\times 6$)-lattices. Using the ($24^3\times 6$)-lattice, the critical 
value of $\beta$ is between 5.8875 and 5.90. \label{ST3lat}}      
\end{figure}
The lattice results for $\rho$ in \cite{EnKaSaMo1982} differ drastically from 
our results for temperatures close the first confinement, that is, the electric-magnetic 
transition. (The lattice is doomed to produce incorrect results for infrared sensitive quantities 
close to the electric-magnetic transition and in the magnetic phase: A finite spatial 
lattice size $L$ cuts off 
{\sl physical} correlations on length scales $>L$ since the correlation length 
$l_M=M_{m}^{-1}>L$ close to the electric-magnetic transition 
and $l_M=\infty$ in the magnetic phase. Here $M_{m}$ denotes the 
mass of a magnetic monopole.) 

\noindent We obtain
\eqb
\label{rhooverrhoSBc}
\left.\frac{\rho}{\rho_{\tiny\mbox{SB}}}\right|_{T\sim 1.5\,T_{c,E}}\sim 1.27\,,
\eqe
where $\rho_{\tiny\mbox{SB}}\equiv\frac{\pi^2}{5}T^4$ denotes the Stefan-Boltzmann limit 
(ideal gas of three species of massless gluons with two polarizations each). On 
the lattice this ratio is measured to be
smaller than unity: $\left.\frac{\rho}{\rho_{\tiny\mbox{SB}}}\right|_{T\sim 1.5\,T_{c,E}}=0.84$. 
At $T\sim 5T_{c,E}$ we obtain 
\eqb
\label{rhooverrhoSB5}
\left.\frac{\rho}{\rho_{\tiny\mbox{SB}}}\right|_{T\sim 5\,T_{c,E}}\sim 1.33\,
\eqe
while the lattice measures $\left.\frac{\rho}{\rho_{SB}}\right|_{T\sim 5\,T_{c,E}}=0.85$.   

Our results for the pressure $P$ are {\sl negative} for $T$ 
close to $T_{c,E}$ (see Fig.\, \ref{pressure}) -- much in 
contrast to the positive values obtained 
in \cite{EnKaSaMo1982}. At $T\sim 5\,T_{c,E}$ we obtain
\eqb
\label{PoverrhoSB5}
\left.\frac{P}{P_{\tiny\mbox{SB}}}\right|_{T\sim 5\,T_{c,E}}\sim 1.31\,
\eqe
while the lattice measures $\left.\frac{P}{P_{\tiny\mbox{SB}}}\right|_{T\sim 5\,T_{c,E}}\sim 0.88$. (On the lattice 
$P$ is extracted from the measured values of $\Delta$ and $\rho$, and 
$P_{\tiny\mbox{SB}}\equiv\frac{\pi^2}{15}T^4$ denotes the 
Stefan-Boltzmann limit.)

Notice that the results in Eq.\,(\ref{rhooverrhoSB5}) and Eq.\,(\ref{PoverrhoSB5}) are very close 
to the ratio $\frac{2\times 3+2}{6}\frac{4}{3}$ of the number of degrees of freedom in a gas of 
two species of (nearly) massless gluons (three polarizations per species) and one massless species 
and a gas of massless gluons. (At $\lambda_E\sim 75$ the value of the mass 
parameter is $a\sim 2\pi\frac{5.5}{650}\sim 0.086$. Thus the Boltzmann suppression is small for TLH modes, 
compare also with Fig.\,\ref{IM}.) At extremely high temperatures a TLH mode `remembers' its massiveness 
at low temperatures in terms of an extra polarization. The latter originates from a 
tiny mass which solves the infrared problem of loop expansions, for formal arguments see \cite{HofmannSept2006}. 

The peak-value of the specific heat is about 
$\left.\frac{c_V}{T^3}\right|_{T_{c,E}}\sim 20$ while 
it is measured to be $\sim 8$ on the lattice. Moreover, we have 
$\left.\frac{\Delta}{T^4}\right|_{T_{c,E}}\sim 4.8$ while the 
lattice obtains a value $\sim 0.85$. The much lower values obtained on the lattice are 
not surprising: Finite lattice sizes cut off existing long-range correlations at $T_{c,E}$. 

No result for the entropy density was directly reported in \cite{EnKaSaMo1982}.\vspace{0.1cm}\\ 
\noindent\underline{SU(3) case:}\vspace{0.1cm}\\ 
Here we discuss the results obtained in \cite{Bielfeld1996} with a 
Wilson action on the lattice of the largest 
time extension, $N_\beta=8$, and the results obtained 
in \cite{Deng1988,Brown1988}.  

In the vicinity of the transition point $T_{c,E}$ 
the situation for both $\rho$ and $P$ is similar to the SU(2) case: 
Drastic differences between the lattice measurements and 
our results occur. Again, $P$ is negative close to $T_{c,E}$ 
contradicting the positive values obtained with the integral method in 
\cite{Bielfeld1996}. A lattice simulation \cite{Deng1988} of $P$, which used the differential method, 
has reported negative pressure for $T$ shortly above the transition 
already in 1988. The most negative value of $P/T^4\sim -0.5$ obtained 
in \cite{Deng1988} very close to the phase transition is down by a factor 
of about $0.19$ as compared to our 
result at the electric-magnetic transition, 
see Figs.\,\ref{pressure} and \ref{pressureLat}. Again, this is 
explained by the finite lattice-size cutoff on physical long-range correlations. 
The lattice-result 
obtained with the integral method \cite{Bielfeld1996} is by construction 
positive definite, see Sec.\,\ref{DVI}, and thus 
it is {\sl unphysical}. For that reason we 
renounce a (useless) comparison of our results for pressure, 
energy density, and interaction measure with those obtained in \cite{Bielfeld1996}. In 
\cite{Deng1988} only the dependence of $\frac{P}{T^4}$ on the 
lattice coupling was presented. One can use Fig.\,\ref{pressureLat} and the temperature 
dependence of $\frac{P}{T^4}$ in Fig.\,\ref{B1} to 
gauge particular values of this quantity against temperature. (In both simulations 
\cite{Deng1988} and \cite{Bielfeld1996} the universal part 
of the two-loop perturbative $\beta$ function was used to relate lattice 
coupling to lattice spacing.) For example, a value of $\frac{P}{T^4}\sim 1.6$ in 
\cite{Deng1988} corresponds to a value $\frac{P}{T^4}\sim 1.5$ in \cite{Bielfeld1996}. 
The latter is associated with a temperature $T=3.2\,T_{c,E}$ by virtue of Fig.\,\ref{B1}. 
We have
\eqb
\label{rhoN3}
\left.\frac{P}{T^4}\right|_{T\sim 3.2\,T_{c,E}}\sim 2.2\,.
\eqe
This is larger than the result obtained in \cite{Deng1988} and explained by the 
insufficient account of infrared correlations in the lattice simulation. These correlations generate masses for six 
out of eight gluon species, thus extra polarizations, and therefore a larger 
value for $\frac{P}{T^4}$.  

Our asymptotic ($\lambda_E=35$)
values for $P$ and $\rho$ are  
\eqb
\label{PoverrhoSB3a}
\left.\frac{P}{P_{SB}}\right|_{as}\sim 1.30\,,\ \ \ \ \ \ \left.\frac{\rho}{\rho_{SB}}\right|_{as}\sim 1.37\,,
\eqe
where $\rho_{SB}=\frac{8}{15}\pi^2\,T^4=3\,P_{SB}$. 
Both values in Eq.\,(\ref{PoverrhoSB3a}) are close to the ratio $R=\frac{11}{8}=1.375$ 
of the numbers of polarization in a free gluon gas, where six gluon species have a tiny mass, and in a free 
gluon gas where all gluon species are massless.

The entropy density approaches zero for $T\searrow T_{c,M}$, see Fig.\,(\ref{ST3}). 
This expresses the fact that dual gauge modes
are decoupled (infinite masses): The ground state strongly dominates the thermodynamics.

\subsubsection{Differential versus integral method\label{DVI}}

What are the reasons for the qualitative difference 
between the pressure-results obtained 
in \cite{Bielfeld1996,EngelsKarschScheideler1999} using the integral method and in 
\cite{Brown1988,Deng1988} using the differential method? 
While the differential method is based on the definition 
\eqb
\label{ptd}
P=T\frac{\pd\ln Z}{\pd V}\,,
\eqe
which is proper for a lattice of {\sl finite} volume $V$, the integral method assumes 
the thermodynamical limit $V\to\infty$ from the start. 
In this limit one has
\eqb
\label{ptdL}
P=T\frac{\ln Z}{V}\,,
\eqe
and thus the pressure equals minus the free energy density. 
In Eqs.\,(\ref{ptd}) and (\ref{ptdL}) $Z$ denotes the partition function. 

The official reason for the introduction of the 
integral method, see for example \cite{EngelsFingberg1990}, was 
that one wanted to avoid the use of the 
imprecisely known $\beta$ 
function in the strong-coupling regime of the theory. (Based on the definition in Eq.\,(\ref{ptd}),  
the $\beta$ function multiplies the sum of spatial and time plaquette averages in the 
expression for the pressure.) When using the definition in Eq.\,(\ref{ptdL}), 
the derivative of the pressure with respect to the bare coupling $\bar{\beta}$ 
($\bar{\beta}=\frac{6}{\bar{g}^2}$ for SU(3)) 
can be expressed as an expectation over minus the sum of spatial and time-like plaquettes 
without the beta-function prefactor. Thus the pressure is, up to an unknown 
integration constant, obtained 
in terms of an integral of a sum of 
plaquette averages over $\beta$. The integration constant 
is chosen in such a way that the pressure vanishes at a temperature well 
below $T_c$. Instead of only integrating over minus the sum of spatial and 
time-like plaquette expectations an extra term was added to the 
{\sl integrand} \cite{Bielfeld1996,EngelsKarschScheideler1999} 
to assure that the pressure vanishes at $T=0$. The added term equals twice 
the plaquette expectation taken on a symmetric lattice (the expectation at $T=0$ ). 
We stress that this prescription does not 
follow from the definition in Eq.\,(\ref{ptdL}). 
Moreover, the assumption that $P=0$ for $T\sim 0.8\,T_{c}$ or 
so is a strong bias. (There are massless fermionic particles 
in the center phase which keep the total pressure 
positive for temperatures comparable to this value.) 

The results for $P(T)$ obtained when using the 
integral method show a rather large dependence on the spatial size and the time extent $N_\tau$ of the lattice 
\cite{Bielfeld1996}. We believe that this 
reflects the considerable deviation from the assumed 
thermodynamical limit for realistic lattice sizes. The problem 
was addressed in \cite{EngelsKarschScheideler1999} where a 
correction factor $r$ was introduced to relate $P$, 
obtained with the integral method, to $P$, obtained 
with the differential method. For a given value of $N_\tau$ 
the factor $r$ was determined from the pressure-ratio at $\bar{g}=0$. 
Subsequently, this value of $r$ was used at {\sl finite} coupling $\bar{g}$ 
to extract the spatial anisotropy 
coefficient $c_\sigma$ (essentially the $\beta$ function) 
by demanding the equality of the pressure obtained 
with the integral and the differential method. 
In doing so, twice the plaquette expectation at $T=0$ was, again, added to minus 
the sum of spatial and time-like plaquette expectations in the differential-method expression 
for the pressure. It may be questioned 
whether a simple correction factor $r$ does correctly 
account for finite-size effects and, if yes, 
whether it is justified to determine $r$ in the limit 
of noninteracting gluons. (The $c_\sigma$-values obtained in this way 
do not coincide with those obtained in \cite{Klassen1998}.) 
In addition, it seems that the imprecise knowledge of the $\beta$ 
function, which contains information about fluctuations in the 
ultraviolet, is much less of a problem for a lattice simulation of the pressure 
than the missing infrared physics is (finite lattice size) \cite{Blum2004}.

Using the universal part of the two-loop perturbative beta function in the differential method, 
negative values for the pressure were obtained for $T$ close to 
$T_c$ in \cite{Deng1988}. Moreover, a rapid approach of $\rho$ and
$P$ to their respective free-gas limits was observed. This is in 
qualitative (but not quantitative) agreement 
with our results, see Figs.\,\ref{IM}, \ref{pressure}, and \ref{rho}.

\section{Implications for particle physics and cosmology\label{Apps}}

In this section we provide outlooks on the implications 
of the nonperturbative approach to SU(2) and 
SU(3) Yang-Mills thermodynamics in view of so-far unexplained phenomena 
in particle physics and cosmology. The way of 
how selected problems are addressed in this section 
is preliminary, mostly qualitative and thus should not be understood 
as the final say on the matter. Rather, we try to 
provide a certain amount of stimulus for 
future developments.     

\subsection{A Planck-scale axion: Cosmic coincidence today and $CP$ violation\label{PSA}}

Among the gauge groups SU($N$) ($N$ finite) we regard SU(2) and 
SU(3) as particular due to their unique phase 
diagrams. We have come to appreciate that nature seems to prefer situations with a 
unique outcome. Thus we tend to believe that dynamics subject to a 
finite gauge symmetry, that is, dynamics below 
the Planck scale $M_P\sim 1.2\times 10^{19}\,$GeV, 
obeys the SU(2) or SU(3) gauge principle. A possible scenario would 
be that at $M_P$ an SU($N$) gauge symmetry ($N=\infty$) is 
dynamically broken into a four-dimensional low-energy manifestation 
involving several SU(2) and SU(3) factor groups, which can behave 
in an electric-magnetically dual way to one another, and into 
nonfluctuating gravity. 

This set-up may, in fact, be described by 
the low-energy sector of a bosonic string theory whose vacuum instability 
is resolved in terms of tachyon 
condensation in the presence of a D-brane 
\cite{Sen2000}. A low-energy Kaluza-Klein \cite{Kaluza1921,Klein1926} 
compactification of the Weyl-invariant and thus 
$d$-dimensional bosonic string theory (in a flat background $d=26$) to four 
dimensions yields gauge symmetries which are associated with the isometries of the $d-4$ dimensional 
compactification manifold: We assume that these isometries are products of 
SU(2) and SU(3) corresponding to a compactification 
manifold which, locally, is $S_3\times (S_3\times S_5)\times \cdots$ \cite{Steenrod1951,Aguilar1999}. 
In addition, there are 
a low-mass scalar dilaton field $\varphi$ and a massless 
antisymmetric tensor $B_{\mu\nu}$. While the former may drive the 
monopole condensation process within, say, an SU(2) theory of 
Yang-Mills scale $\Lambda_1\sim M_P$ and thus may trigger the 
inflation of the Universe (making it spatially flat) the latter may be 
responsible for adiabatically generated density 
perturbations \cite{Prokopec2005}.  

The scales of the SU(2) and SU(3) 
factors would dynamically bet set into a certain hierarchy: 
$\La_1\sim M_P>\La_2>\cdots>\La_{\tiny\mbox{CMB}}$. The scale $\La_{\tiny\mbox{CMB}}$ 
is associated with an SU(2) theory that is not confining at the present temperature of the Universe: 
Being {\sl at} the electric-magnetic transition this theory 
generates the photon as its only massless excitation, 
see Sec.\,\ref{EWSB}. Being in its center phase at a temperature $T\sim \La_1\sim M_P$, 
$SU(2)_1$ generates fermions by re-heating (single and self-intersecting 
center-vortex loops). Because $SU(2)_1$ used to be part of SU($N=\infty$) and thus was mixing 
with the other factors these fermions couple to the gauge fields of these factors. 
Since higher-charge states (self-intersecting center-vortex loops) with opposite charges are generated in 
equal amounts we would expect that they quickly annihilate 
into charge-zero states (no self-intersections). (Re-creation of higher-charge states 
after annihilation is less likely due to the redshift of the spectrum by the 
rapid power-law expansion of the Universe furnished by a free-gas 
equation of state. The latter originates from the gauge-mode 
excitations of the other factors). Although the 
massless fermions (single center-vortex loops) 
do not couple to the propagating gauge fields of the other factors by naked gauge charges 
they do so by their dipole moments. Considering these massless fermions 
to be fundamental, there is a global, axial U(1) 
symmetry which, however, is dynamically broken \cite{AtiyahSinger1984} and anomalous 
due to the calorons of the other factors 
\cite{BellJackiw1969,Fujikawa1979}. 

Integrating out SU(2)$_1$'s massless fermions, the relevant composite field 
is a (canonically normalized) axion field $a$ whose 
coupling to the gauge fields of the other factors is
\eqb
\label{axialanomaly}
{\cal L}_{a.a.}=\frac{a}{F}\, \sum_{i=2}^{\tiny{\mbox{CMB}}}\mbox{tr}\,\tilde{F}_{\mu\nu,i}F_{\mu\nu,i}
\eqe
where $F\sim M_P$ denotes the Peccei-Quinn scale \cite{PecceiQuinn1977}: The scale at 
which $SU(2)_1$'s fermions come into existence and which measure the magnitude of 
the Cooper-pair condensate involving the massless species. The sum is over the 
other factors, and $\tilde{F}_{\mu\nu,i}=\frac12\,\epsilon_{\mu\nu\alpha\beta}\,F_{\alpha\beta,i}$ 
is the dual field strength. Eq.\,(\ref{axialanomaly}) represents 
a term in the action for the other factors (deconfining at $T\sim M_P$) 
which violates parity (P) and charge conjugation (C) symmetries.   

While $a$ would be a massless phase if the axial anomaly was absent and 
the axial U(1) only was broken dynamically the anomaly-induced coupling in 
Eq.\,(\ref{axialanomaly}) gives rise to an axion 
potential $V_a=\sum_{i=2}^{\tiny{\mbox{CMB}}}V_{a,i}$. 
The operator $\mbox{tr}\,\tilde{F}_{\mu\nu,i}F_{\mu\nu,i}$ measures the 
average topological charge density carried by 
{\sl propagating} gauge fields. This is a conserved quantity for $T\gg \La_i$, and 
thus it is independent of temperature. Integrating 
over topologies and accounting for dimensional transmutation 
we have 
\eqb
\label{axmass}
V_{a,i}\sim \left\{\begin{array}{c}\left(1-\cos\frac{a}{F}\right)\,\La_i^4\,,\ \ 
\ \ \ (\mbox{theory $i$ in electric or magnetic phase})\,,\\ 
0\,\ \ \ \ \ (\mbox{theory $i$ in center phase})\,.\end{array}\right.
\eqe
For $a$ smaller (but not much smaller) than $F$ the cosine in Eq.\,(\ref{axmass}) can be expanded, 
and the axion mass-squared at $T\sim M_P$ reads
\eqb
\label{axmassreally}
m_a^2\sim\frac{1}{F^2}\,\sum_{i=2}^{\tiny{\mbox{CMB}}}
\La_i^4\sim \sum_{i=2}^{\tiny{\mbox{CMB}}}\frac{\La_i^4}{M_P^2}\,.
\eqe
As the temperature of the Universe falls below $\La_2$ the 
associated theory fails to contribute to the 
axion mass-squared and so forth. 

Now in an expanding Friedmann-Robertson-Walker Universe the (spatially homogeneous) 
axion field $a$ satisfies the following equation 
of motion
\eqb
\label{hubbleexp}
\ddot{a}+3\,H\dot{a}+m_a^2\,a=0\,.
\eqe
The Hubble parameter $H\equiv\frac{\dot{R}}{R}=\sqrt{\frac{8\pi\rho(T)}{3\,M_P^2}}$ 
is determined by the energy density $\rho(T)=\sum_{i=2}^{\tiny{\mbox{CMB}}}\rho_i(T)$ of 
the Universe. At $T\sim M_P$ this energy density is given by the Stefan-Boltzmann 
limit of the theories $i=2,\cdots,\mbox{CMB}$ if the hierarchy between $M_P$ and $\Lambda_2$ 
is sufficiently large. (The SU(2) theory with $i=1$ went 
center, subsequently experienced a strong dilution of 
its massive excitations, and does not produce a ground-state 
contribution to $\rho$ at $T\sim M_P$.) We have $\rho_i(T)=\frac{4\pi^2}{15}\,T^4$ (SU(2)) and 
$\rho_i(T)=\frac{11\pi^2}{15}\,T^4$ (SU(3)). Thus $H$ is radiation-dominated 
at $T\sim M_P$ and $H^2\gg m_a^2$. But this means 
that $a$ is frozen to the slope of its potential with an 
amplitude $a\sim M_P$. For $T\sim \La_2$ the Universe's energy density remains 
radiation-dominated. The axion mass-squared, however, is reduced by the value 
$\frac{\La_2^4}{M_P^2}$ since the theory with $i=2$ went center. 
If $\La_2\ll M_P$ then the mass-squared $\sim 
\sum_{i=3}^{\tiny{\mbox{CMB}}}\frac{\La_i^4}{\La_2^2}$ of the axion generated by the theory with 
$i=2$ is much larger than $H^2$. By Eq.\,(\ref{hubbleexp}) this means 
that this axion rapidly relaxes to the minimum of its potential 
and thus is irrelevant for subsequent cosmology. Moreover, the 
fermions generated by the center transition of the theory 
with $i=2$ exhibit a large asymmetry in fermion number. 
This is qualitatively true since 
all three Sakharov conditions \cite{Sakharov1967} are satisfied: (i) the center transition is nonthermal (Hagedorn), 
(ii) there is a local violation of fermion number since fermions 
are nonlocal objects, and (iii) the generation of 
fermions takes place in the presence of 
CP violation (the frozen-in Planck-scale axion $a$). 

\begin{figure}
\begin{center}
\leavevmode
\leavevmode
\vspace{5.0cm}
\includegraphics{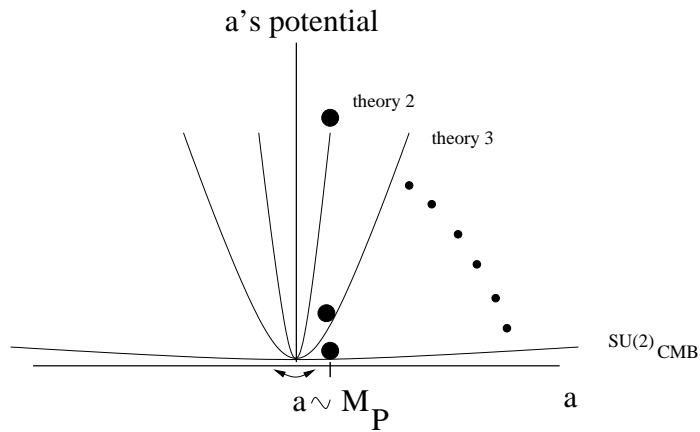}
\end{center}
\caption{The fate of a Planck-scale axion along the Universe's evolution. At temperatures sizably 
larger than $T_{\tiny\mbox{CMB}}$ the axion is frozen to the slope of 
its potential by cosmological friction ($H\gg m_a$), for $T\sim T_{\tiny\mbox{CMB}}$ 
axion mass and Hubble parameter become comparable: The axion starts to roll 
down its potential.\label{Fig40}}      
\end{figure}

This goes on until $T=T_{\tiny\mbox{CMB}}$ 
where the last theory in the chain, SU(2)$_{\tiny\mbox{CMB}}$, is close to its 
center transition (more specifically, at the 
electric-magnetic phase boundary today), see Fig.\,\ref{Fig40}. 
Here $H^2$ is dominated by the ground-state 
energy. We have 
\eqb
\label{coinc}
H^2\sim \frac{8\pi}{3}\frac{4\pi T_{\tiny\mbox{CMB}}\La_{\tiny\mbox{CMB}}^3+\rho_V+\rho_K}{M_P^2}
\stackrel{\sim}>m_a^2\sim \frac{\La_{\tiny\mbox{CMB}}^4}{M_P^2}\,.
\eqe
In Eq.\,(\ref{coinc}) $\rho_V$ denotes the energy density 
associated with the value of the axion potential at $T_{\tiny\mbox{CMB}}$, and 
$\rho_K$ is an energy density due to axion rolling. 
Both contributions are comparable since $H$ is not much larger than $m_a$, 
compare with Eq.\,(\ref{hubbleexp}). The formerly frozen-in axion field $a$ 
slowly starts to roll down its potential.
While $\rho_V$ has an equation of state $\rho_V=-P_V$ the kinetic contribution 
$\rho_K$ is associated with an equation of state $P_K=0$ which is 
the same as that of nonrelativistic matter. 

Cosmic coincidence may have an explanation in terms of a Planck-scale axion and an SU(2) Yang-Mills 
theory of scale $\La_{\tiny\mbox{CMB}}$ comparable to the present 
temperature of the CMB. For related ideas see \cite{Wilczek2004}. 
The alert reader may object that the 
$\theta$ angle in Quantum Chromodynamics (QCD) is constrained 
to be an extremely small number by a measurement 
of the neutron's electric dipole moment. On the other hand, the mass 
of the $\eta^\prime$ is much larger than the pion mass. Thus one would 
conclude that the Planck-scale axion is 
irrelevant in the former while it is relevant 
in the latter case. What is 
the resolution of this puzzle? The electric dipole moment of the 
neutron is measured with a photon of momentum much smaller than 
the QCD confinement scale. Thus this photon does not probe a 
phase of QCD with propagating gauge bosons: The operator $\tilde{F}_{\mu\nu} 
F_{\mu\nu}$ has a vanishing expectation. The $\eta^\prime$, on the other hand, 
is generated in a scattering process involving propagating gluons: 
Inside the vertex the operator $\tilde{F}_{\mu\nu}F_{\mu\nu}$ has a finite 
expectation.   

We will see in Sec.\,\ref{EWSB} that the ground-state 
contribution of SU(2)$_{\tiny\mbox{CMB}}$ in the {\sl absence} of the Planck-scale 
axion is small in comparison with the measured value of today's cosmological constant. 
The scenario outlined above does not yet explain the origin of {\sl clustering} dark matter as it is 
observed in the anomalous rotation curves of galaxies but we 
will see below that the decoupled TLH modes of SU(2)$_{\tiny\mbox{CMB}}$ are 
candidates for this form of matter.

\subsection{Electroweak sector of the Standard Model: Nature of leptons, 
electroweak symmetry breaking, masses of intermediary vector bosons, 
intergalactic magnetic fields, and solar wind\label{EWSB}}
Here we would like to propose a formulation of the electroweak 
sector of the Standard Model in terms of pure SU(2) Yang-Mills theories. 
\vspace{0.1cm}\\ 
\noindent\underline{SU(2)$_{\tiny\mbox{CMB}}$:}\vspace{0.1cm}\\ 
Let us first discuss the U(1)$_Y$ factor of the electroweak gauge 
group SU(2)$_W\times$U(1)$_Y$. We claim that this factor 
is the unbroken subgroup of an SU(2) Yang-Mills 
theory of scale comparable to that of the CMB 
temperature: $T_{\tiny\mbox{CMB}}=2.728\,K=2.351\times 10^{-4}\,$eV. 
Only one point in the phase diagram of this theory exists, the boundary between the electric and magnetic phases, 
where this claim is in accord with observations: The photon 
is unscreened and practically massless ($m_\gamma<10^{-14}\,$eV from a precision 
measurement of the Coulomb potential \cite{Williams1971}, see \cite{Dvali2003} for a 
discussion on why stronger bounds are unreliable), see Fig.\,\ref{looppress}. 
Thus we identify $T_{\tiny\mbox{CMB}}=T_{c,E}$. Notice that 
isolated charges in the electric phase of SU(2)$_{\tiny\mbox{CMB}}$ 
have a dual interpretation: What is a magnetic monopole 
in SU(2) is an electrically charged particle w.r.t. U(1)$_Y$.   

The energy density $\rho^{gs}$ of the ground-state at $T_{\tiny\mbox{CMB}}$ 
is $\rho^{gs}=4\pi\,T_{\tiny\mbox{CMB}}\La_E^3=4\pi\times (2.351\times 10^{-4}\,\mbox{eV})\,\La_E^3=
2\lambda_{c,E}\La_E^4=27.7\,\La_E^4$. Moreover, we have 
$\La_E=1.065\times 10^{-4}\,$eV. Substituting this into $\rho^{gs}$, we have 
$\rho^{gs}=\left(2.444\times 10^{-4}\,\mbox{eV}\right)^4$. This is about 0.36\% of the 
commonly accepted value of today's dark energy density $(10^{-3}\,\mbox{eV})^4$. 
The dominating, missing part would be generated by a 
slowly rolling Planck-scale axion, see Sec.\,\ref{PSA}. 

An immediate question to answer is why the masslessness and 
the unscreened propagation of the photon is a singled-out situation. The answer to this 
question is encoded in Fig.\,\ref{rho}: The energy density of an SU(2) Yang-Mills theory 
dips at the electric-magnetic phase boundary. On the electric side this is explained by 
the thermodynamical decoupling of TLH modes, on the 
magnetic side an extra polarization, which costs energy, needs to be generated 
for the photon. To facilitate the jump in energy density for SU(2)$_{\tiny{CMB}}$ 
to reach the magnetic phase thermal equilibrium needs to be violated by the 
eventually fast rolling Planck-scale axion. The photon acquires 
mass and the ground state becomes superconducting (electrically charged monopoles condense). 
It is suggestive that the occurrence of intergalactic magnetic fields is 
related to the Universe being slightly out of thermal 
equilibrium due to (slow) axion rolling today. 

At $T_{\tiny\mbox{CMB}}$ TLH modes decouple 
thermodynamically. Recall that their mass is given 
by $m_{\tiny\mbox{TLH}}=2e\,|\phi|$ and $e_{dec}=\infty$ at 
$T_{c,E}=T_{\tiny\mbox{CMB}}$. In the real world $e_{dec}$ is 
large but not infinite because SU(2)$_{\tiny\mbox{CMB}}$ is not 
the only Yang-Mills theory in the Universe. Taking 
$e_{dec}\sim 10^6$, say, the mass of a decoupled TLH mode is 
$m_{\tiny\mbox{TLH}}\sim 57\,$eV. (The reason why we chose this 
value for $e_{dec}$ is mildly justified by our 
discussion of the gauge group SU(2)$_e$ below.) Since the two TLH modes 
do not interact and thus are stable (no decay 
into light fermions is possible because SU(2)$_{\tiny\mbox{CMB}}$ 
is not in its center phase yet) they yield a tiny contribution to clustering dark matter. 

Finally, we would like to make a remark concerning the observed 
large-angle anomaly in the temperature-(electric)polarization 
cross correlation seen by WMAP \cite{WMAP2003}. The standard explanation 
is that this effect is generated by CMB photons scattering off electric charges which are 
released by an early re-ionization of the interstellar medium at 
redshift $z\sim 10-20$. We would like to propose that CMB photons scatter off electrically 
charged and dilute monopoles at temperatures 
$>T_{\tiny\mbox{CMB}}$ which are condensed at $T_{\tiny\mbox{CMB}}$. 
Since static magnetic (electric with respect to SU(2)) 
fields are completely screened in the 
photon propagator deep inside the electric phase of SU(2)$_{\tiny\mbox{CMB}}$, see Eq.\,(\ref{Pi00limit}), 
the effect should be weaker in the temperature-(magnetic)polarization 
cross correlation.        
\vspace{0.1cm}\\ 
\noindent\underline{SU(2)$_e \times$ SU(2)$_\mu\times$ SU(2)$_{\tau}$:}\vspace{0.1cm}\\       
To relate the existence and the interactions of leptons, as they are described by the 
electroweak sector of the Standard Model (SM), of which Quantum Electrodynamics (QED) 
is an integral part, with 
pure SU(2) Yang-Mills dynamics is motivated by the following observations. 
(i) The masses of charged leptons are unexplained parameters in the SM. 
In particular, their small values on the scale of their apparent pointlikeness is 
unexplained. (ii) No deeper explanation 
for the value of the magnetic dipole moment of a charged 
lepton other than that following from the Dirac equation and small radiative 
corrections is given. (iii) There are experimental indications 
that scattering processes involving the electron or the positron do not obey the 
QED predictions if the momentum transfer is close to the mass of a 
charged lepton. (iv) No Higgs particle has been observed experimentally up to a hypothetical 
Higgs mass of $\sim 115$\,GeV suggesting that electroweak 
symmetry breaking takes place by a different mechanism than assumed in the SM. 
(v) The naive ground state of the SM generates a 
cosmological constant which is many orders of magnitude 
larger than the observed value. 

\begin{figure}
\begin{center}
\leavevmode
\leavevmode
\vspace{6.4cm}
\includegraphics{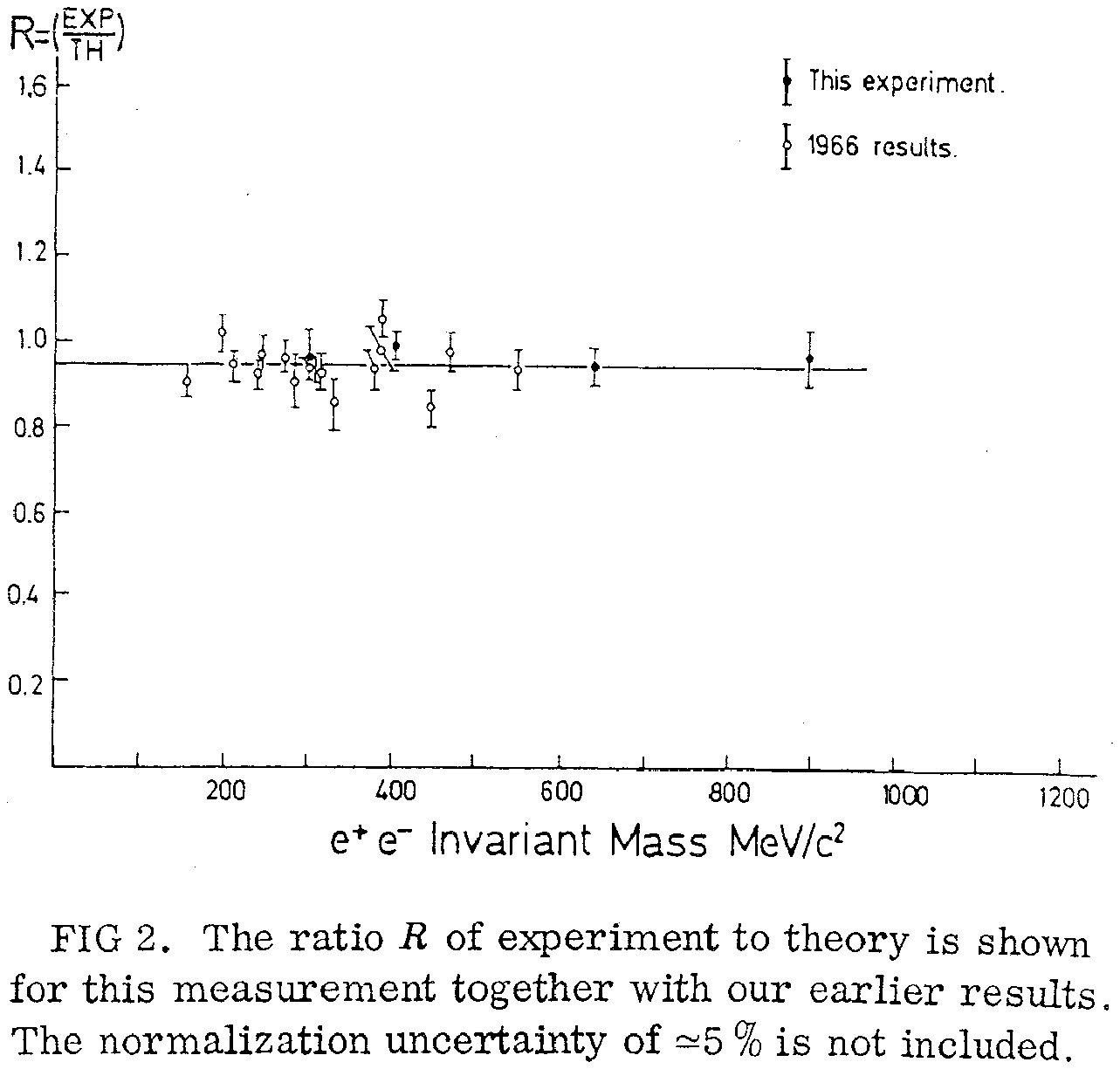}
\end{center}
\caption{Ratio of experiment 
to theory of the wide-angle $e^+e^-$ pair-creation cross section by $\gamma$ scattering off the 
field of a carbon nucleus as a function of the invariant mass $M$ of 
the lepton pair. Plot taken from \protect\cite{Alvensleben1968}.\label{Fig45}}      
\end{figure}

Points (i), (ii), (iv), and (v) are undisputed facts. To see that there is some truth to point (iii) 
we present experimental results. In Fig.\,\ref{Fig45} a plot of the ratio of experiment 
to theory of the wide-angle $e^+e^-$ pair-creation cross section through $\gamma$ scattering off of the 
field of a carbon nucleus is shown as a function 
of the invariant mass $M$ of the created lepton pair.  
Notice the substantial deviation from unity for 
$M\sim 2\,m_\mu\cdots 4\,m_\mu\sim (210\cdots 420)$\,MeV. Notice also that for $M>500\,$MeV theory 
and experiment do agree. A much more drastic deviation within 
the same kinematic regime was seen earlier in \cite{Blumenthal1966}. This, however, was not 
confirmed in \cite{Alvensleben1968}. 

\begin{figure}
\begin{center}
\leavevmode
\leavevmode
\vspace{5.4cm}
\includegraphics{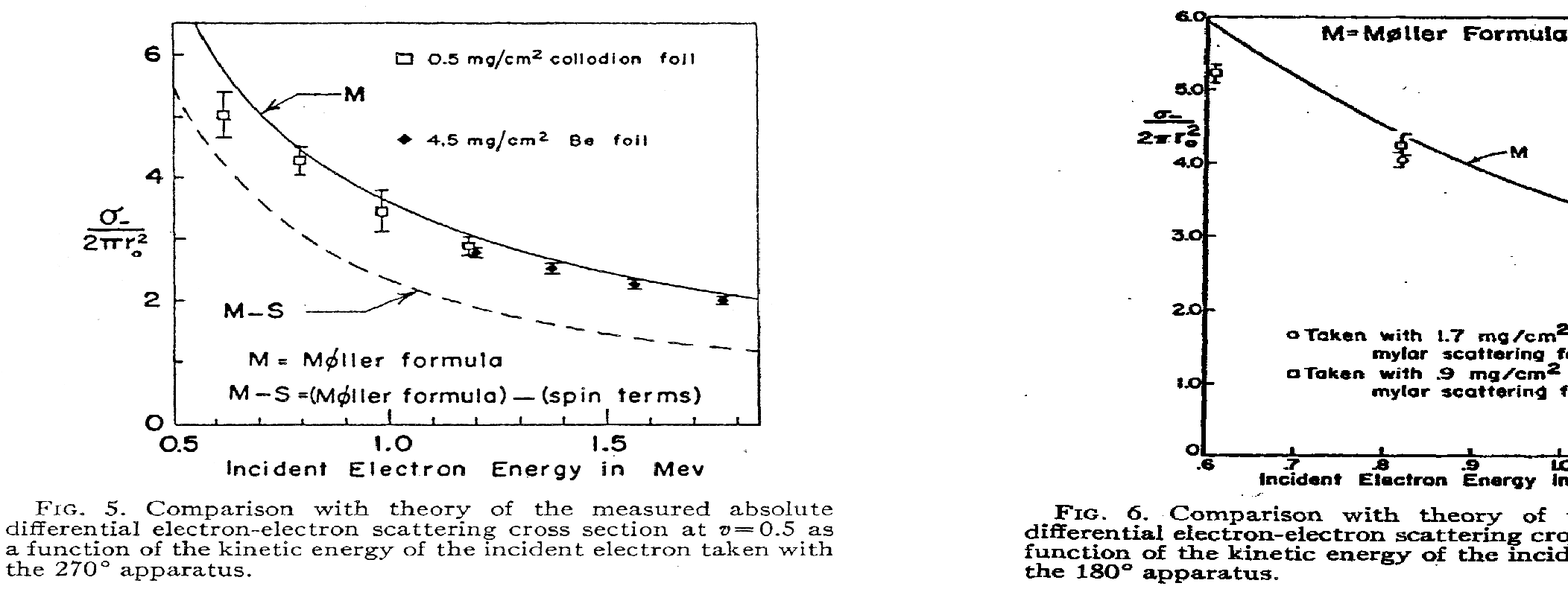}
\end{center}
\caption{M\o ller scattering of electrons, for an explanation see text. Taken from \protect\cite{Ashkin1953}.\label{Fig50}}      
\end{figure}

In Fig.\,\ref{Fig50} 
a plot of the differential cross section for $e^-e^-$ scattering (electron incident on an atomic 
electron of a target) at a fixed fraction $\nu=0.5$ of the incident kinetic energy $E_{\tiny\mbox{kin}}$ 
transferred in the collision is shown for $0.6\,\mbox{MeV}\le E_{\tiny\mbox{kin}}\le 1.7\,\mbox{MeV}$ taken with 
a $270^o$ apparatus (left panel) and for $0.6\,\mbox{MeV}\le E_{\tiny\mbox{kin}}\le 1.2\,\mbox{MeV}$ taken with 
a $180^o$ apparatus (right panel) \cite{Ashkin1953}. The solid line indicates the theoretical result 
obtained by using the M\o ller formula.   
Notice the agreement with QED for large values of $E_{\tiny\mbox{kin}}$ (in particular in the 
left panel of Fig.\,\ref{Fig50}). In \cite{Scott1951} the differential cross 
section for M\o ller scattering was measured for $E_{\tiny\mbox{kin}}=15.7\,$MeV 
as a function of the scattering angle and found to be in agreement with QED on the 
0.4\% error level. The disagreement at the lowest 
value $E_{\tiny\mbox{kin}}=0.6\,$MeV in Fig.\,\ref{Fig50}, which corresponds to a center-of-mass 
energy of about $2.5\,m_e$, is conspicuous. 

If the Yang-Mills scales of the factors SU(2)$_e$, SU(2)$_\mu$, and SU(2)$_\tau$ are about $m_e$, 
$m_\mu$, and $m_\tau$, respectively, then the masses 
of the charge-one states in the center phase of each theory, see Fig.\,\ref{intersect}, are determined to 
be these values. Since $T_{\tiny\mbox{CMB}}\ll m_e, m_\mu, m_\tau$ these theories 
are in their center phases. By looking at Fig.\,\ref{intersect} a $g$-factor of two 
is imperative by the asignment of angular momentum one half in the presence of one unit of (electric) center 
flux in the vortex loop. The latter generates the lowest nonvanishing quantum of 
magnetic moment (Bohr magneton). Notice that Fig.\,\ref{intersect} provides for an 
intuitive manifestation of the concept of spin-1/2: Inside the intersection core the center flux 
is diverted to the right above and to the left below such that an eddy is generated. The latter carries the electric 
charge of the soliton. 

Let us now give some qualitative arguments why the gauge group 
SU(2)$_{\tiny\mbox{CMB}}\times$SU(2)$_e\times$SU(2)$_\mu\times$SU(2)$_\tau$ 
together with a Planck-scale axion may be a viable candidate to describe the phenomenology 
of electroweak interactions. Why do the photon (SU(2)$_{\tiny\mbox{CMB}}$) and the massive intermediate 
vector bosons (magnetic and electric phase of SU(2)$_e$) couple 
to the charge and/or the magnetic moment of charge-one and charge-zero states? 
The answer to this question is rooted in the symmetry 
breakdown at $T\sim M_P$ where the factors SU(2)$_{\tiny\mbox{CMB}}$, 
SU(2)$_e$, SU(2)$_\mu$, and SU(2)$_\tau$ were generated out 
of one large gauge group whose gauge bosons where mixing. (In the SM the mixing of the 'photon' of U(1)$_Y$ with that 
of SU(2)$_W$ is parametrized by the Weinberg 
angle $\theta$ with $\sin^2\theta=0.23$.) The interaction of the photon of 
SU(2)$_{\tiny\mbox{CMB}}$ with the electrically charged soliton of, say, 
SU(2)$_e$ thus is furnished by an adiabatic rotation into the 
(massive) photon of SU(2)$_e$ when approaching the charge of the latter and 
an adiabatic back-rotation into the photon of 
SU(2)$_{\tiny\mbox{CMB}}$ after the interaction has taken place. Where are the higher 
charge states? These states are instable by repulsion mediated by the photon of 
SU(2)$_{\tiny\mbox{CMB}}$, and thus they are very broad. The density of these states, however, 
is over-exponentially rising. Why do we only see the structure of a charged lepton 
in scattering experiments with a center-of-mass energy $\sqrt{s}$ comparable to 
the mass of the lepton, see Figs.\,\ref{Fig45},\ref{Fig50}? Radial excitations of a 't 
Hooft monopole have been investigated in \cite{ForgasVolkov2003}. The first 
excited level is comparable to twice the mass of the monopole ground state. 
This must semi-quantitatively also hold for a $Z_2$ monopole (self-intersection of a center-vortex loop). 
For $\sqrt{s}\ll m_e, m_\mu, m_\tau$ the $Z_2$ monopole is not excitable, a QED 
point-particle description holds. For $m_Z\gg\sqrt{s}\gg m_e, m_\mu, m_\tau$, 
where $m_Z\sim 90\,$GeV is the mass 
of the $Z$ boson, the energy deposited 
into the vertex is converted into a large entropy carried by the Hagedorn spectrum 
of instable states. The latter protect the $Z_2$ monopole against radial excitations, a QED 
point-particle description again holds. For $\sqrt{s}\sim 2\,m_e, 2\,m_\mu, 2\,m_\tau$ the $Z_2$ 
monopole is excited radially: A QED point-particle description fails. 
For $\sqrt{s}\gg m_Z$ the Hagedorn phase boundary of SU(2)$_e$ is locally 
overcome (the $Z$ boson is interpreted as the decoupled dual gauge mode on the magnetic 
side of the magnetic-center phase boundary, the $W^\pm$ bosons as the decoupled gauge modes on 
the electric side of the electric-magnetic phase boundary): The multiplicity of final states should be in stark 
contradiction to the SM prediction (we expect charge 
nonconservation in such processes). Where does the parity violation come from? This is an intermediate 
consequence of the existence of a Planck-scale axion. What is the nature of the charge-zero state 
(single center-vortex loop)? The mass of this state for SU(2)$_e$ 
is roughly given by $m_\nu\sim \frac{m_e}{g_{dec}}$, compare with Eq.\,(\ref{EANOvortex}). 
Moreover, the mass of the $Z$ boson is given as 
$m_Z\sim g_{dec}\,m_e$. From the experimentally known values $m_e\sim 5\times 10^5$\,eV and 
$m_Z\sim 9\times 10^{10}$\,eV we thus have $g_{dec}\sim \frac{9}{5}\times 10^5$ and therefore 
$m_\nu\sim \frac{25}{9}\,$eV. This is close to the upper bound for the mass of the 
electron neutrino obtained from a tritium $\beta$ decay 
experiment: $m_\nu<2.3\,$eV \cite{Krauss2004}. Thus a single 
center-vortex loop of SU(2)$_e$ viably is a candidate for the electron neutrino. Notice that 
this soliton has no antiparticle: Neutrinos need to be of 
the Majorana type in accord with the successful search for 
neutrinoless double beta decay \cite{Klapdor2004}. A similar 
situation holds for SU(2)$_\mu$ and SU(2)$_\tau$. We expect the masses of their 
intermediary vector bosons $m_{Z^\prime}, m_{W^{\prime,\pm}}$ and 
$m_{Z^{\prime\prime}}, m_{W^{\prime\prime,\pm}}$ to scale with 
their Yang-Mills scales $m_\mu$ and $m_\tau$ and large values of the 
gauge couplings at the respective phase boundaries. 
Thus there are very weak and very, very weak interactions in addition 
to the weak interactions which, however, will be very hard to detect experimentally. 
(To detect, say, $m_{Z^\prime}$ directly would need a 
center-of-mass energy in $e^+e^-$ annihilation which should at least be 
two-hundred times $m_Z$.) A remark concerning point (v) is in order: 
Since SU(2)$_e \times$ SU(2)$_\mu\times$ SU(2)$_{\tau}$ are in their 
center phases at present their contribution 
to the ground-state energy density and pressure of the Universe is nil. Above, we have computed 
the contribution arising from SU(2)$_{\tiny\mbox{CMB}}$ when assuming the 
Planck-scale axion to be absent.

Let us make a short remark on the solar wind. This particle flux is mainly composed 
of protons (about $3\times10^{43}$ protons depart 
annually from the solar surface \cite{Manuel2004}). If the conservation of 
electric charge, which is a built-in feature of the SM, 
would hold then the sun would continuously acquire 
negative charge: A disastrous implication 
for earth's orbit would arise. The problem is resolved by the 
observation that in the solar core temperatures are 
greater than $m_e$. Electronic charge, however, is absent in the magnetic 
or electric phase of SU(2)$_e$. According to the phase diagram of 
SU(2)$_e$ the solar core contains a superconducting mantel 
(magnetic phase, Bose condensate of electric monopole-antimonopole pairs) 
whose negative pressure together 
with gravity balances the positive thermal pressure of the 
innermost core (electric phase) where fusion takes place. 
Within the core region there is a clear dominance of 
positive charge which the sun deposes off by means 
of the solar wind. A superconducting core of the sun is also 
demanded in \cite{Manuel2004} for other reasons.   

We conclude this section by stressing that no fundamental Higgs field is needed 
to break the weak symmetry SU(2)$_W$ or SU(2)$_e$. In contrast to the SM, 
where this symmetry breaking is complete by a nonvanishing expectation of a 
fundamental Higgs field, the dynamical breakdown of SU(2)$_e$ proceeds in a two-stage, Higgs-particle free 
way: SU(2)$_e\to$U(1) (electric phase; adjoint nonfluctuating Higgs field) 
and U(1)$_D\to 1$ (magnetic phase; complex nonfluctuating Higgs field). Moreover, 
the electron and its neutrino are stable solitons in the center phase of SU(2)$_e$.

\subsection{Quantum chromodynamics: Quark confinement and fractional quantum Hall 
effect\label{QCD}}

In this section we pursue an admittedly speculative and not very matured idea 
about the nature of quarks and their interactions. 

Quantum chromodynamics (QCD) is an integral part of the SM. QCD is the gauge 
theory of strong interactions: Pointlike current quarks, which are spin-1/2 fermions of to-be-measured 
masses, are fundamentally charged under the gauge group SU(3)$_C$ and 
interact by the exchanges of massless gluons, 
the gauge bosons of SU(3)$_C$. The latter interact with one another according 
to a pure Yang-Mills action. The electric charges of 
quarks are 2/3 or $-1/3$. Let us only discuss the three quark flavors 
of lowest mass $m_u=(3\cdots 5)\,$MeV (charge 2/3), $m_d=(5\cdots 7)\,$MeV (charge $-1/3$), 
and $m_s=(100\cdots 140)\sim \,$MeV (charge $-1/3$) (all $\overline{\mbox{MS}}$ scheme, results depend on 
the renormalization point). 

Since leptons are likely 
to be the stable solitons in the center phases of pure SU(2) Yang-Mills theories 
it is tempting to speculate that quarks are related to the 
charge-one solitons (center-vortex loops with one self-intersection, 
spin-1/2 fermions) in the center phases of 
pure SU(3) Yang-Mills theories. If we assign an SU(3)$_u$ and SU(3)$_d$ theory of Yang-Mills 
scale $\sim m_u$ and $m_d$ to the quark flavors $u$ and $d$, respectively, then we need 
to understand why these quarks are confined and why the 
electric charge appears to be $2/3$ or $-1/3$. 
A plausible way of generating quark confinement would be to 
add an additional SU(3) Yang-Mills theory of scale, say $\Lambda=140\,$MeV 
which, however, is a magnetic dual to the other SU(3) theories. 
A center-vortex condensate of this theory constrains 
the (color)electric flux between a $u$ or $d$ quark and a $u$ or $d$ 
antiquark into a tube and thus confines. The additional SU(3) 
theory also has charge-one solitons in its center phase 
which are confined by the center-vortex condensates of the 
theories SU(3)$_u$ and SU(3)$_d$. 
It is tempting to interpret these solitons as 
strange quarks $s$ and thus to invoke 
the label SU(3)$_s$. 

What about the electric quark charges? Due to confinement 
the trajectories of each quark flavor are forced 
onto a more or less two-dimensional 
spherical surface if the ground state 
of a given hadron is considered. A flux dual 
to the flux in the confining tube is readily available in the 
center-vortex condensate being responsible for confinement. This is the set-up for the occurrence of the 
fractional quantum Hall effect: Quarks that would be integer 
charged spin-1/2 fermions in the absence of the dual fluxes form 
bound states with these fluxes. Bound states with three dual flux 
quanta are bosons, and thus they condense. Excitations above this condensate 
have fractional electric charge. For a thorough discussion 
of this phenomenon, see \cite{Laughlin1998}. 

Again, all of what was said in this section is preliminary. We would like to stress though that 
the above scenario has the potential to explain why the equation of state in hydrodynamical simulations 
of the elliptic flow measured in ultra-relativistic heavy-ion collisions at RHIC seems to be 
so close to the free-gas limit despite the fact 
that strong correlations thermalize the system very rapidly 
\cite{Shuryak2004}. (At $T_c\sim 170\,$MeV the two theories SU(3)$_u$ and SU(3)$_d$ are deep 
inside their electric phases while SU(3)$_s$ is just above the 
electric-magnetic transition.)

\section{Conclusions\label{CO}}

We have developed a nonperturbative approach to SU(2) and SU(3) Yang-Mills 
thermodynamics. The formation of a macroscopic, adjoint, and nonfluctuating 
Higgs field in the deconfining (electric) phase of each theory, which involves the (admissible part of the) 
moduli space of a caloron-anticaloron system, the Bose 
condensation of thermalized magnetic monopole-antimonopole systems into a macroscopic, 
nonfluctuating complex field in a preconfining (magnetic) phase, and the 
nonthermal condensation of systems composed of a center-vortex and an 
anti-center-vortex in a confining (center) phase were shown. A change of 
the statistics of the excitations from bosonic to fermionic was observed across 
the last phase transition. The degeneracy of the ground state with 
respect to a (global) electric $Z_2$ (SU(2)) and $Z_3$ (SU(3)) 
symmetry was observed in the electric phase, and the uniqueness 
of the ground state with respect to these symmetries was derived in 
the magnetic phase. Moreover, the nature of the phase transitions, 
electric-magnetic and magnetic-center, was clarified. The evolution of 
thermodynamical quantities with temperature 
was computed within the electric phase and the magnetic phase, and the 
density of fermionic states was estimated in the center phase.    

It did not escape the author's attention that the results obtained may resolve a number of 
long-standing problems in particle physics and cosmology.

\section*{Acknowledgments}
The author would like to thank B. Garbrecht, H. Gies, Th. Konstandin, T. Prokopec, H. Rothe, K. Rothe, 
M. Schmidt, I.-O. Stamatescu, and W. Wetzel for very helpful, continuing discussions. 
Important support for numerical calculations was provided by J. Rohrer and is 
thankfully acknowledged. The author acknowledges vivid discussion with Francesco Giacosa 
and Markus Schwarz. In particular, I am grateful to Francesco Giacosa for pointing out the need for 
a modification of the evolution equations for the effective gauge couplings which 
was overlooked in the pervious version. Very useful discussions with P. van Baal, 
E. Gubankova, J. Moffat, J. Polonyi, D. Rischke, and F. Wilczek and 
illuminating conversations with D. B\"odeker, R. Brandenberger, G. Dunne, Ph. de Forcrand, A. Guth, 
F. Karsch, A. Kovner, M. Laine, H. Liu, C. Nunez, J. Pawlowski, R. D. Pisarski, K. Rajagopal, K. Redlich, E. Shuryak, 
D. T. Son, A. Vainshtein, J. Verbaarschot, and F. Wilczek are gratefully acknowledged. 
A very useful correspondence with O. Manuel 
on solar models is thankfully acknowledged. The warm hospitality of the Center for 
Theoretical Physics at M.I.T, where part of this research was carried out 
(sponsored by Deutsche Forschungsgemeinschaft), is thankfully acknowledged.\\  
This paper is dedicated to my family and in particular to my 
wife Karin Thier. Thank you for your continuing encouragement, your persistent 
moral support, and your unconditional love.

\bibliographystyle{prsty}

\end{document}